\documentclass[journal,comsoc]{IEEEtran}

\usepackage{amssymb,cite}
\usepackage{amsmath,multirow,threeparttable,graphicx,times,subfigure,url,verbatim,color}
\usepackage{rotating}
\usepackage[normalem]{ulem}
\usepackage{booktabs}
\usepackage{array}
\usepackage{hyperref}
\usepackage{adjustbox}
\hypersetup{hidelinks}
\newcommand{\tabincell}[2]{\begin{tabular}{@{}#1@{}}#2\end{tabular}}

\definecolor{red}{rgb}{0,0,0} 

\begin{document}

\title{Convergence of Edge Computing and Deep Learning: A Comprehensive Survey}

\author{
    Xiaofei Wang,~\IEEEmembership{Senior Member,~IEEE,}
    Yiwen Han,~\IEEEmembership{Student Member,~IEEE,}
    Victor C.M. Leung,~\IEEEmembership{Fellow,~IEEE,}
    Dusit Niyato,~\IEEEmembership{Fellow,~IEEE,}
    Xueqiang Yan,
    Xu Chen,~\IEEEmembership{Member,~IEEE}
    \thanks{Xiaofei Wang and Yiwen Han are with the College of Intelligence and Computing, Tianjin University, Tianjin, China. E-mails: xiaofeiwang@tju.edu.cn, hanyiwen@tju.edu.cn.}
    \thanks{V. C. M. Leung is with the College of Computer Science and Software Engineering, Shenzhen University, Shenzhen, China, and also with the Department of Electrical and Computer Engineering, the University of British Columbia, Vancouver, Canada. E-mail: vleung@ieee.org.}
    \thanks{Dusit Niyato is with School of Computer Science and Engineering, Nanyang Technological University, Singapore. E-mail: dniyato@ntu.edu.sg.}
    \thanks{Xueqiang Yan is with 2012 Lab of Huawei Technologies, Shenzhen, China. Email: yanxueqiang1@huawei.com.}
    \thanks{Xu Chen is with the School of Data and Computer Science, Sun Yat-sen University, Guangzhou, China. E-mail: chenxu35@mail.sysu.edu.cn.}
    \thanks{Corresponding author: Yiwen Han (hanyiwen@tju.edu.cn)}
}

\markboth{To be appeared in IEEE Communications Surveys \& Tutorials}{}
\maketitle

\begin{abstract}

Ubiquitous sensors and smart devices from factories and communities are generating massive amounts of data, and ever-increasing computing power is driving the core of computation and services from the cloud to the edge of the network. 
As an important enabler broadly changing people's lives, from face recognition to ambitious smart factories and cities, {\color{red}developments of artificial intelligence (especially deep learning, DL) based applications and services are thriving.} 
However, due to efficiency and latency issues, the current cloud computing service architecture hinders the vision of ``providing artificial intelligence for every person and every organization at everywhere''. 
{\color{red}Thus, unleashing DL services using resources at the network edge near the data sources has emerged as a desirable solution.}
Therefore, \textit{edge intelligence}, aiming to facilitate the deployment of DL services by edge computing, has received significant attention. 
In addition, DL, as the representative technique of artificial intelligence, can be integrated into edge computing frameworks to build \textit{intelligent edge} for dynamic, adaptive edge maintenance and management. 
With regard to mutually {\color{red}beneficial} \textit{edge intelligence} and \textit{intelligent edge}, this paper introduces and discusses: 
1) the application scenarios of both; 
2) the practical implementation methods and enabling technologies, namely DL training and inference in the customized edge computing framework; 
3) challenges and future trends of more pervasive and fine-grained intelligence. 
{\color{red}We believe that by consolidating information scattered across the communication, networking, and DL areas, this survey can help readers to} understand the connections between enabling technologies while promoting further discussions on the fusion of \textit{edge intelligence} and \textit{intelligent edge}, i.e., Edge DL.

\end{abstract}

\begin{keywords}
Edge computing, deep learning, wireless communication, computation offloading, artificial intelligence 
\end{keywords}

\section{Introduction}
\label{sec:introduction}

With the proliferation of computing and storage devices, from server clusters in cloud data centers (the cloud) to personal computers and smartphones, further, to wearable and other Internet of Things (IoT) devices, we are now in an information-centric era in which computing is ubiquitous and computation services are overflowing from the cloud to the edge. According to a Cisco white paper \cite{ciscofcandiots}, $50$ billion IoT devices will be connected to the Internet by 2020. On the other hand, Cisco estimates that nearly $850$ Zettabytes (ZB) of data will be generated each year outside the cloud by 2021, while global data center traffic is only $20.6$ ZB \cite{ciscoglobalcloudindex}. This indicates that data sources for big data are also undergoing a transformation: from large-scale cloud data centers to an increasingly wide range of edge devices. However, existing cloud computing is gradually unable to manage these massively distributed computing power and analyze their data: 1) a large number of computation tasks need to be delivered to the cloud for processing \cite{Barbera2013}, which undoubtedly poses serious challenges on network capacity and the computing power of cloud computing infrastructures; 2) many new types of applications, e.g., cooperative autonomous driving, have strict or tight delay requirements that the cloud would have difficulty meeting since it may be far away from the users \cite{Hu}.

Therefore, edge computing \cite{ETSI2, Shi2016} emerges as an attractive alternative, especially to host computation tasks as close as possible to the data sources and end users. Certainly, edge computing and cloud computing are not mutually exclusive \cite{Mudassar2018, Yousefpour2018a}. Instead, the edge complements and extends the cloud. Compared with cloud computing only, the main advantages of edge computing combined with cloud computing are three folds: 1) \textbf{backbone network alleviation}, distributed edge computing nodes can handle a large number of computation tasks without exchanging the corresponding data with the cloud, thus alleviating the traffic load of the network; 2) \textbf{agile service response}, services hosted at the edge can significantly reduce the delay of data transmissions and improve the response speed; 3) \textbf{powerful cloud backup}, the cloud can provide powerful processing capabilities and massive storage when the edge cannot afford.


As a typical and more widely used new form of applications \cite{Redmon2016}, various deep learning-based intelligent services and applications have changed many aspects of people's lives due to the great advantages of Deep Learning (DL) in the fields of Computer Vision (CV) and Natural Language Processing (NLP) \cite{Schmidhuber2015}. These achievements are not only derived from the evolution of DL but also inextricably linked to increasing data and computing power. Nevertheless, for a wider range of application scenarios, such as smart cities, Internet of Vehicles (IoVs), etc., there are only a limited number of intelligent services offered due to the following factors. 
\begin{itemize}
	\item \textbf{\textit{Cost}}: training and inference of DL models in the cloud requires devices or users to transmit massive amounts of data to the cloud, thus consuming a large amount of network bandwidth;
	\item \textbf{\textit{Latency}}: the delay to access cloud services is generally not guaranteed and might not be short enough to satisfy the requirements of many time-critical applications such as cooperative autonomous driving \cite{Khelifi2018};
	\item \textbf{\textit{Reliability}}: most cloud computing applications relies on wireless communications and backbone networks for connecting users to services, but for many industrial scenarios, intelligent services must be highly reliable, even when network connections are lost;
	\item \textbf{\textit{Privacy}}: the data required for DL might carry a lot of private information, and privacy issues are critical to areas such as smart home and cities.
\end{itemize}
 
\begin{table*}[htbp!!!!!!!!!]
  \centering
  \scriptsize
  \caption{{\color{red}List of Important Abbreviations in Alphabetical Order}}
    \begin{tabular}{clclcl}
    \toprule
    \textbf{Abbr.} & \multicolumn{1}{c}{\textbf{Definition}} & \textbf{Abbr.} & \multicolumn{1}{c}{\textbf{Definition}} & \textbf{Abbr.} & \multicolumn{1}{c}{\textbf{Definition}} \\
    \midrule
    A-LSH & Adaptive Locality Sensitive Hashing & DVFS  & Dynamic Voltage and Frequency Scaling & NLP   & Natural Language Processing \\
    \midrule
    AC    & Actor-Critic & ECSP  & Edge Computing Service Provider & NN    & Neural Network \\
    \midrule
    A3C   & Asynchronous Advantage Actor-Critic & EEoI  & Early Exit of Inference & NPU   & Neural Processing Unit \\
    \midrule
    AE    & Auto-Encoder & EH    & Energy Harvesting & PPO   & Proximate Policy Optimization \\
    \midrule
    AI    & Artificial Intelligence & FAP   & Fog radio Access Point & QoE   & Quality of Experience \\
    \midrule
    APU   & AI Processing Unit & FCNN  & Fully Connected Neural Network & QoS   & Quality of Service \\
    \midrule
    AR    & Augmented Reality & FL    & Federated Learning & RAM   & Random Access Memory \\
    \midrule
    ASIC  & Application-Specific Integrated Circuit & FPGA  & Field Programmable Gate Array & RNN   & Recurrent Neural Network \\
    \midrule
    BS    & Base Station & FTP   & Fused Tile Partitioning & RoI   & Region-of-Interest \\
    \midrule
    C-RAN & Cloud-Radio Access Networks & GAN   & Generative Adversarial Network & RRH   & Remote Radio Head \\
    \midrule
    CDN   & Content Delivery Network & GNN   & Graph Neural Network & RSU   & Road-Side Unit \\
    \midrule
    CNN   & Convolutional Neural Network & IID   & Independent and Identically Distributed & SDN   & Software-Defined Network \\
    \midrule
    CV    & Computer Vision & IoT   & Internet of Things & SGD   & Stochastic Gradient Descent \\
    \midrule
    DAG   & Directed Acyclic Graph & IoV   & Internet of Vehicles & SINR  & Signal-to-Interference-plus-Noise Ratio \\
    \midrule
    D2D   & Device-to-Device & KD    & Knowledge Distillation & SNPE  & Snapdragon Neural Processing Engine \\
    \midrule
    DDoS  & Distributed Denial of Service & $k$NN & $k$-Nearest Neighbor & TL    & Transfer Learning \\
    \midrule
    DDPG  & Deep Deterministic Policy Gradient & MAB   & Multi-Armed Bandit & UE    & User Equipment \\
    \midrule
    DL    & Deep Learning & MEC   & Mobile (Multi-access) Edge Computing & VM   & Virtual Machine \\
    \midrule
    DNN   & Deep Neural Networks & MDC   & Micro Data Center & VNF    & Virtual Network Function \\
    \midrule
    DQL   & Deep Q-Learning & MDP   & Markov Decision Process & V2V   & Vehicle-to-Vehicle \\
    \midrule
    DRL   & Deep Reinforcement Learning & MLP   & Multi-Layer Perceptron & WLAN  & Wireless Local Area Network \\
    \midrule
    DSL   & Domain-specific Language & NFV   & Network Functions Virtualizatio & ZB    & Zettabytes \\
    \bottomrule
    \end{tabular}%
  \label{tab:abbreviation}%
\end{table*}%


\begin{figure}[!!!!!!!!!!!!!!hhhhhhhhhht]
    \centering
    \includegraphics[width=8.85 cm]{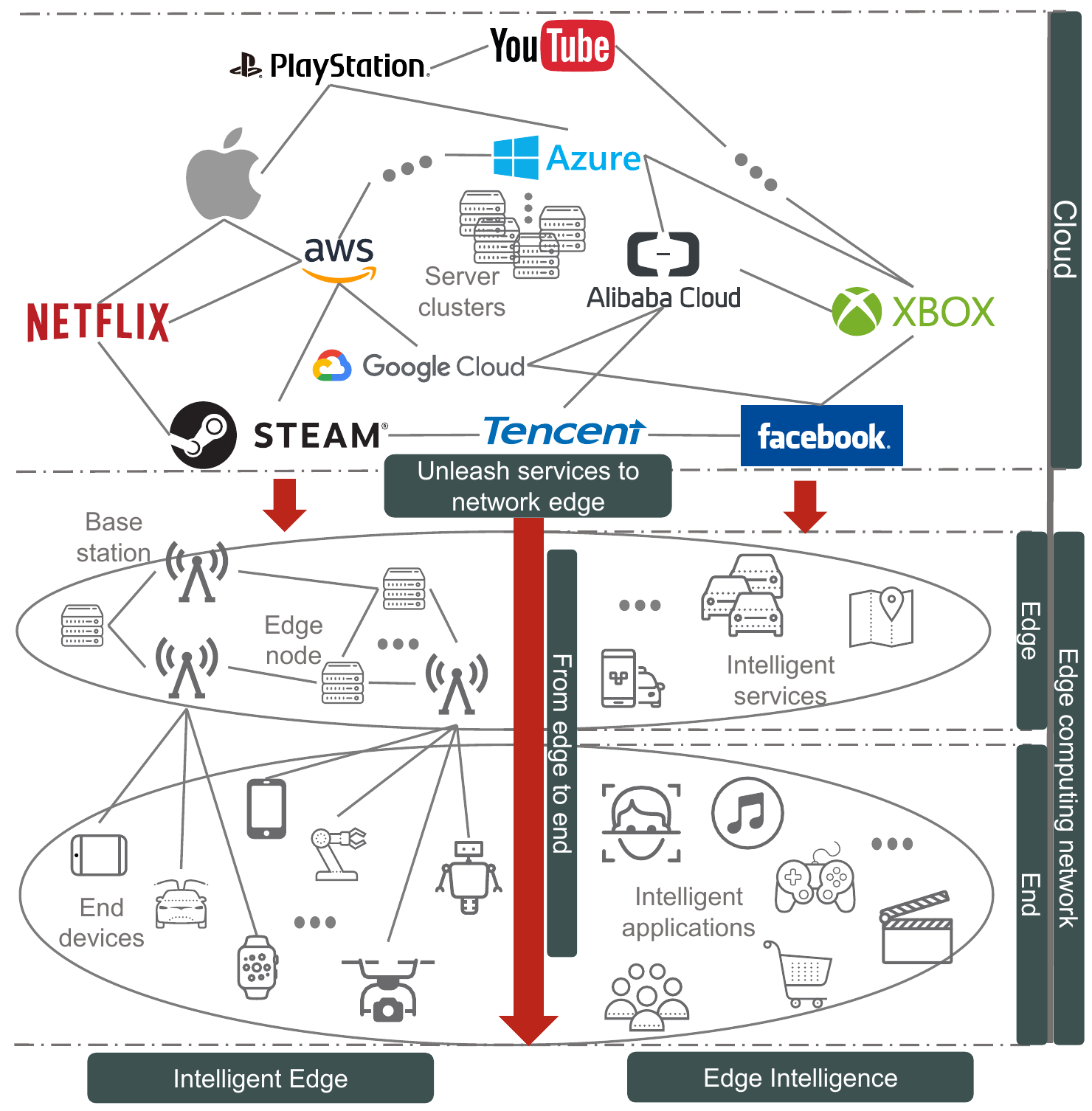}
    \caption{Edge intelligence and intelligent edge.}
    \label{fig:EIandIE}
\end{figure}

Since the edge is closer to users than the cloud, edge computing is expected to solve many of these issues. In fact, edge computing is gradually being combined with Artificial Intelligence (AI), benefiting each other in terms of the realization of \textit{edge intelligence} and \textit{intelligent edge} as depicted in Fig. \ref{fig:EIandIE}. Edge intelligence and intelligent edge are not independent of each other. Edge intelligence is the goal, and the DL services in intelligent edge are also a part of edge intelligence. In turn, intelligent edge can provide higher service throughput and resource utilization for edge intelligence.

\begin{figure}[!!!!!!!!!!!!!!hhhhhhhhhht]
    \centering
    \includegraphics[width=5.5 cm]{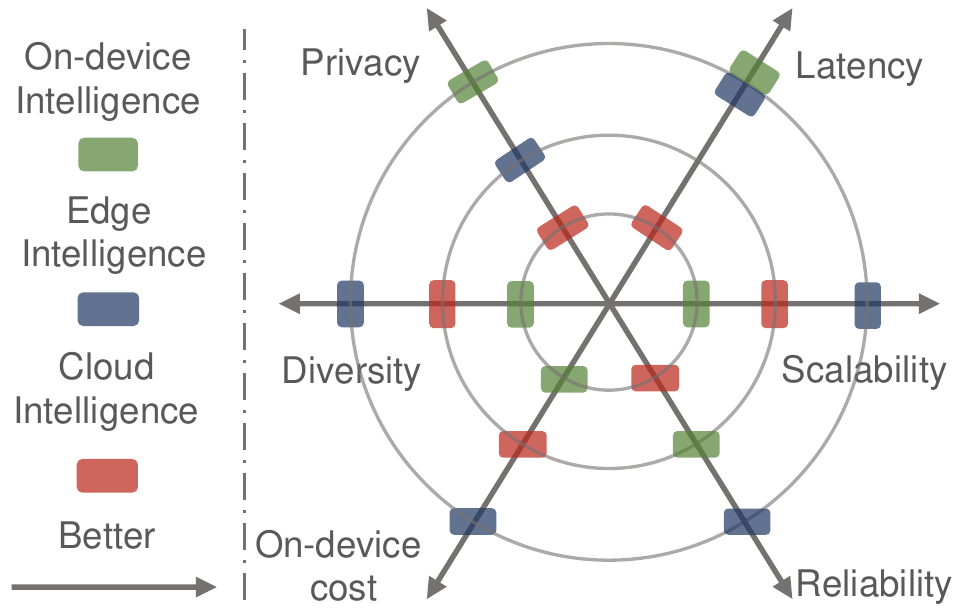}
    \caption{Capabilities comparison of cloud, on-device and edge intelligence.}
    \label{fig:CloudEdgeDLAbility}
\end{figure}

To be specific, on one hand, edge intelligence is expected to push DL computations from the cloud to the edge as much as possible, thus enabling various distributed, low-latency and reliable intelligent services. As shown in Fig. \ref{fig:CloudEdgeDLAbility}, the advantages include: 1) DL services are deployed close to the requesting users, and the cloud only participates when additional processing is required \cite{Neurosurgeon}, hence significantly reducing the latency and cost of sending data to the cloud for processing; 2) since the raw data required for DL services is stored locally on the edge or user devices themselves instead of the cloud, protection of user privacy is enhanced; 3) the hierarchical computing architecture provides more reliable DL computation; 4) with richer data and application scenarios, edge computing can promote the pervasive application of DL and realize the prospect of ``providing AI for every person and every organization at everywhere'' \cite{DemocratizingAI}; 5) diversified and valuable DL services can broaden the commercial value of edge computing and accelerate its deployment and growth.

On the other hand, intelligent edge aims to incorporate DL into the edge for dynamic, adaptive edge maintenance and management. With the development of communication technology, network access methods are becoming more diverse. At the same time, the edge computing infrastructure acts as an intermediate medium, making the connection between ubiquitous end devices and the cloud more reliable and persistent \cite{Yang2019a}. Thus the end devices, edge, and cloud are gradually merging into a community of shared resources. However, the maintenance and management of such a large and complex overall architecture (community) involving wireless communication, networking, computing, storage, etc., is a major challenge \cite{Li2018a}. Typical network optimization methodologies rely on fixed mathematical models; however, it is difficult to accurately model rapidly changing edge network environments and systems. DL is expected to deal with this problem: when faced with complex and cumbersome network information, DL can rely on its powerful learning and reasoning ability to extract valuable information from data and make adaptive decisions, achieving intelligent maintenance and management accordingly.

\begin{figure}[!!!!!!!!!!!!!!hhhhhhhhhht]
    \centering
    \includegraphics[width=8.85 cm]{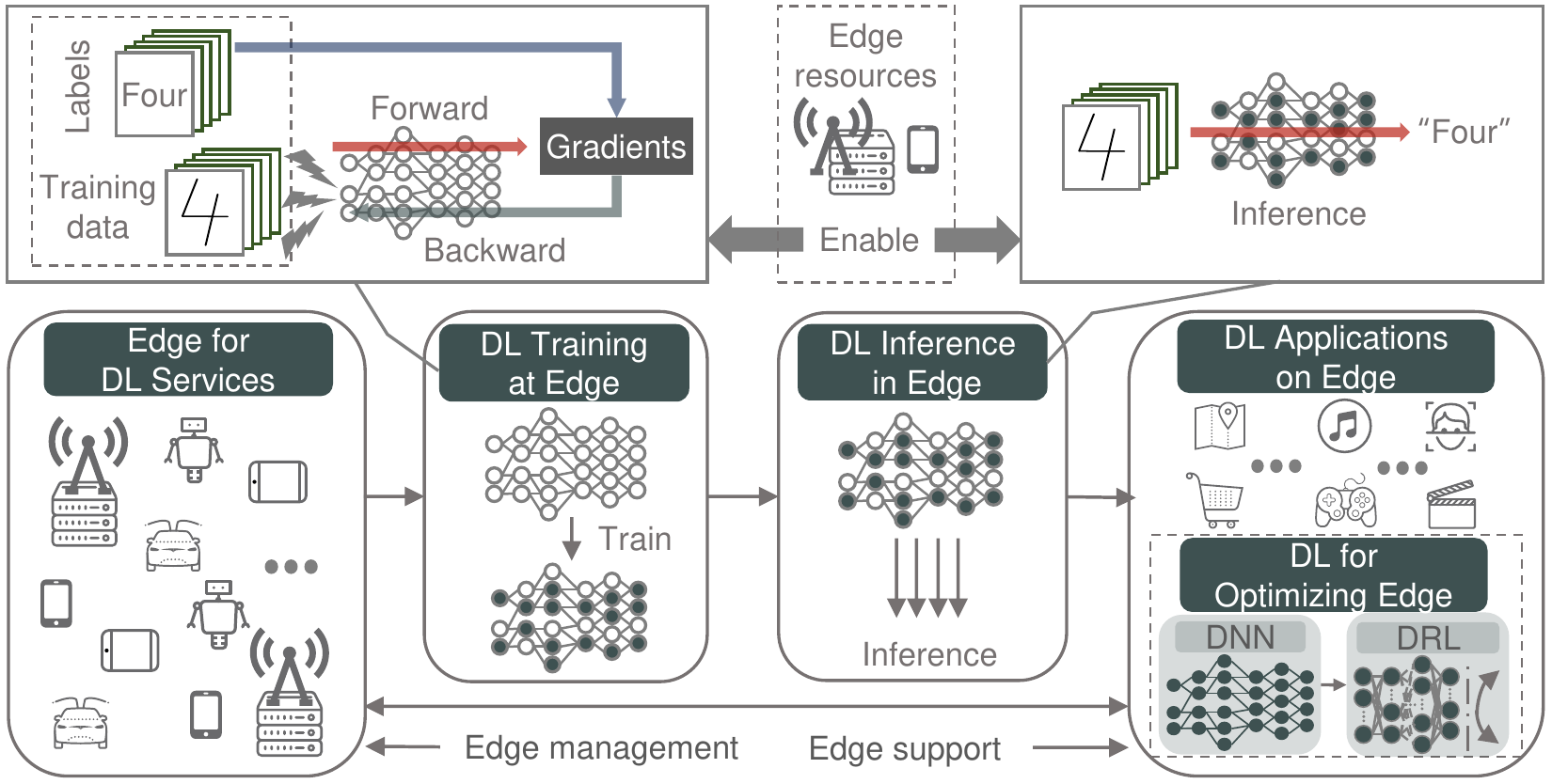}
    \caption{{\color{red}Landscape of Edge DL according to the proposed taxonomy.}}
    \label{fig:FiveEnablersAbstract}
\end{figure}

{\color{red}Therefore, considering that edge intelligence and intelligent edge, i.e., Edge DL, together face some of the same challenges and practical issues in multiple aspects, we identify the following five technologies that are essential for Edge DL: 
\begin{itemize}
	\item [1)] \textit{DL applications on Edge}, technical frameworks for systematically organizing edge computing and DL to provide intelligent services;
	\item [2)] \textit{DL inference in Edge}, focusing on the practical deployment and inference of DL in the edge computing architecture to fulfill different requirements, such as accuracy and latency;
	\item [3)] \textit{Edge computing for DL}, which adapts the edge computing platform in terms of network architecture, hardware and software to support DL computation;
	\item [4)] \textit{DL training at Edge}, training DL models for edge intelligence at distributed edge devices under resource and privacy constraints;
	\item [5)] \textit{DL for optimizing Edge}, application of DL for maintaining and managing different functions of edge computing networks (systems), e.g., edge caching \cite{Wang2017e}, computation offloading\cite{Tran2017a}.
\end{itemize}}
{\color{red}As illustrated in Fig. \ref{fig:FiveEnablersAbstract}, ``DL applications on Edge'' and ``DL for optimizing edge'' correspond to the theoretical goals of edge intelligence and intelligent edge, respectively. 
To support them, various DL models should be trained by intensive computation at first. 
In this case, for the related works leveraging edge computing resources to train various DL models, we classify them as ``DL training at Edge''. 
Second, to enable and speed up Edge DL services, we focus on a variety of techniques supporting the efficient inference of DL models in edge computing frameworks and networks, called ``DL inference in Edge''. 
At last, we classify all techniques, which adapts edge computing frameworks and networks to better serve Edge DL, as ``Edge computing for DL''. }

{\color{red}To the best of our knowledge, existing articles that are most related to our work include \cite{Park, Zhou2019, Chen2019a, Lim2019}.
Different from our more extensive coverage of Edge DL, \cite{Park} is focussed on the use of machine learning (rather than DL) in edge intelligence for wireless communication perspective, i.e., training machine learning at the network edge to improve wireless communication. 
Besides, discussions about DL inference and training are the main contribution of \cite{Zhou2019, Chen2019a, Lim2019}. 
Different from these works, this survey focuses on these respects: 
1) comprehensively consider deployment issues of DL by edge computing, spanning networking, communication, and computation; 
2) investigate the holistic technical spectrum about the convergence of DL and edge computing in terms of the five enablers; 
3) point out that DL and edge computing are beneficial to each other and considering only deploying DL on the edge is incomplete.}



\begin{figure}[!!!!!!!!!!!!!!hhhhhhhhhht]
    \centering
    \includegraphics[width=8.85 cm]{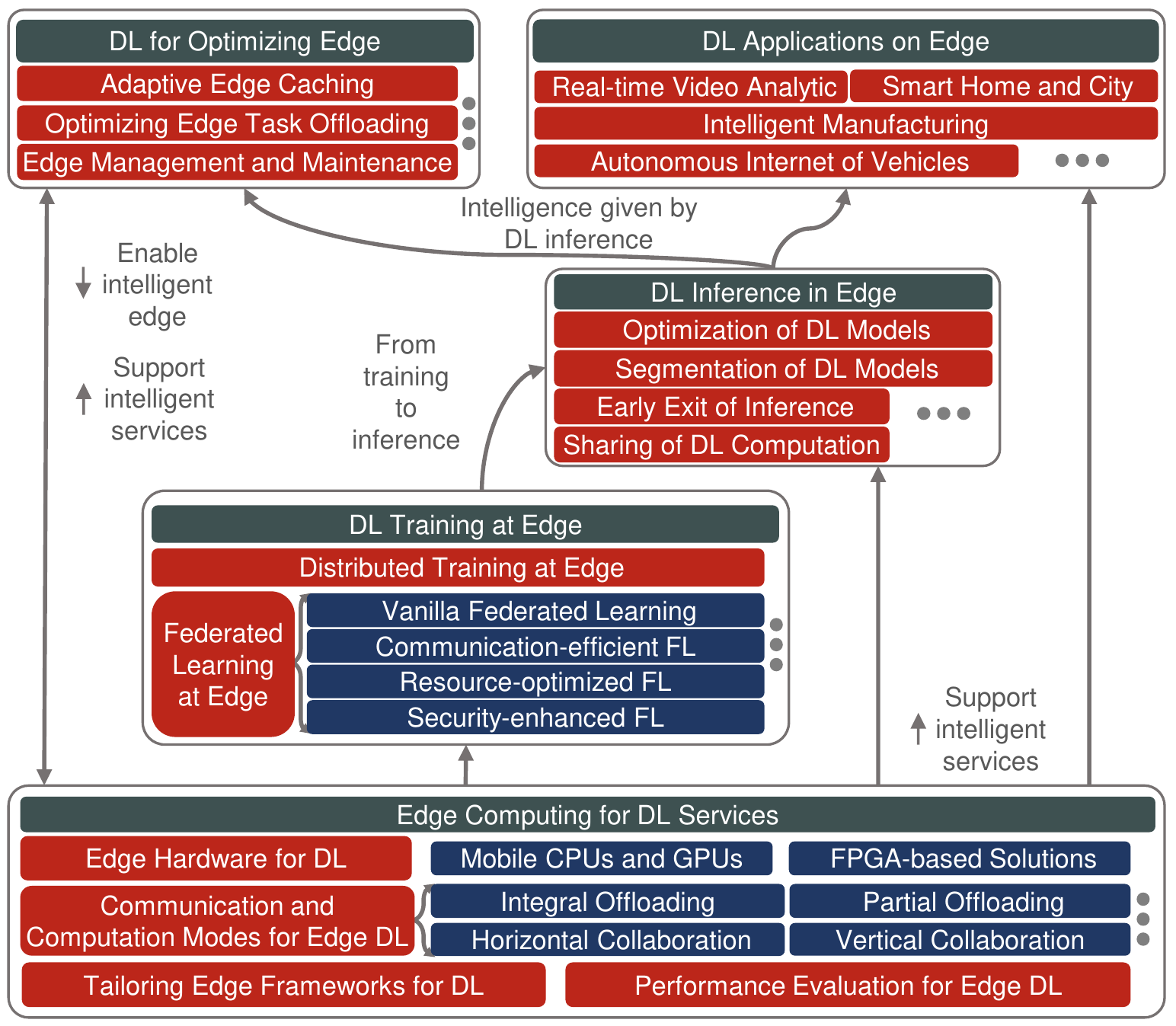}
    \caption{{\color{red}Conceptual relationships of edge intelligence and intelligent edge.}}
    \label{fig:FiveEnablers}
\end{figure}

This paper is organized as follows (as abstracted in Fig. \ref{fig:FiveEnablers}). We have given the background and motivations of this survey in the current section.  Next, we provide some fundamentals related to edge computing and DL in Section \ref{sec:edgefundamentals} and Section \ref{sec:dlfundamentals}, respectively. 
The following sections introduce the five enabling technologies, i.e., DL applications on edge (Section \ref{sec:AIonEdge}), DL inference in edge (Section \ref{sec:AIinEdge}), edge computing for DL services (Section \ref{sec:EdgeforAI}), DL training at edge (Section \ref{sec:AIatEdge}), and DL for optimizing edge (Section \ref{sec:AIforEdge}). 
Finally, we present lessons learned and discuss open challenges in Section \ref{sec:challenges} and conclude this paper in Section \ref{sec:conclusion}. 
All related acronyms are listed in Table \ref{tab:abbreviation}.

\section{Fundamentals of Edge Computing}
\label{sec:edgefundamentals}

Edge computing has become an important solution to break the bottleneck of emerging technologies by virtue of its advantages of reducing data transmission, improving service latency and easing cloud computing pressure. The edge computing architecture will become an important complement to the cloud, even replacing the role of the cloud in some scenarios. More detailed information can be found in \cite{Mouradian2018, Yousefpour2018a, Bilal2018}.

\subsection{Paradigms of Edge Computing}

In the development of edge computing, there have been various new technologies aimed at working at the edge of the network, with the same principles but different focuses, such as Cloudlet \cite{Satyanarayanan2009The}, Micro Data Centers (MDCs) \cite{7098039}, Fog Computing \cite{bonomi2012fog}\cite{Bonomi2014} and Mobile Edge Computing \cite{ETSI2} (viz., Multi-access Edge Computing \cite{ETSI3} now). However, the edge computing community has not yet reached a consensus on the standardized definitions, architectures and protocols of edge computing \cite{Bilal2018}. We use a common term ``edge computing'' for this set of emerging technologies. In this section, different edge computing concepts are introduced and differentiated. 

\subsubsection{Cloudlet and Micro Data Centers}

Cloudlet is a network architecture element that combines mobile computing and cloud computing. It represents the middle layer of the three-tier architecture, i.e., mobile devices, the micro cloud, and the cloud. Its highlights are efforts to 1) define the system and create algorithms that support low-latency edge cloud computing, and 2) implement related functionality in open source code as an extension of Open Stack cloud management software \cite{Satyanarayanan2009The}. Similar to Cloudlets, MDCs \cite{7098039} are also designed to complement the cloud. The idea is to package all the computing, storage, and networking equipment needed to run customer applications in one enclosure, as a stand-alone secure computing environment, for applications that require lower latency or end devices with limited battery life or computing abilities.

\subsubsection{Fog Computing}


One of the highlights of fog computing is that it assumes a fully distributed multi-tier cloud computing architecture with billions of devices and large-scale cloud data centers \cite{bonomi2012fog}\cite{Bonomi2014}. While cloud and fog paradigms share a similar set of services, such as computing, storage, and networking, the deployment of fog is targeted to specific geographic areas. In addition, fog is designed for applications that require real-time responding with less latency, such as interactive and IoT applications. Unlike Cloudlet, MDCs and MEC, fog computing is more focused on IoTs.

\subsubsection{Mobile (Multi-access) Edge Computing (MEC)}

Mobile Edge Computing places computing capabilities and service environments at the edge of cellular networks \cite{ETSI2}. It is designed to provide lower latency, context and location awareness, and higher bandwidth. Deploying edge servers on cellular Base Stations (BSs) allows users to deploy new applications and services flexibly and quickly. The European Telecommunications Standards Institute (ETSI) further extends the terminology of MEC from Mobile Edge Computing to Multi-access Edge Computing by accommodating more wireless communication technologies, such as Wi-Fi \cite{ETSI3}.


\subsubsection{Definition of Edge Computing Terminologies}


The definition and division of edge devices are ambiguous in most literature (the boundary between edge nodes and end devices is not clear). For this reason, as depicted in Fig. \ref{fig:EIandIE}, we further divide common edge devices into end devices and edge nodes: the ``end devices'' (end level) is used to refer to mobile edge devices (including smartphones, smart vehicles, etc.) and various IoT devices, and the ``edge nodes'' (edge level) include Cloudlets, Road-Side Units (RSUs), Fog nodes, edge servers, MEC servers and so on, namely servers deployed at the edge of the network.

\begin{figure}[!!!!!!!!!!!!!!hhhhhhhhhht]
    \centering
    \includegraphics[width=7.5 cm]{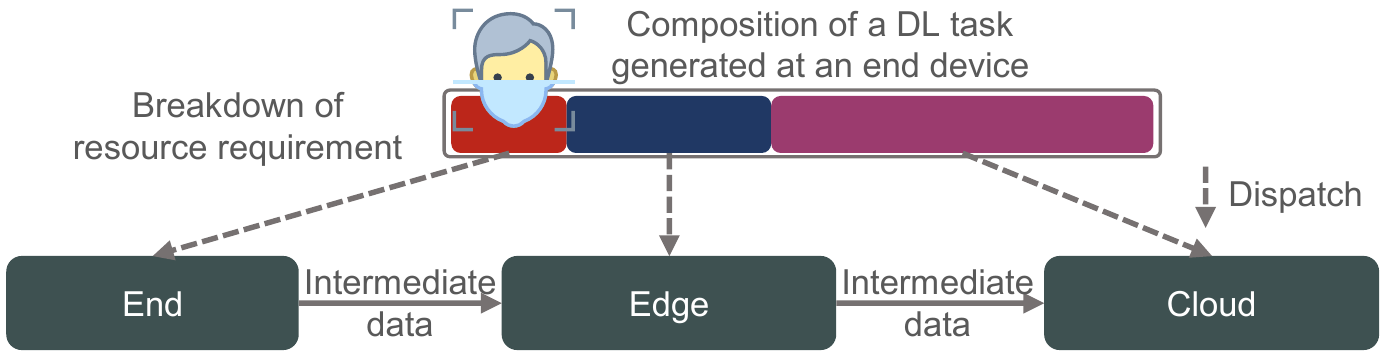}
    \caption{A sketch of collaborative end-edge-cloud DL computing.}
    \label{fig:EndEdgeCloudComputing}
\end{figure}

\begin{figure}[!!!!!!!!!!!!!!hhhhhhhhhht]
    \centering
    \includegraphics[width=8.85 cm]{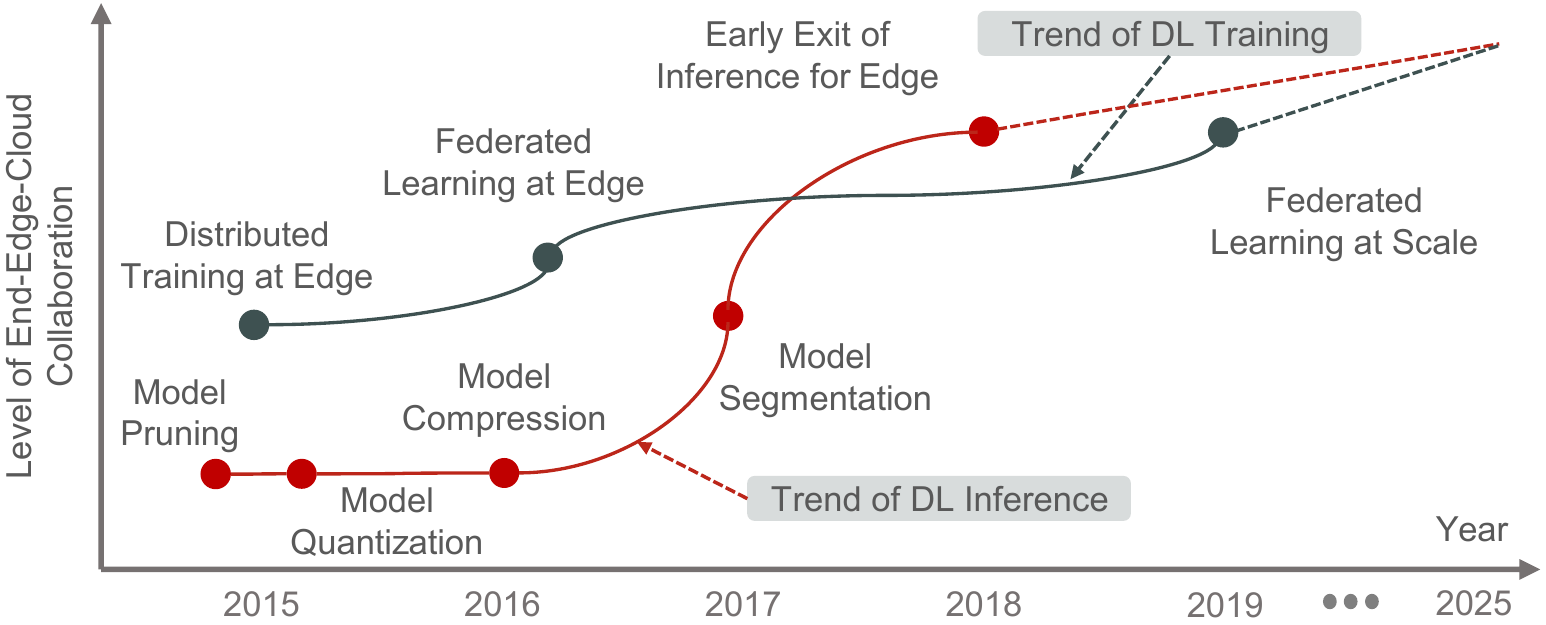}
    \caption{Computation collaboration is becoming more important for DL with respect to both training and inference.}
    \label{fig:CollaborationTrend}
\end{figure}

\begin{table*}[htbp]
    \centering
    \scriptsize
    \caption{Summary of Edge Computing AI Hardwares and Systems}
    \label{tab:HardwareSummary}%
      \begin{tabular}{cccl}
      \toprule
            & \textbf{Owner} & \textbf{Production} & \multicolumn{1}{c}{\textbf{Feature}} \\
      \midrule
      \multicolumn{1}{c}{\multirow{3}[1]{*}[-2em]{\textbf{\tabincell{c}{Integrated \\ Commodities}}}} & Microsoft & Data Box Edge \cite{databox} & Competitive in data preprocessing and data transmission \\
  \cmidrule{2-4}          & Intel & \tabincell{c}{Movidius Neural \\ Compute Stick \cite{Movidius}} & Prototype on any platform with plug-and-play simplicity \\
  \cmidrule{2-4}          & NVIDIA & Jetson \cite{jetsonnano} & Easy-to-use platforms that runs in as little as 5 Watts  \\
  \cmidrule{2-4}          & Huawei & Atlas Series \cite{atlas} & An all-scenario AI infrastructure solution that bridges ``device, edge, and cloud'' \\
      \midrule
      \multicolumn{1}{c}{\multirow{7}[1]{*}[-4em]{\textbf{\tabincell{c}{AI Hardware \\ for \\Edge Computing}}}} & Qualcomm  & Snapdragon 8 Series \cite{Snapdragon8} & Powerful adaptability to major DL frameworks \\
  \cmidrule{2-4}          & HiSilicon &  Kirin 600/900 Series \cite{hisilicon}  & Independent NPU for DL computation \\
  \cmidrule{2-4}          & HiSilicon &  Ascend Series \cite{ascend} & \tabincell{l}{Full coverage – from the ultimate low energy consumption scenario \\ to high computing power scenario} \\
  \cmidrule{2-4}          & MediaTek & Helio P60 \cite{mediatekp60} & Simultaneous use of GPU and NPU to accelerate neural network computing \\
  \cmidrule{2-4}          & NVIDIA  & Turing GPUs \cite{turing}  & Powerful capabilities and compatibility but with high energy consumption \\
  \cmidrule{2-4}          & Google & TPU \cite{Jouppi2017In} & Stable in terms of performance and power consumption \\
  \cmidrule{2-4}          & Intel  & Xeon D-2100 \cite{D2100} & Optimized for power- and space-constrained cloud-edge solutions \\
  \cmidrule{2-4}          & Samsung  & Exynos 9820 \cite{Exynos9820} &  Mobile NPU for accelerating AI tasks \\
      \midrule
      \multicolumn{1}{c}{\multirow{6}[1]{*}[-3.5em]{\textbf{\tabincell{c}{Edge  \\  Computing \\ Frameworks}}}} & Huawei  & KubeEdge \cite{Xiong2018}  & Native support for edge-cloud collaboration \\
  \cmidrule{2-4}          & Baidu & OpenEdge \cite{openedge} & Computing framework shielding and application production simplification \\
  \cmidrule{2-4}          & Microsoft  & Azure IoT Edge \cite{Azureiotedge} & Remotely edge management with zero-touch device provisioning \\
  \cmidrule{2-4}          & Linux Foundation & EdgeX \cite{edgexfoundry} & IoT edge across the industrial and enterprise use cases \\
  \cmidrule{2-4}          & Linux Foundation & Akraino Edge Stack \cite{Akraino} & Integrated distributed cloud edge platform \\
  \cmidrule{2-4}          & NVIDIA  & NVIDIA EGX \cite{Nvidiaegx} & Real-time perception, understanding, and processing at the edge \\
  \cmidrule{2-4}          & Amazon & AWS IoT Greengrass \cite{AWSIoTGreengrass} & Tolerance to edge devices even with intermittent connectivity \\
  \cmidrule{2-4}          & Google & Google Cloud IoT \cite{GoogleCloudIoT} & Compatible with Google AI products, such as TensorFlow Lite and Edge TPU
  \\
      \bottomrule
      \end{tabular}%
\end{table*}%

\subsubsection{Collaborative End-Edge-Cloud Computing}
\label{subsec:endedgecloudcomputing}

While cloud computing is created for processing computation-intensive tasks, such as DL, it cannot guarantee the delay requirements throughout the whole process from data generation to transmission to execution. Moreover, independent processing on the end or edge devices is limited by their computing capability, power consumption, and cost bottleneck. Therefore, collaborative end-edge-cloud computing for DL \cite{Neurosurgeon}, abstracted in Fig. \ref{fig:EndEdgeCloudComputing}, is emerging as an important trend as depicted in Fig. \ref{fig:CollaborationTrend}. In this novel computing paradigm, computation tasks with lower computational intensities, generated by end devices, can be executed directly at the end devices or offloaded to the edge, thus avoiding the delay caused by sending data to the cloud. For a computation-intensive task, it will be reasonably segmented and dispatched separately to the end, edge and cloud for execution, reducing the execution delay of the task while ensuring the accuracy of the results \cite{Neurosurgeon, Li, Huang2018b}. The focus of this collaborative paradigm is not only the successful completion of tasks but also achieving the optimal balance of equipment energy consumption, server loads, transmission and execution delays.

\subsection{Hardware for Edge Computing}


In this section, we discuss potential enabling hardware of edge intelligence, i.e., customized AI chips and commodities for both end devices and edge nodes. Besides, edge-cloud systems for DL are introduced as well (listed in Table \ref{tab:HardwareSummary}). 


\subsubsection{AI Hardware for Edge Computing}
\label{subsubsec:aichips}


Emerged edge AI hardware can be classified into three categories according to their technical architecture: 
1) Graphics Processing Unit (GPU)-based hardware, which tend to have good compatibility and performance, but generally consume more energy, e.g., NVIDIA' GPUs based on Turing architecture \cite{turing}; 
2) Field Programmable Gate Array (FPGA)-based hardware \cite{Nurvitadhi:2017:FBG:3020078.3021740, Jiang2018}, which are energy-saving and require less computation resources, but with worse compatibility and limited programming capability compared to GPUs; 
3) Application Specific Integrated Circuit (ASIC)-based hardware, such as Google's TPU \cite{Jouppi2017In} and HiSilicon's Ascend series \cite{ascend}, usually with a custom design that is more stable in terms of performance and power consumption. 

As smartphones represent the most widely-deployed edge devices, chips for  smartphones have undergone rapid developments, and their capabilities have been extended to the acceleration of AI computing. 
To name a few, Qualcomm first applies AI hardware acceleration \cite{Snapdragon8} in Snapdragon and releases Snapdragon Neural Processing Engine (SNPE) SDK \cite {Snapdragonsdk}, which supports almost all major DL frameworks. 
Compared to Qualcomm, HiSilicon's 600 series and 900 series chips \cite{hisilicon} do not depend on GPUs. 
Instead, they incorporate an additional Neural Processing Unit (NPU) to achieve fast calculation of vectors and matrices, which greatly improves the efficiency of DL. 
Compared to HiSilicon and Qualcomm, MediaTek's Helio P60 not only uses GPUs but also introduces an AI Processing Unit (APU) to further accelerate neural network computing \cite{mediatekp60}. 
Performance comparison of most commodity chips with respect to DL can be found in \cite{Ignatov2018}, and more customized chips of edge devices will be discussed in detail later.

\begin{figure*}[!!!!!!!!!!!!!!hhhhhhhhhht]
    \centering
    \includegraphics[width=17 cm]{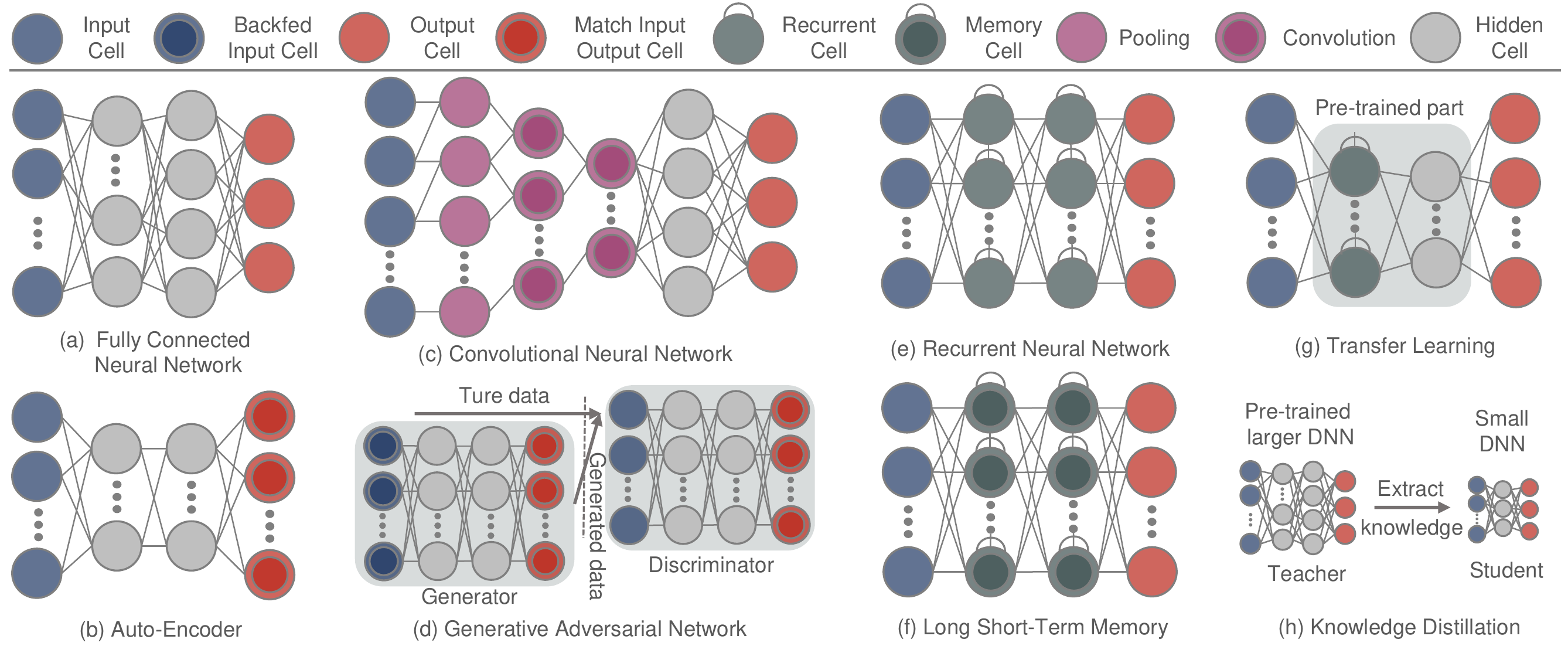}
    \caption{{\color{red}Basic structures and functions of typical DNNs and DL.}}
    \label{fig:DNNStructure}
\end{figure*}

\subsubsection{Integrated Commodities Potentially for Edge Nodes}

Edge nodes are expected to have computing and caching capabilities and to provide high-quality network connection and computing services near end devices. Compared to most end devices, edge nodes have more powerful computing capability to process tasks. On the other side, edge nodes can respond to end devices more quickly than the cloud. Therefore, by deploying edge nodes to perform the computation task, the task processing can be accelerated while ensuring accuracy. In addition, edge nodes also have the ability to cache, which can improve the response time by caching popular contents. For example, practical solutions including Huawei' Atlas modules \cite{atlas} and Microsoft's Data Box Edge \cite{databox} can carry out preliminary DL inference and then transfer to the cloud for further improvement. 

\subsubsection{Edge Computing Frameworks}
\label{subsubsec:EdgeComputingPlatformsforDL}

Solutions for edge computing systems are blooming. For DL services with complex configuration and intensive resource requirements, edge computing systems with advanced and excellent microservice architecture are the future development direction. Currently, Kubernetes is as a mainstream container-centric system for the deployment, maintenance, and scaling of applications in cloud computing \cite{7036275}. Based on Kubernetes, Huawei develops its edge computing solution ``KubeEdge'' \cite{Xiong2018} for networking, application deployment and metadata synchronization between the cloud and the edge (also supported in Akraino Edge Stack \cite{Akraino}). ``OpenEdge'' \cite{openedge} focus on shielding computing framework and simplifying application production. For IoT, Azure IoT Edge \cite{Azureiotedge} and  EdgeX \cite{edgexfoundry} are devised for delivering cloud intelligence to the edge by deploying and running AI on cross-platform IoT devices.

\begin{table}[htbp!!!!!!!!!!!!!!!]
  \centering
  \scriptsize
  \caption{{\color{red}Potential DL libraries for edge computing}}
  \label{Table:DLFrameworks}
    \begin{tabular}{lm{0.4em}<{\centering}m{0.4em}<{\centering}m{0.4em}<{\centering}m{0.4em}<{\centering}m{0.4em}<{\centering}m{0.4em}<{\centering}m{0.4em}<{\centering}m{0.4em}<{\centering}m{0.4em}<{\centering}m{0.4em}<{\centering}m{0.4em}<{\centering}m{0.4em}<{\centering}m{0.4em}<{\centering}m{0.4em}<{\centering}}
    \toprule
    \textbf{Library} & \rotatebox{90}{CNTK \cite{CNTK}}  & \rotatebox{90}{Chainer \cite{tokui2015chainerpaper}} & \rotatebox{90}{TensorFlow \cite{Abadi2016paper}} & \rotatebox{90}{DL4J \cite{D4J}}  & \rotatebox{90}{TensorFlow Lite \cite{tflite}} & \rotatebox{90}{MXNet \cite{Chenbpaper}} & \rotatebox{90}{(Py)Torch \cite{pytorch}} & \rotatebox{90}{CoreML \cite{coreml}} & \rotatebox{90}{SNPE \cite{Snapdragonsdk}}  & \rotatebox{90}{NCNN \cite{ncnn}}  & \rotatebox{90}{MNN \cite{MNN}}   & \rotatebox{90}{Paddle-Mobile \cite{PaddleMobile}} & \rotatebox{90}{MACE \cite{MACE}}  & \rotatebox{90}{FANN \cite{FANN}} \\
    \midrule
    \textbf{Owner} & \rotatebox{90}{Microsoft} & \rotatebox{90}{Preferred Networks} & \rotatebox{90}{Google} & \rotatebox{90}{Skymind} & \rotatebox{90}{Google} & \rotatebox{90}{Apache Incubator} & \rotatebox{90}{Facebook} & \rotatebox{90}{Apple} & \rotatebox{90}{Qualcomm} & \rotatebox{90}{Tencent} & \rotatebox{90}{Alibaba} & \rotatebox{90}{Baidu} & \rotatebox{90}{XiaoMi} & \rotatebox{90}{ETH Zürich} \\
    \midrule
    \textbf{\tabincell{l}{Edge \\ Support}} & $\times$ & $\times$ & \checkmark & \checkmark & \checkmark & \checkmark & \checkmark & \checkmark & \checkmark & \checkmark & \checkmark & \checkmark & \checkmark & \checkmark \\
    \midrule
    \textbf{Android} & $\times$ & $\times$ & $\times$ & \checkmark & \checkmark & \checkmark & \checkmark & $\times$ & \checkmark & \checkmark & \checkmark & \checkmark & \checkmark & $\times$ \\
    \midrule
    \textbf{iOS} & $\times$ & $\times$ & $\times$ & $\times$ & $\times$ & \checkmark & \checkmark & \checkmark & $\times$ & \checkmark & \checkmark & \checkmark & \checkmark & $\times$ \\
    \midrule
    \textbf{Arm} & $\times$ & $\times$ & \checkmark & \checkmark & \checkmark & \checkmark & \checkmark & $\times$ & \checkmark & \checkmark & \checkmark & \checkmark & \checkmark & \checkmark \\
    \midrule
    \textbf{FPGA} & $\times$ & $\times$ & $\times$ & $\times$ & $\times$ & $\times$ & \checkmark & $\times$ & $\times$ & $\times$ & $\times$ & \checkmark & $\times$ & $\times$ \\
    \midrule
    \textbf{DSP} & $\times$ & $\times$ & $\times$ & $\times$ & $\times$ & $\times$ & $\times$ & $\times$ & \checkmark & $\times$ & $\times$ & $\times$ & $\times$ & $\times$ \\
    \midrule
    \textbf{GPU} & \checkmark & \checkmark & \checkmark & \checkmark & \checkmark & \checkmark & \checkmark & $\times$ & $\times$ & $\times$ & $\times$ & $\times$ & $\times$ & $\times$ \\
    \midrule
    \textbf{\tabincell{l}{Mobile \\ GPU}} & $\times$ & $\times$ & $\times$ & $\times$ & \checkmark & $\times$ & $\times$ & \checkmark & \checkmark & \checkmark & \checkmark & \checkmark & \checkmark & $\times$ \\
    \midrule
    \textbf{\tabincell{l}{Training \\ Support}} & \checkmark & \checkmark & \checkmark & \checkmark & $\times$ & \checkmark & \checkmark & $\times$ & $\times$ & $\times$ & $\times$ & $\times$ & $\times$ & \checkmark \\
    \bottomrule
    \end{tabular}%
\end{table}%

\subsection{{\color{red}Virtualizing the Edge}}
\label{subsec:VirtualizingtheEdge}
{\color{red}
The requirements of virtualization technology for integrating edge computing and DL reflect in the following aspects: 
1) The resource of edge computing is limited. 
Edge computing cannot provide that resources for DL services as the cloud does. 
Virtualization technologies should maximize resource utilization under the constraints of limited resources;
2) DL services rely heavily on complex software libraries.
The versions and dependencies of these software libraries should be taken into account carefully.
Therefore, virtualization catering to Edge DL services should be able to isolate different services. 
Specifically, the upgrade, shutdown, crash, and high resource consumption of a single service should not affect other services;
3) The service response speed is critical for Edge DL. 
Edge DL requires not only the computing power of edge devices but also the agile service response that the edge computing architecture can provide.

\begin{figure}[!!!!!!!!!!!!!!hhhhhhhhhht]
    \centering
    \includegraphics[width=8.85 cm]{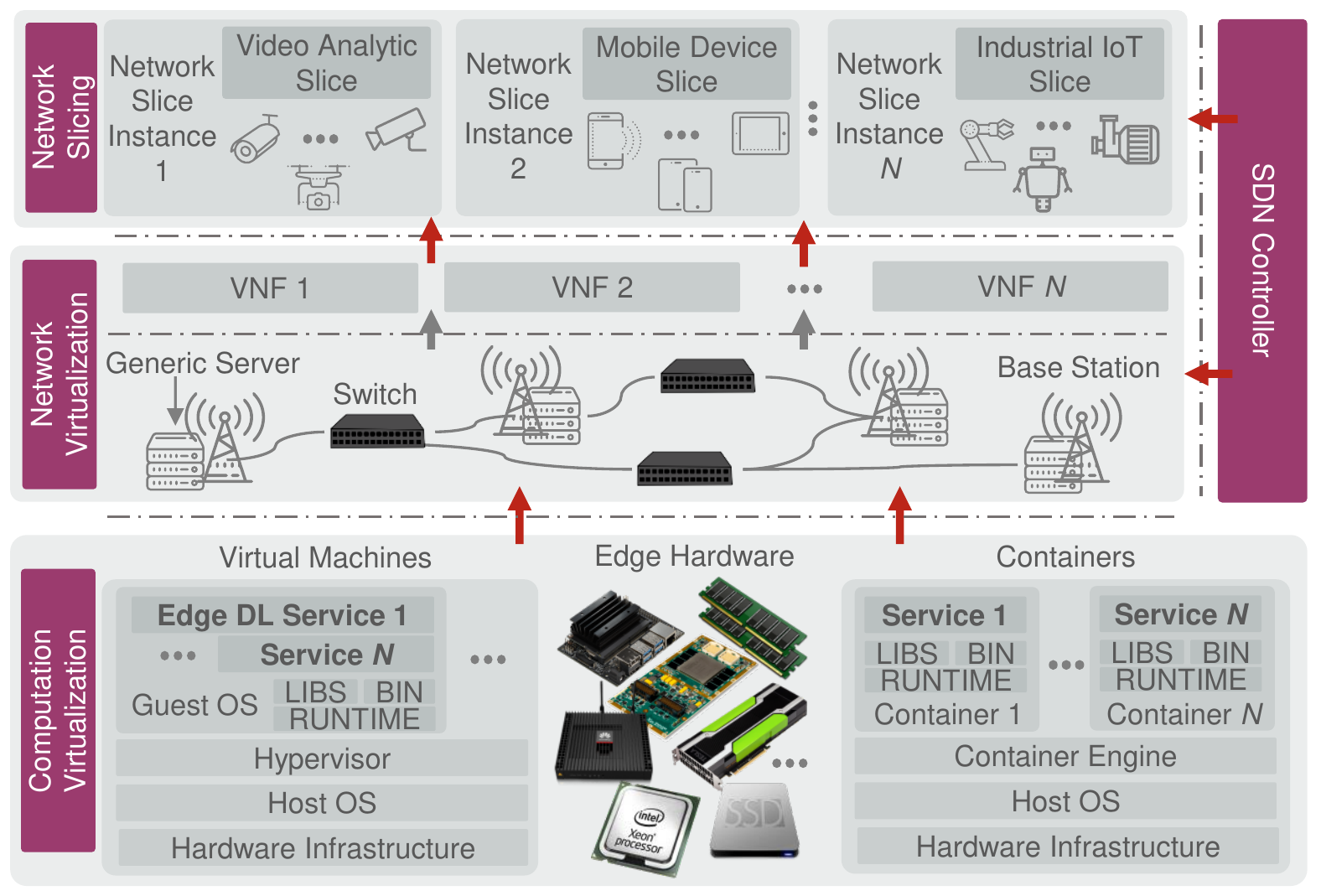}
    \caption{{\color{red}Virtualizing edge computing infrastructure and networks.}}
    \label{fig:EdgeVirtualization}
\end{figure}

The combination of edge computing and DL to form high-performance Edge DL services requires the coordinated integration of computing, networking and communication resources, as depicted in Fig. \ref{fig:EdgeVirtualization}. 
Specifically, both the computation virtualization and the integration of network virtualization, and management technologies are necessary.
In this section, we discuss potential virtualization technologies for the edge.

\subsubsection{Virtualization Techniques}
\label{subsec:VirtualizationTechniques}

Currently, there are two main virtualization strategies: Virtual Machine (VM) and container. 
In general, VM is better at isolating while container provides easier deployment of repetitive tasks \cite{VirtualMachineEdgeSurvey}.
With VM virtualization at operating system level, a VM hypervisor splits a physical server into one or multiple VMs, and can easily manage each VM to execute tasks in isolation. 
Besides, the VM hypervisor can allocate and use idle computing resources more efficiently by creating a scalable system that includes multiple independent virtual computing devices.

In contrast to VM, container virtualization is a more flexible tool for packaging, delivering, and orchestrating software infrastructure services and applications. 
Container virtualization for edge computing can effectively reduce the workload execution time with high performance and storage requirements, and can also deploy a large number of services in a scalable and straightforward fashion \cite{7930383}. 
A container consists of a single file that includes an application and execution environment with all dependencies, which makes it enable efficient service handoff to cope with user mobility \cite{Ma2018}. 
Owning to that the execution of applications in the container does not depend on additional virtualization layers as in VM virtualization, the processor consumption and the amount of memory required to execute the application are significantly reduced.

\subsubsection{Network Virtualization}
\label{subsubsec:NetworkVirtualization}

Traditional networking functions, combined with specific hardware, is not flexible enough to manage edge computing networks in an on-demand fashion. 
In order to consolidate network device functions into industry-standard servers, switches and storage, Network Functions Virtualization (NFV) enables Virtual Network Functions (VNFs) to run in software, by separating network functions and services from dedicated network hardware. 
Further, Edge DL services typically require high bandwidth, low latency, and dynamic network configuration, while Software-defined Networking (SDN) allows rapid deployment of services, network programmability and multi-tenancy support, through three key innovations \cite{SDNEdgeSurvey}: 
1) Decoupling of control planes and data planes; 
2) Centralized and programmable control planes;
3) Standardized application programming interface. 
With these advantages, it supports a highly customized network strategy that is well suited for the high bandwidth, dynamic nature of Edge DL services.

Network virtualization and edge computing benefit each other. 
On the one hand, NFV/SDN can enhance the interoperability of edge computing infrastructure. 
For example, with the support of NFV/SDN, edge nodes can be efficiently orchestrated and integrated with cloud data centers \cite{Lin2018}. 
On the other hand, both VNFs and Edge DL services can be hosted on a lightweight NFV framework (deployed on the edge) \cite{DeepNFV}, thus reusing the infrastructure and infrastructure management of NFV to the largest extent possible \cite{ETSI4}.

\subsubsection{Network Slicing}
\label{subsubsec:NetworkSlicing}

Network slicing is a form of agile and virtual network architecture, a high-level abstraction of the network that allows multiple network instances to be created on top of a common shared physical infrastructure, each of which optimized for specific services. 
With increasingly diverse service and QoS requirements, network slicing, implemented by NFV/SDN, is naturally compatible with distributed paradigms of edge computing.
To meet these, network slicing can be coordinated with joint optimization of computing and communication resources in edge computing networks \cite{Chien2019b}. 
Fig. \ref{fig:EdgeVirtualization} depicts an example of network slicing based on edge virtualization. 
In order to implement service customization in network slicing, virtualization technologies and SDN must be together to support tight coordination of resource allocation and service provision on edge nodes while allowing flexible service control. 
With network slicing, customized and optimized resources can be provided for Edge DL services, which can help reduce latency caused by access networks and support dense access to these services \cite{TalebMECSurvey}.
}

\section{Fundamentals of Deep Learning}
\label{sec:dlfundamentals}

With respect to CV, NLP, and AI, DL is adopted in a myriad of applications and corroborates its superior performance \cite{LeCun2015}. Currently, a large number of GPUs, TPUs, or FPGAs are required to be deployed in the cloud to process DL service requests. Nonetheless, the edge computing architecture, on account of it covers a large number of distributed edge devices, can be utilized to better serve DL. Certainly, edge devices typically have limited computing power or power consumption compared to the cloud. Therefore, the combination of DL and edge computing is not straightforward and requires a comprehensive understanding of DL models and edge computing features for design and deployment. In this section, we compendiously introduce DL and related technical terms, paving the way for discussing the integration of DL and edge computing (more details can be found in \cite{haykin2009neural}).

\subsection{Neural Networks in Deep Learning}
\label{subsec:nndl}


DL models consist of various types of Deep Neural Networks (DNNs) \cite{haykin2009neural}. Fundamentals of DNNs in terms of basic structures and functions are introduced as follows. 


\subsubsection{Fully Connected Neural Network (FCNN)}
\label{subsubsec:fcnn}

The output of each layer of FCNN, i.e., Multi-Layer Perceptron (MLP), is fed forward to the next layer, as in Fig. \ref{fig:DNNStructure}(a). Between contiguous FCNN layers, the output of a neuron (cell), either the input or hidden cell, is directly passed to and activated by neurons belong to the next layer \cite{Collobert:2004:LPM:1015330.1015415}. FCNN can be used for feature extraction and function approximation, however with high complexity, modest performance, and slow convergence.


\subsubsection{Auto-Encoder (AE)}
\label{subsubsec:autoencoder}

AE, as in Fig. \ref{fig:DNNStructure}(b), is actually a stack of two NNs that replicate input to its output in an unsupervised learning style. The first NN learns the representative characteristics of the input (encoding). The second NN takes these features as input and restores the approximation of the original input at the match input output cell, used to converge on the identity function from input to output, as the final output (decoding). Since AEs are able to learn the low-dimensional useful features of input data to recover input data, it is often used to classify and store high-dimensional data \cite{manning1999foundations}.


\subsubsection{Convolutional Neural Network (CNN)}
\label{subsubsec:cnn}

By employing pooling operations and a set of distinct moving filters, CNNs seize correlations between adjacent data pieces, and then generate a successively higher level abstraction of the input data, as in Fig. \ref{fig:DNNStructure}(c). Compared to FCNNs, CNNs can extract features while reducing the model complexity, which mitigates the risk of overfitting \cite{10.1007/978-3-319-10590-1_53}. These characteristics make CNNs achieve remarkable performance in image processing and also useful in processing structural data similar to images.

\subsubsection{Generative Adversarial Network (GAN)}
\label{subsubsec:gan}

GAN originates from game theory. As illustrated in Fig. \ref{fig:DNNStructure}(d), GAN is composed of \textit{generator} and \textit{discriminator}. The goal of the \textit{generator} is to learn about the true data distribution as much as possible by deliberately introducing feedback at the back-fed input cell, while the \textit{discriminator} is to correctly determine whether the input data is coming from the true data or the \textit{generator}. These two participants need to constantly optimize their ability to generate and distinguish in the adversarial process until finding a Nash equilibrium \cite{NIPS2014_5423_GAN}. According to the features learned from the real information, a well-trained \textit{generator} can thus fabricate indistinguishable information. 


\subsubsection{Recurrent Neural Network (RNN)}
\label{subsubsec:rnn}

RNNs are designed for handling sequential data. As depicted in Fig. \ref{fig:DNNStructure}(e), each neuron in RNNs not only receives information from the upper layer but also receives information from the previous channel of its own \cite{Schmidhuber2015}. In general, RNNs are natural choices for predicting future information or restoring missing parts of sequential data. However, a serious problem with RNNs is the gradient explosion. LSTM, as in Fig. \ref{fig:DNNStructure}(f), improving RNN with adding a gate structure and a well-defined memory cell, can overcome this issue by controlling (prohibiting or allowing) the flow of information \cite{Hochreiter:1997:LSM:1246443.1246450}.



{\color{red}
\subsubsection{Transfer Learning (TL)}
\label{subsubsec:TransferLearning}

TL can transfer knowledge, as shown in Fig. \ref{fig:DNNStructure}(g), from the source domain to the target domain so as to achieve better learning performance in the target domain \cite{5288526TransferLearning}. By using TL, existing knowledge learned by a large number of computation resources can be transferred to a new scenario, and thus accelerating the training process and reducing model development costs. Recently, a novel form of TL emerges, viz., Knowledge Distillation (KD) \cite{hinton2015distilling} emerges. As indicated in Fig. \ref{fig:DNNStructure}(h), KD can extract implicit knowledge from a well-trained model (teacher), inference of which possess excellent performance but requires high overhead. Then, by designing the structure and objective function of the target DL model, the knowledge is ``transferred'' to a smaller DL model (student), so that the significantly reduced (pruned or quantized) target DL model achieves high performance as possible.


}

\subsection{Deep Reinforcement Learning (DRL)}
\label{subsec:DRL}

\begin{figure}[!!!!!!!!!!!!!!hhhhhhhhhht]
    \centering
    \includegraphics[width=7 cm]{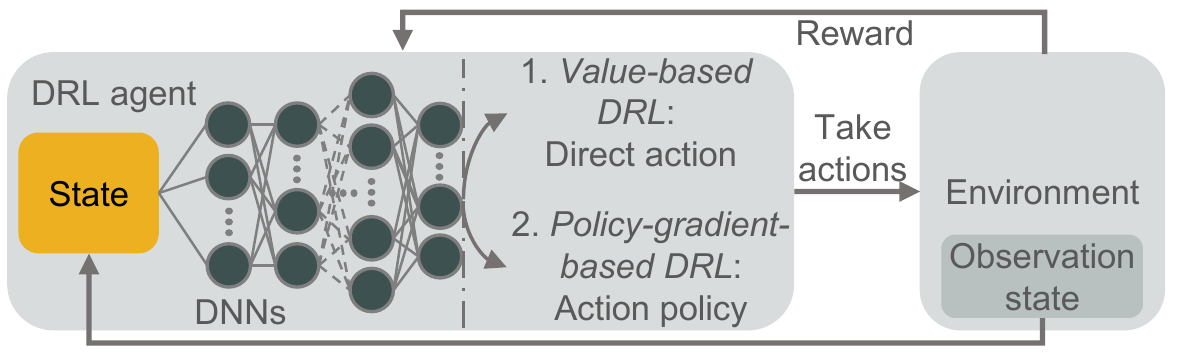}
    \caption{Value-based and policy-gradient-based DRL approaches.}
    \label{fig:DRL}
\end{figure}

As depicted in Fig. \ref{fig:DRL}, the goal of RL is to enable an agent in the environment to take the best action in the current state to maximize long-term gains, where the interaction between the agent's action and state through the environment is modeled as a Markov Decision Process (MDP). DRL is the combination of DL and RL, but it focuses more on RL and aims to solve decision-making problems. The role of DL is to use the powerful representation ability of DNNs to fit the value function or the direct strategy to solve the explosion of state-action space or continuous state-action space problem. By virtue of these characteristics, DRL becomes a powerful solution in robotics, finance, recommendation system, wireless communication, etc \cite{Mousavi2018, Park}.

\subsubsection{Value-based DRL}
\label{subsubsec:ValuebasedDRL}

As a representative of value-based DRL, Deep $Q$-Learning (DQL) uses DNNs to fit action values, successfully mapping high-dimensional input data to actions \cite{Mnih2015}. In order to ensure stable convergence of training, experience replay method is adopted to break the correlation between transition information and a separate target network is set up to suppress instability. Besides, Double Deep $Q$-Learning (Double-DQL) can deal with that DQL generally overestimating action values \cite{Hasselt20016}, and Dueling Deep $Q$-Learning (Dueling-DQL) \cite{Wang2015Dueling} can learn which states are (or are not) valuable without having to learn the effect of each action at each state.

\subsubsection{Policy-gradient-based DRL}
\label{subsubsec:PolicygradientbasedDRL}

Policy gradient is another common strategy optimization method, such as Deep Deterministic Policy Gradient (DDPG) \cite{lillicrap2015continuous}, Asynchronous Advantage Actor-Critic (A3C) \cite{pmlr-v48-mniha16}, Proximate Policy Optimization (PPO) \cite{Schulman2017Proximal}, etc. It updates the policy parameters by continuously calculating the gradient of the policy expectation reward with respect to them, and finally converges to the optimal strategy \cite{Sutton:1999:PGM:3009657.3009806}. Therefore, when solving the DRL problem, DNNs can be used to parameterize the policy, and then be optimized by the policy gradient method. Further, Actor-Critic (AC) framework is widely adopted in policy-gradient-based DRL, in which the policy DNN is used to update the policy, corresponding to the Actor; the value DNN is used to approximate the value function of the state action pair, and provides gradient information, corresponding to the Critic.

\subsection{Distributed DL Training}
\label{subsec:DistributedTraining}

\begin{figure}[!!!!!!!!!!!!!!hhhhhhhhhht]
    \centering
    \includegraphics[width=8.85 cm]{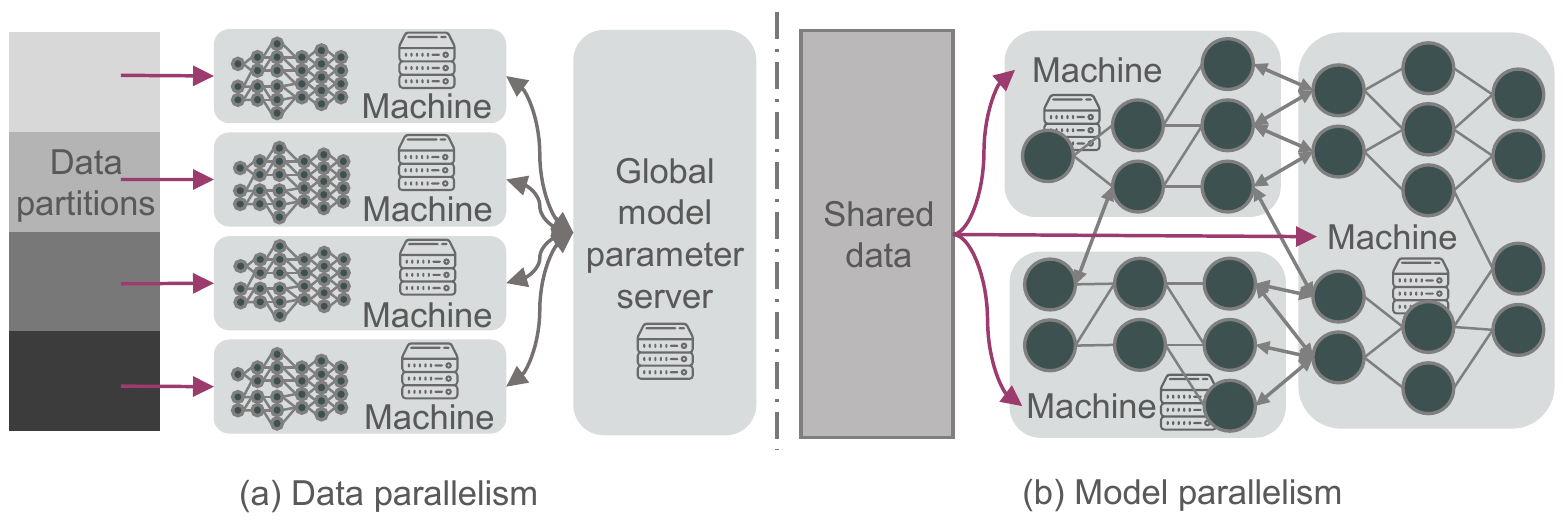}
    \caption{{\color{red}Distributed training in terms of data and model parallelism.}}
    \label{fig:DistributedLearning}
\end{figure}

At present, training DL models in a centralized manner consumes a lot of time and computation resources, hindering further improving the algorithm performance. Nonetheless, distributed training can facilitate the training process by taking full advantage of parallel servers. There are two common ways to perform distributed training, i.e., \textit{data parallelism} and \textit{model parallelism} \cite{Monin1981, zou2014mariana, Chen2012, Seide2014} as illustrated in Fig. \ref{fig:DistributedLearning}.

Model parallelism first splits a large DL model into multiple parts and then feeds data samples for training these segmented models in parallel. This not only can improve the training speed but also deal with the circumstance that the model is larger than the device memory. Training a large DL model generally requires a lot of computation resources, even thousands of CPUs are required to train a large-scale DL model. In order to solve this problem, distributed GPUs can be utilized for model parallel training \cite{Ltifi2010}. Data parallelism means dividing data into multiple partitions, and then respectively training copies of the model in parallel with their own allocated data samples. By this means, the training efficiency of model training can be improved \cite{Nishihara2016}.

Coincidentally, a large number of end devices, edge nodes, and cloud data centers, are scattered and envisioned to be connected by virtue of edge computing networks. These distributed devices can potentially be powerful contributors once the DL training jumps out of the cloud. 

{\color{red}
\subsection{Potential DL Libraries for Edge}
\label{subsec:DeepLearningLibraries}

Development and deployment of DL models rely on the support of various DL libraries. 
However, different DL libraries have their own application scenarios. 
For deploying DL on and for the edge, efficient lightweight DL libraries are required. 
Features of DL frameworks potentially supporting future edge intelligence are listed in Table \ref{Table:DLFrameworks} (excluding libraries unavailable for edge devices, such as Theano \cite{bergstra2010theano}).
}
\section{Deep Learning Applications on Edge}
\label{sec:AIonEdge}

In general, DL services are currently deployed in cloud data centers (the cloud) for handling requests, due to the fact that most DL models are complex and hard to compute their inference results on the side of resource-limited devices. 
However, such kind of ``end-cloud'' architecture cannot meet the needs of real-time DL services such as real-time analytics, smart manufacturing and etc. 
Thus, deploying DL applications on the edge can broaden the application scenarios of DL especially with respect to the low latency characteristic.
{\color{red}In the following, we present edge DL applications and highlight their advantages over the comparing architectures without edge computing. }



\subsection{Real-time Video Analytic}
\label{subsec:ssva}

\begin{figure}[!!!!!!!!!!!!!!hhhhhhhhhht]
    \centering
    \includegraphics[width=6.5 cm]{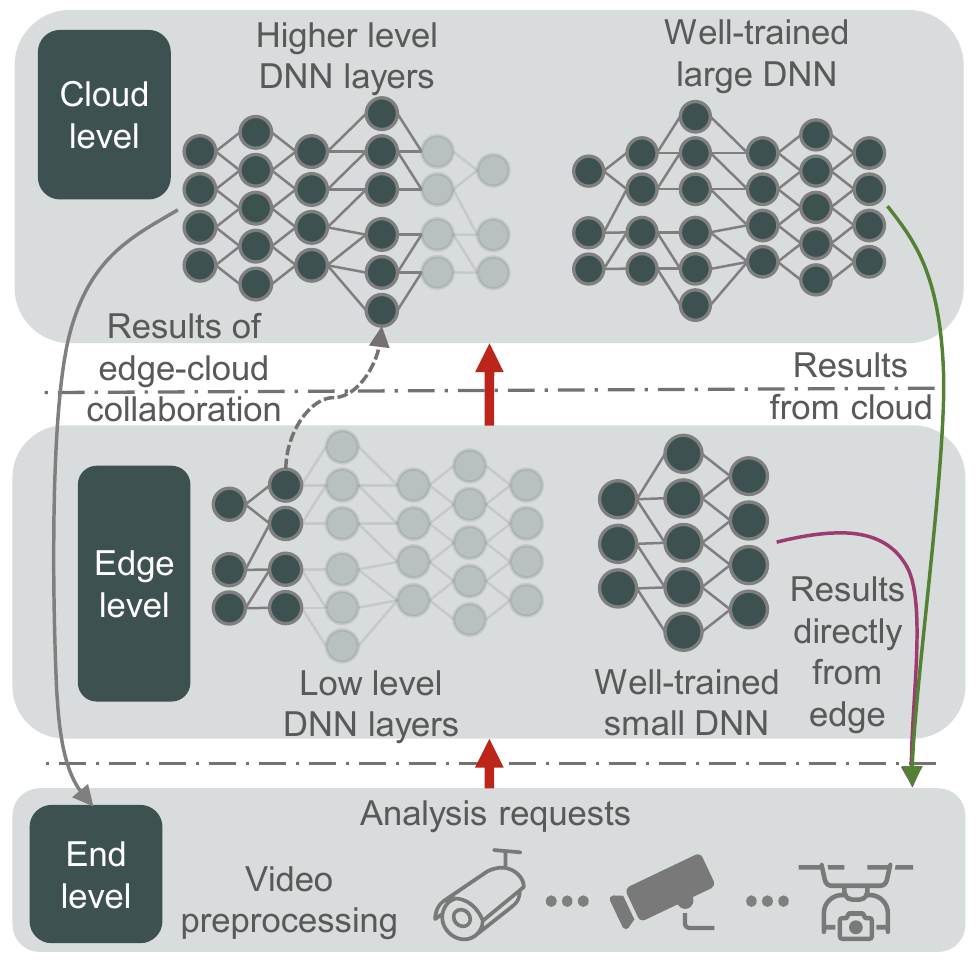}
    \caption{The collaboration of the end, edge and cloud layer for performing real-time video analytic by deep learning.}
    \label{fig:DLVideoAnalytic}
\end{figure}

Real-time video analytic is important in various fields, such as automatic pilot, VR and Augmented Reality (AR), smart surveillance, etc. In general, applying DL for it requires high computation and storage resources. Unfortunately, executing these tasks in the cloud often incurs high bandwidth consumption, unexpected latency, and reliability issues. With the development of edge computing, those problems tend to be addressed by moving video analysis near to the data source, viz., end devices or edge nodes, as the complementary of the cloud. In this section, as depicted in Fig. \ref{fig:DLVideoAnalytic}, we summarize related works as a hybrid hierarchical architecture, which is divided into three levels: end, edge, and cloud.

%

\subsubsection{End Level}
\label{subsubsec:FirstLayerOfArch}

At the end level, video capture devices, such as smartphones and surveillance cameras are responsible for video capture, media data compression \cite{Ren2018}, image pre-processing, and image segmentation \cite{Liu2018a}. 
By coordinating with these participated devices, collaboratively training a domain-aware adaptation model can lead to better object recognition accuracy when used together with a domain-constrained deep model \cite{DeepCham}. 
Besides, in order to appropriately offload the DL computation to the end devices, the edge nodes or the cloud, end devices should comprehensively consider tradeoffs between video compression and key metrics, e.g., network condition, data usage, battery consumption, processing delay, frame rate and accuracy of analytics, and thus determine the optimal offloading strategy \cite{Ren2018}. 

{\color{red}If various DL tasks are executed at the end level independently, enabling parallel analytics requires a solution that supports efficient multi-tenant DL. 
With the model pruning and recovery scheme, \textit{NestDNN} \cite{NestDNN} transforms the DL model into a set of descendant models, in which the descendant model with fewer resource requirements shares its model parameters with the descendant model requiring more resources, making itself nested inside the descendent model requiring more resources without taking extra memory space. 
In this way, the multi-capacity model provides variable resource-accuracy trade-offs with a compact memory footprint, hence ensuring efficient multi-tenant DL at the end level.}

\subsubsection{Edge Level}
\label{subsubsec:SecondLayerOfArch}

Numerous distributed edge nodes at the edge level generally cooperate with each other to provide better services. 
For example, \textit{LAVEA} \cite{LAVEA} attaches edge nodes to the same access point or BS as well as the end devices, which ensure that services can be as ubiquitous as Internet access. 
In addition, compressing the DL model on the edge can improve holistic performance. 
The resource consumption of the edge layer can be greatly reduced while ensuring the analysis performance, by reducing the unnecessary filters in CNN layers \cite{Nikouei2018a}. 
{\color{red}Besides, in order to optimize performance and efficiency, \cite{EdgeEye} presents an edge service framework, i.e., \textit{EdgeEye}, which realizes a high-level abstraction of real-time video analytic functions based on DL. 
To fully exploit the bond function of the edge, \textit{VideoEdge} \cite{VideoEdge} implements an end-edge-cloud hierarchical architecture to help achieve load balancing concerning analytical tasks while maintaining high analysis accuracy.}

\subsubsection{Cloud Level}
\label{subsubsec:ThirdLayerOfArch}

At the cloud level, the cloud is responsible for the integration of DL models among the edge layer and updating parameters of distributed DL models on edge nodes \cite{Ren2018}. Since the distributed model training performance on an edge node may be significantly impaired due to its local knowledge, the cloud needs to integrate different well-trained DL models to achieve global knowledge. When the edge is unable to provide the service confidently (e.g., detecting objects with low confidence), the cloud can use its powerful computing power and global knowledge for further processing and assist the edge nodes to update DL models.

\subsection{Autonomous Internet of Vehicles (IoVs)}
\label{subsec:InternetofVehicles}

It is envisioned that vehicles can be connected to improve safety, enhance efficiency, reduce accidents, and decrease traffic congestion in transportation systems \cite{He2018e}. There are many information and communication technologies such as networking, caching, edge computing which can be used for facilitating the IoVs, though usually studied respectively. On one hand, edge computing provides low-latency, high-speed communication and fast-response services for vehicles, making automatic driving possible. On the other hand, DL techniques are important in various smart vehicle applications. Further, they are expected to optimize complex IoVs systems.

In \cite{He2018e}, a framework which integrates these technologies is proposed. This integrated framework enables dynamic orchestration of networking, caching and computation resources to meet requirements of different vehicular applications \cite{He2018e}. Since this system involves multi-dimensional control, a DRL-based approach is first utilized to solve the optimization problem for enhancing the holistic system performance. Similarly, DRL is also used in \cite{Qi2018} to obtain the optimal task offloading policy in vehicular edge computing. Besides, Vehicle-to-Vehicle (V2V) communication technology can be taken advantaged to further connect vehicles, either as an edge node or an end device managed by DRL-based control policies \cite{Le2018a}.

\subsection{Intelligent Manufacturing}
\label{subsec:IntelligentManufacturing}

Two most important principles in the intelligent manufacturing era are automation and data analysis, the former one of which is the main target and the latter one is one of the most useful tools \cite{Li2018i}. In order to follow these principles, intelligent manufacturing should first address response latency, risk control, and privacy protection, and hence requires DL and edge computing. In intelligent factories, edge computing is conducive to expand the computation resources, the network bandwidth, and the storage capacity of the cloud to the IoT edge, as well as realizing the resource scheduling and data processing during manufacturing and production \cite{Hu2019}. For autonomous manufacturing inspection, \textit{DeepIns} \cite{Li2018i} uses DL and edge computing to guarantee performance and process delay respectively. The main idea of this system is partitioning the DL model, used for inspection, and deploying them on the end, edge and cloud layer separately for improving the inspection efficiency. 

Nonetheless, with the exponential growth of IoT edge devices, 1) how to remotely manage evolving DL models and 2) how to continuously evaluate these models for them are necessary. In \cite{Soto2016}, a framework, dealing with these challenges, is developed to support complex-event learning during intelligent manufacturing, thus facilitating the development of real-time application on IoT edge devices. Besides, the power, energy efficiency, memory footprint limitation of IoT edge devices \cite{Plastiras2018} should also be considered. Therefore, caching, communication with heterogeneous IoT devices, and computation offloading can be integrated \cite{Hao2019} to break the resource bottleneck.

\subsection{Smart Home and City}
\label{subsec:SmartHomeandCity}

The popularity of IoTs will bring more and more intelligent applications to home life, such as intelligent lighting control systems, smart televisions, and smart air conditioners. But at the same time, smart homes need to deploy numerous wireless IoT sensors and controllers in corners, floors, and walls. For the protection of sensitive home data, the data processing of smart home systems must rely on edge computing. Like use cases in \cite{Liu2017c, Dhakal2017}, edge computing is deployed to optimize indoor positioning systems and home intrusion monitoring so that they can get lower latency than using cloud computing as well as the better accuracy. Further, the combination of DL and edge computing can make these intelligent services become more various and powerful. For instance,  it endows robots the ability of dynamic visual servicing \cite{Tian2018} and enables efficient music cognition system \cite{Lu2018}. 

If the smart home is enlarged to a community or city, public safety, health data, public facilities, transportation, and other fields can benefit. The original intention of applying edge computing in smart cities is more due to cost and efficiency considerations. The natural characteristic of geographically distributed data sources in cities requires an edge computing-based paradigm to offer location-awareness and latency-sensitive monitoring and intelligent control. For instance, the hierarchical distributed edge computing architecture in \cite{Tang2017a} can support the integration of massive infrastructure components and services in future smart cities. This architecture can not only support latency-sensitive applications on end devices but also perform slightly latency-tolerant tasks efficiently on edge nodes, while large-scale DL models responsible for deep analysis are hosted on the cloud. Besides, DL can be utilized to orchestrate and schedule infrastructures to achieve the holistic load balancing and optimal resource utilization among a region of a city (e.g., within a campus \cite{Chang2018}) or the whole city. 

\section{Deep Learning Inference in Edge}
\label{sec:AIinEdge}

In order to further improve the accuracy, DNNs become deeper and require larger-scale dataset. By this means, dramatic computation costs are introduced. Certainly, the outstanding performance of DL models is inseparable from the support of high-level hardware, and it is difficult to deploy them in the edge with limited resources. Therefore, large-scale DL models are generally deployed in the cloud while end devices just send input data to the cloud and then wait for the DL inference results. However, the cloud-only inference limits the ubiquitous deployment of DL services. Specifically, it can not guarantee the delay requirement of real-time services, e.g., real-time detection with strict latency demands. Moreover, for important data sources, data safety and privacy protection should be addressed. To deal with these issues,  DL services tend to resort to edge computing. Therefore, DL models should be further customized to fit in the resource-constrained edge, while carefully treating the trade-off between the inference accuracy and the execution latency of them.

\subsection{Optimization of DL Models in Edge}
\label{subsec:DeepLearningOptimization}


DL tasks are usually computationally intensive and requires large memory footprints. But in the edge, there are not enough resources to support raw large-scale DL models. Optimizing DL models and quantize their weights can reduce resource costs. In fact, model redundancies are common in DNNs \cite{Denton2014, Chen2015a} and can be utilized to make model optimization possible. The most important challenge is how to ensure that there is no significant loss in model accuracy after being optimized. In other words, the optimization approach should transform or re-design DL models and make them fit in edge devices, with as little loss of model performance as possible. In this section, optimization methods for different scenarios are discussed: 1) general optimization methods for edge nodes with relatively sufficient resources; 2) fine-grained optimization methods for end devices with tight resource budgets.
 
\subsubsection{General Methods for Model Optimization}
\label{subsubsec:GeneralModelOptimizationMethods}

On one hand, increasing the depth and width of DL models with nearly constant computation overhead is one direction of optimization, such as inception \cite{Szegedy2015} and deep residual networks \cite{He2016a} for CNNs. On the other hand, for more general neural network structures, existing optimization methods can be divided into four categories \cite{Cheng2017}: 1) parameter pruning and sharing \cite{Han2015, Alwani2016}, including also weights quantization \cite{DBLP:conf/nips/CourbariauxBD15, 10.1007/978-3-319-46493-0_32, Mcdanel2017}; 2) low-rank factorization \cite{Denton2014}; 3) transferred/compact convolution filters \cite{Iandola2017, Howard2012, Nikouei2018a}; 4) knowledge distillation \cite{Sharma2018}. These approaches can be applied to different kinds of DNNs or be composed to optimize a complex DL model for the edge.

\subsubsection{Model Optimization for Edge Devices}
\label{subsubsec:ModelOptimizationforEdgeDevices}


In addition to limited computing and memory footprint, other factors such as network bandwidth and power consumption also need to be considered. In this section, efforts for running DL on edge devices are differentiated and discussed.

\begin{itemize}
    \item \textit{Model Input}: 
    Each application scenario has specific optimization spaces. 
    {\color{red}Concerning object detection, \textit{FFS-VA} uses two prepositive stream-specialized filters and a small full-function tiny-YOLO model to filter out vast but non-target-object frames \cite{FFSVA}. 
    In order to adjust the configuration of the input video stream (such as frame resolution and sampling rate) online with low cost, \textit{Chameleon} \cite{Chameleon} greatly saves the cost of searching the best model configuration by leveraging temporal and spatial correlations of the video inputs, and allows the cost to be amortized over time and across multiple video feeds.
    Besides, as depicted in Fig. \ref{fig:ReduceModelInput}, narrowing down the classifier's searching space \cite{Nikouei2018} and dynamic Region-of-Interest (RoI) encoding \cite{EdgeAssistedRealTimeObjectDetection} to focus on target objects in video frames can further reduce the bandwidth consumption and data transmission delay. }
    Though this kind of methods can significantly compress the size of model inputs and hence reduce the computation overhead without altering the structure of DL models, it requires a deep understanding of the related application scenario to dig out the potential optimization space.
    \begin{figure}[!!!!!!!!!!!!!!hhhhhhhhhht]
        \centering
        \includegraphics[width=8 cm]{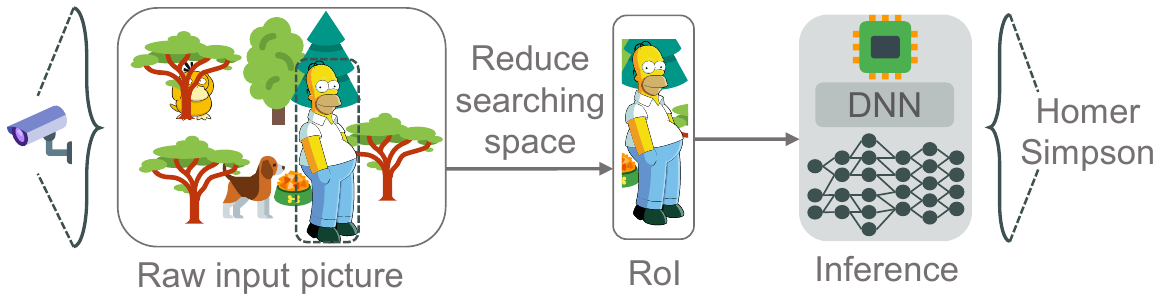}
        \caption{Optimization for model inputs, e.g., narrowing down the searching space of DL models (pictures are with permission from \cite{FoxHomer}).}
        \label{fig:ReduceModelInput}
    \end{figure}
    \item \textit{Model Structure}: 
    Not paying attention to specific applications, but focusing on the widely used DNNs' structures is also feasible. 
    For instance, point-wise group convolution and channel shuffle \cite{ZhangCVPR}, paralleled convolution and pooling computation \cite{Du2018}, depth-wise separable convolution \cite{Nikouei2018a} can greatly reduce computation cost while maintaining accuracy. 
    {\color{red}\textit{NoScope} \cite{NoScope} leverages two types of models rather than the standard model (such as YOLO \cite{Redmon2016}): specialized models that waive the generality of standard models in exchange for faster inference, and difference detectors that identify temporal differences across input data. 
    After performing efficient cost-based optimization of the model architecture and thresholds for each model, \textit{NoScope} can maximize the throughput of DL services and by cascading these models.}
    Besides, as depicted in Fig. \ref{fig:ModelPrune}, parameters pruning can be applied adaptively in model structure optimization as well \cite{Han2017, Mao2016b, Bhattacharya}. 
    Furthermore, the optimization can be more efficient if across the boundary between algorithm, software and hardware. 
    Specifically, general hardware is not ready for the irregular computation pattern introduced by model optimization. 
    Therefore, hardware architectures should be designed to work directly for optimized models \cite{Han2017}.  
    \begin{figure}[!!!!!!!!!!!!!!hhhhhhhhhht]
        \centering
        \includegraphics[width=7 cm]{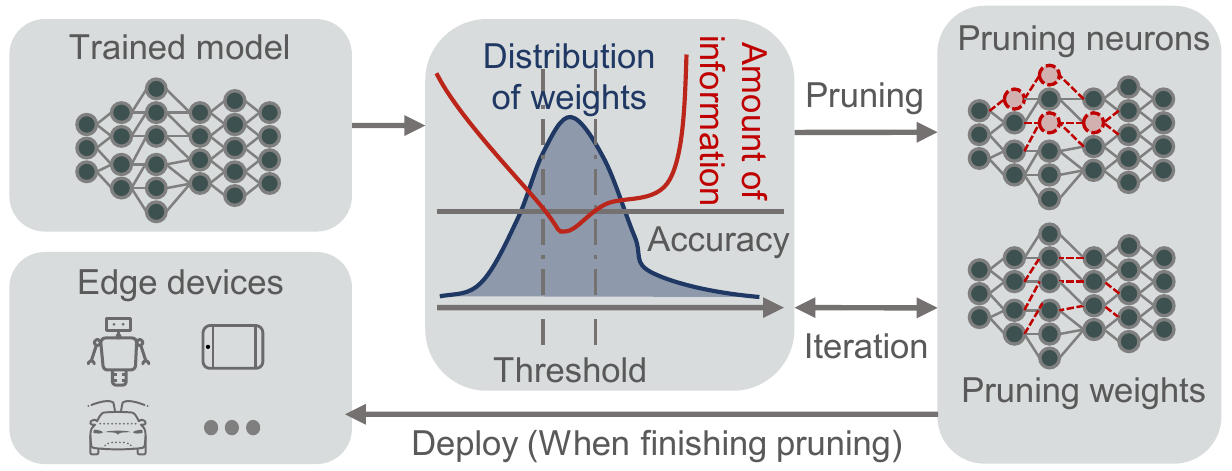}
        \caption{Adaptive parameters pruning in model structure optimization.}
        \label{fig:ModelPrune}
    \end{figure}
     {\color{red}\item \textit{Model Selection}: 
    With various DL models, choosing the best one from available DL models in the edge requires weighing both precision and inference time. 
    In \cite{Taylor2018}, the authors use $k$NN to automatically construct a predictor, composed of DL models arranged in sequence. 
    Then, the model selection can be determined by that predictor along with a set of automatically tuned features of the model input.
    Besides, combining different compression techniques (such as model pruning), multiple compressed DL models with different tradeoffs between the performance and the resource requirement can be derived. 
    \textit{AdaDeep} \cite{AdaDeep} explores the desirable balance between performance and resource constraints, and based on DRL, automatically selects various compression techniques (such as model pruning) to form a compressed model according to current available resources, thus fully utilizing the advantages of them.}

    \item \textit{Model Framework}: 
    Given the high memory footprint and computational demands of DL, running them on edge devices requires expert-tailored software and hardware frameworks. 
    A software framework is valuable if it 1) provides a library of optimized software kernels to enable deployment of DL \cite{Lai2018}; 2) automatically compresses DL models into smaller dense matrices by finding the minimum number of non-redundant hidden elements \cite{Yao2017}; 3) performs quantization and coding on all commonly used DL structures \cite{Mao2016b, Shen2014, Yao2017}; 4) specializes DL models to contexts and shares resources across multiple simultaneously executing DL models \cite{Shen2014}. 
    With respect to the hardware, running DL models on Static Random Access Memory (SRAM) achieves better energy savings compared to Dynamic RAM (DRAM) \cite{Mao2016b}. 
    Hence, DL performance can be benefited if underlying hardware directly supports running optimized DL models \cite{Han2016} on the on-chip SRAM.    
\end{itemize}

\subsection{Segmentation of DL Models}
\label{subsec:SegmentationofDLModelsinEdge}

In \cite{Neurosurgeon}, the delay and power consumption of the most advanced DL models are evaluated on the cloud and edge devices, finding that uploading data to the cloud is the bottleneck of current DL servicing methods (leading to a large overhead of transmitting). Dividing the DL model and performing distributed computation can achieve better end-to-end delay performance and energy efficiency. In addition, by pushing part of DL tasks from the cloud to the edge, the throughput of the cloud can be improved. Therefore, the DL model can be segmented into multiple partitions and then allocated to 1) heterogeneous local processors (e.g., GPUs, CPUs) on the end device \cite{Lane2016}, 2) distributed edge nodes \cite{Zhang2018, DeepThings}, or 3) collaborative ``end-edge-cloud'' architecture \cite{Zhao2018c, Li2018l, Neurosurgeon, Li}. 

\begin{figure}[!!!!!!!!!!!!!!hhhhhhhhhht]
    \centering
    \includegraphics[width=8.87 cm]{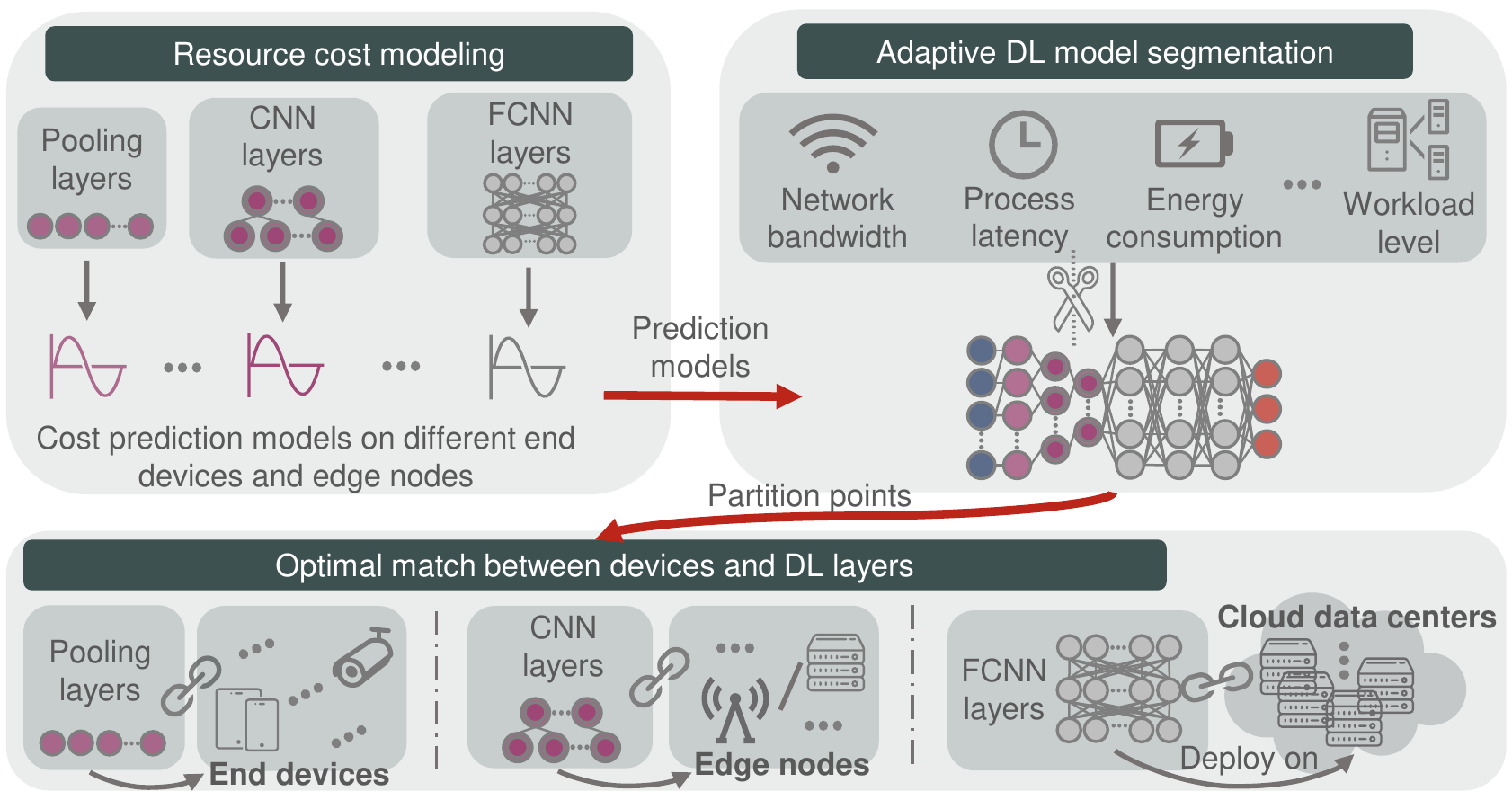}
    \caption{Segmentation of DL models in the edge.}
    \label{fig:ModelSegmentation}
\end{figure}

Partitioning the DL model horizontally, i.e., along the end, edge and cloud, is the most common segmentation method. The challenge lies in how to intelligently select the partition points. As illustrated in Fig. \ref{fig:ModelSegmentation}, a general process for determining the partition point can be divided into three steps \cite{Zhao2018c, Neurosurgeon}: 1) measuring and modeling the resource cost of different DNN layers and the size of intermediate data between layers; 2) predicting the total cost by specific layer configurations and network bandwidth; 3) choosing the best one from candidate partition points according to delay, energy requirements, etc. Another kind of model segmentation is vertically partitioning particularly for CNNs \cite{DeepThings}. In contrast to horizontal partition, vertical partition fuses layers and partitions them vertically in a grid fashion, and thus divides CNN layers into independently distributable computation tasks.

\subsection{Early Exit of Inference (EEoI)}
\label{subsec:EarlyExitofInference}

To reach the best trade-off between model accuracy and processing delay, multiple DL models with different model performance and resource cost can be maintained for each DL service. Then, by intelligently selecting the best model, the desired adaptive inference is achieved \cite{Ogden2018a}. Nonetheless, this idea can be further improved by the emerged EEoI \cite{Mcdanel2016}. 

The performance improvement of additional layers in DNNs is at the expense of increased latency and energy consumption in feedforward inference. As DNNs grow larger and deeper, these costs become more prohibitive for edge devices to run real-time and energy-sensitive DL applications. By additional side branch classifiers, for partial samples, EEoI allows inference to exit early via these branches if with high confidence. For more difficult samples, EEoI will use more or all DNN layers to provide the best predictions. 

\begin{figure}[!!!!!!!!!!!!!!hhhhhhhhhht]
    \centering
    \includegraphics[width=6 cm]{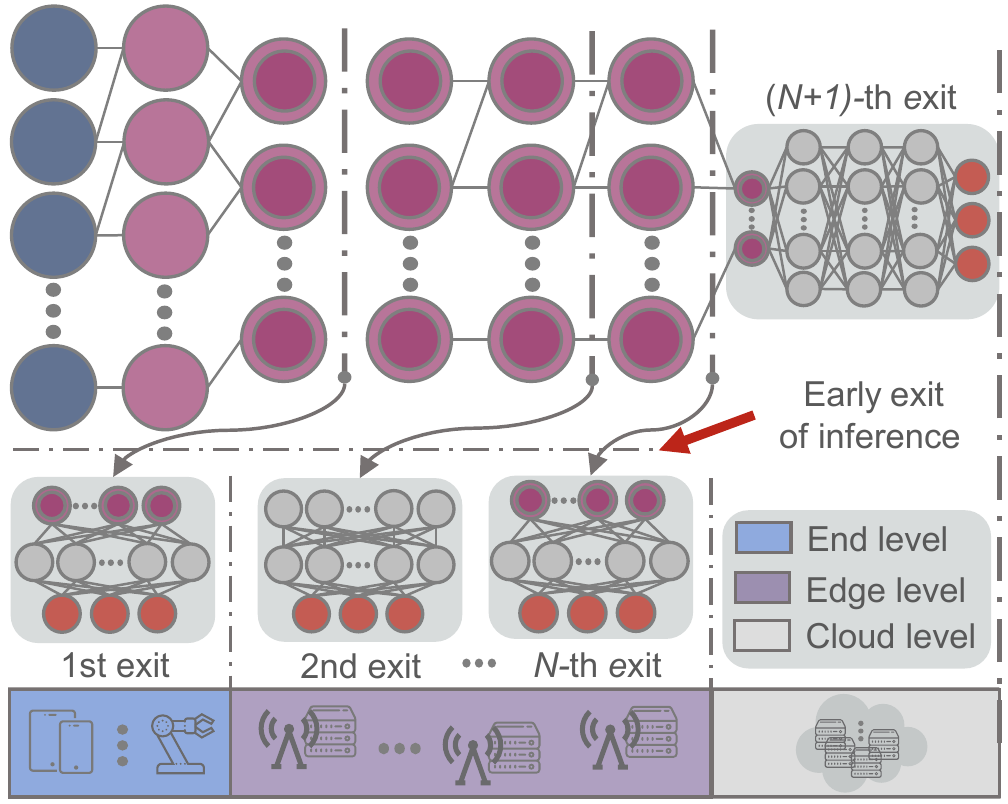}
    \caption{Early exit of inference for DL inference in the edge.}
    \label{fig:EarlyExitofInference}
\end{figure}

As depicted in Fig. \ref{fig:EarlyExitofInference}, by taking advantage of EEoI, fast and localized inference using shallow portions of DL models at edge devices can be enabled. By this means, the shallow model on the edge device can quickly perform initial feature extraction and, if confident, can directly give inference results. Otherwise, the additional large DL model deployed in the cloud performs further processing and final inference. Compared to directly offloading DL computation to the cloud, this approach has lower communication costs and can achieve higher inference accuracy than those of the pruned or quantized DL models on edge devices \cite{Li2018i, Teerapittayanon2017}. In addition, since only immediate features rather than the original data are sent to the cloud, it provides better privacy protection. Nevertheless, EEoI shall not be deemed independent to model optimization (Section \ref{subsubsec:ModelOptimizationforEdgeDevices}) and segmentation (Section \ref{subsec:SegmentationofDLModelsinEdge}). The envision of distributed DL over the end, edge and cloud should take their collaboration into consideration, e.g., developing a collaborative and on-demand co-inference framework \cite{Li2018j} for adaptive DNN partitioning and EEoI.

 {\color{red}
\subsection{Sharing of DL Computation}
\label{subsec:SharingDLInference}

The requests from nearby users within the coverage of an edge node may exhibit spatiotemporal locality \cite{Cachier}. 
For instance, users within the same area might request recognition tasks for the same object of interest, and it may introduce redundant computation of DL inference. 
In this case, based on offline analysis of applications and online estimates of network conditions, \textit{Cachier} \cite{Cachier} proposes to cache related DL models for recognition applications in the edge node and to minimize expected end-to-end latency by dynamically adjusting its cache size. 
Based on the similarity between consecutive frames in first-person-view videos, \textit{DeepMon} \cite{DeepMon} and \textit{DeepCache} \cite{DeepCache} utilize the internal processing structure of CNN layers to reuse the intermediate results of the previous frame to calculate the current frame, i.e., caching internally processed data within CNN layers, to reduce the processing latency of continuous vision applications. 

Nevertheless, to proceed with effective caching and results reusing, accurate lookup for reusable results shall be addressed, i.e., the cache framework must systematically tolerate the variations and evaluate key similarities. 
\textit{DeepCache} \cite{DeepCache} performs cache key lookup to solve this.
Specifically, it divides each video frame into fine-grained regions and searches for similar regions from cached frames in a specific pattern of video motion heuristics.
For the same challenge, \textit{FoggyCache} \cite{FoggyCache} first embeds heterogeneous raw input data into feature vectors with generic representation. 
Then, Adaptive Locality Sensitive Hashing (A-LSH), a variant of LSH commonly used for indexing high-dimensional data, is proposed to index these vectors for fast and accurate lookup. 
At last, Homogenized $k$NN, which utilizes the cached values to remove outliers and ensure a dominant cluster among the $k$ records initially chosen, is implemented based on $k$NN to determine the reuse output from records looked up by A-LSH. 

Differ from sharing inference results, \textit{Mainstream} \cite{Mainstream} proposes to adaptively orchestrate DNN stem-sharing (the common part of several specialized DL models) among concurrent video processing applications. 
By exploiting computation sharing of specialized models among applications trained through TL from a common DNN stem, aggregate per-frame compute time can be significantly decreased. 
Though more specialized DL models mean both higher model accuracy and less shared DNN stems, the model accuracy decreases slowly as less-specialized DL models are employed (unless the fraction of the model specialized is very small). 
This characteristic hence enables that large portions of the DL model can be shared with low accuracy loss in \textit{Mainstream}. 
}

\section{Edge Computing for Deep Learning}
\label{sec:EdgeforAI}


Extensive deployment of DL services, especially mobile DL, requires the support of edge computing. 
This support is not just at the network architecture level, the design, adaptation, and optimization of edge hardware and software are equally important. 
Specifically, 
1) customized edge hardware and corresponding optimized software frameworks and libraries can help DL execution more efficiently; 
2) the edge computing architecture can enable the offloading of DL computation; 
3) well-designed edge computing frameworks can better maintain DL services running on the edge; 
4) fair platforms for evaluating Edge DL performance help further evolve the above implementations.

{\color{red}\subsection{Edge Hardware for DL}}

\subsubsection{Mobile CPUs and GPUs}

DL applications are more valuable if directly enabled on lightweight edge devices, such as mobile phones, wearable devices, and surveillance cameras, near to the location of events. 
Low-power IoT edge devices can be used to undertake lightweight DL computation, and hence avoiding communication with the cloud, but it still needs to face limited computation resources, memory footprint, and energy consumption. 
To break through these bottlenecks, in \cite{Du2018}, the authors focus on ARM Cortex-M micro-controllers and develop \textit{CMSIS-NN}, a collection of efficient NN kernels. By \textit{CMSIS-NN}, the memory footprint of NNs on ARM Cortex-M processor cores can be minimized, and then the DL model can be fitted into IoT devices, meantime achieving normal performance and energy efficiency.  

 {\color{red}With regard to the bottleneck when running CNN layers on mobile GPUs, \textit{DeepMon} \cite{DeepMon} decomposes the matrices used in the CNN layers to accelerate the multiplications between high-dimensional matrices.}
By this means, high-dimensional matrix operations (particularly multiplications) in CNN layers are available in mobile GPUs and can be accelerated. 
In view of this work, various mobile GPUs, already deployed in edge devices, can be potentially explored with specific DL models and play a more important role in enabling edge DL. 

Other than DL inference \cite{Du2018, DeepMon}, important factors that affect the performance of DL training on mobile CPUs and GPUs are discussed in \cite{Chen}. 
Since commonly used DL models, such as VGG \cite{simonyan2014very}, are too large for the memory size of mainstream edge devices, a relatively small Mentee network \cite{venkatesan2016diving} is adopted to evaluate DL training. 
Evaluation results point out that the size of DL models is crucial for training performance and the efficient fusion of mobile CPUs and GPUs is important for accelerating the training process.

\subsubsection{FPGA-based Solutions}

Though GPU solutions are widely adopted in the cloud for DL training and inference, however, restricted by the tough power and cost budget in the edge, these solutions may not be available. Besides, edge nodes should be able to serve multiple DL computation requests at a time, and it makes simply using lightweight CPUs and GPUs impractical. Therefore, edge hardware based on Field Programmable Gate Array (FPGA) is explored to study their feasibility for edge DL. 

FPGA-based edge devices can achieve CNN acceleration with arbitrarily sized convolution and reconfigurable pooling \cite{Du2018}, and they perform faster than the state-of-the-art CPU and GPU implementations \cite{Han2017} with respect to RNN-based speech recognition applications while achieving higher energy efficiency. 
In \cite{Jiang2018}, the design and setup of an FPGA-based edge platform are developed to admit DL computation offloading from mobile devices.
On implementing the FPGA-based edge platform, a wireless router and an FPGA board are combined together. 
Testing this preliminary system with typical vision applications, the FPGA-based edge platform shows its advantages, in terms of both energy consumption and hardware cost, over the GPU (or CPU)-based one. 


Nevertheless, it is still pended to determine whether FPGAs or GPUs/CPUs are more suitable for edge computing, as shown in Table \ref{tab:fpgagpucpu}. 
Elaborated experiments are performed in \cite{Biookaghazadeh2018a} to investigate the advantages of FPGAs over GPUs: 
1) capable of providing workload insensitive throughput; 
2) guaranteeing consistently high performance for high-concurrency DL computation; 
3) better energy efficiency. 
However, the disadvantage of FPGAs lies in that developing efficient DL algorithms on FPGA is unfamiliar to most programmers. 
Although tools such as Xilinx SDSoC can greatly reduce the difficulty \cite{Jiang2018}, at least for now, additional works are still required to transplant the state-of-the-art DL models, programmed for GPUs, into the FPGA platform. 

\begin{table}[htbp]
    \centering
    \scriptsize
    \caption{{\color{red}Comparison of Solutions for Edge nodes}}
    \label{tab:fpgagpucpu}
    \begin{tabular}{lcm{20em}}
    \toprule
    \multicolumn{1}{c}{\textbf{Metrics}} & \textbf{\begin{tabular}[c]{@{}c@{}}Preferred\\ Hardware\end{tabular}} & \multicolumn{1}{c}{\textbf{Analysis}} \\ \midrule
    \tabincell{l}{Resource \\ overhead} & FPGA & FPGA can be optimized by customized designs. \\ \midrule
    \tabincell{l}{DL \\ training} & GPU & Floating point capabilities are better on GPU. \\ \midrule
    \tabincell{l}{DL \\ inference} & FPGA & FPGA can be customized for specific DL models. \\ \midrule
    \tabincell{l}{Interface \\ scalability} & FPGA & It is more free to implement interfaces on FPGAs. \\ \midrule
    \tabincell{l}{Space \\ occupation} & \tabincell{c}{CPU/  \\ FPGA} & Lower power consumption of FPGA leads to smaller space occupation. \\ \midrule
    Compatibility & \tabincell{c}{CPU/ \\ GPU} & CPUs and GPUs have more stable architecture. \\ \midrule
    \tabincell{l}{Development \\ efforts} & \tabincell{c}{CPU/ \\ GPU} & Toolchains and software libraries facilitate the practical development. \\ \midrule
    \tabincell{l}{Energy \\ efficiency} & FPGA & Customized designs can be optimized. \\ \midrule
    \tabincell{l}{Concurrency \\ support} & FPGA & FPGAs are suitable for stream process. \\ \midrule
    \tabincell{l}{Timing \\ latency} & FPGA & Timing on FPGAs can be an order of magnitude faster than GPUs. \\ \bottomrule
    \end{tabular}
\end{table}

\begin{figure}[!!!!!!!!!!!!!!hhhhhhhhhht]
    \centering
    \includegraphics[width=8.3 cm]{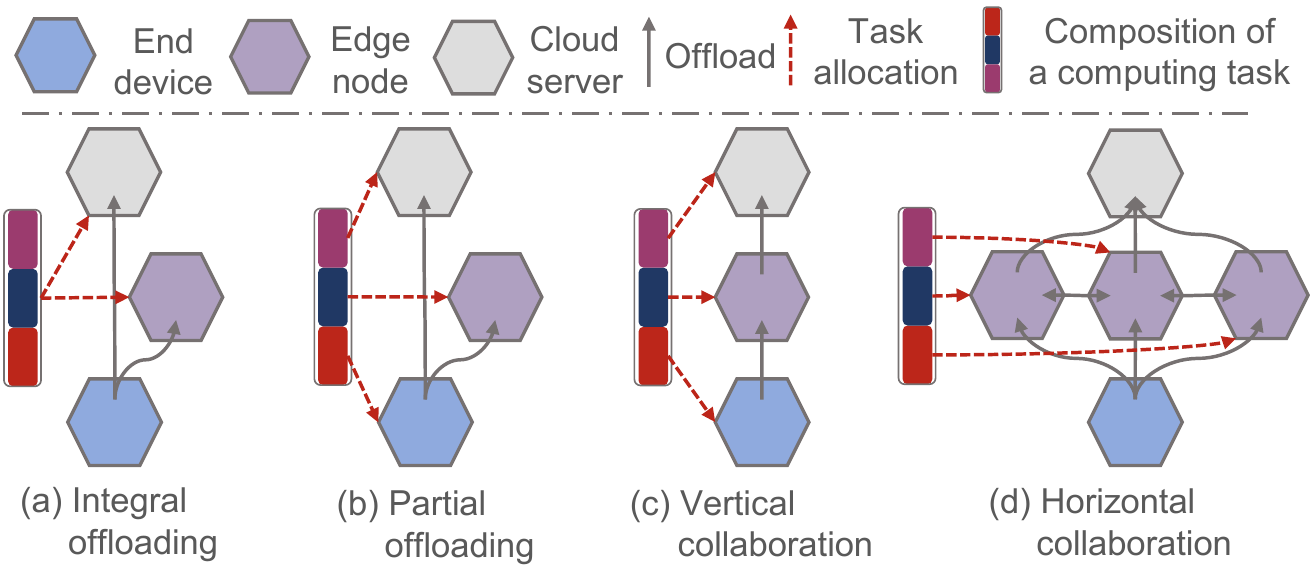}
    \caption{Communication and computation modes for Edge DL.}
    \label{fig:EdgeComputingMode}
\end{figure}

{\color{red}\subsection{Communication and Computation Modes for Edge DL}
\label{subsec:edgecomputingmodefordl}

\begin{table*}[htbp]
    \centering
    \scriptsize
    \label{tab:edgefordlfourmodes}
    \caption{{\color{red}Details about Edge Communication and Computation Modes for DL}}
      \begin{tabular}{m{1em}<{\centering}m{2em}<{\centering}m{5em}<{\centering}m{12em}<{\centering}m{4em}<{\centering}m{5em}<{\centering}m{17em}m{14em}}
      \toprule
        & \textbf{Ref.} & \textbf{DL Model} & \textbf{End/Edge/Cloud} & \textbf{Network} & \textbf{Dependency} & \multicolumn{1}{c}{\textbf{Objective}} & \multicolumn{1}{c}{\textbf{Performance}} \\
      \midrule
      {\multirow{3}[1]{*}[-1.5em]{ {\rotatebox{90}{\textbf{Integral Offloading}}} }} & \rotatebox{90}{\tabincell{c}{\textit{DeepDecision} \\ \cite{DeepDecision}} } & YOLO & Samsung Galaxy S7 / Server with a quad-core CPU at 2.7GHz, GTX970 and 8GB RAM / N/A & {Simulated WLAN \& LAN} & TensorFlow, Darknet & Consider the complex interaction between model accuracy, video quality, battery constraints, network data usage, and network conditions to determine an optimal offloading strategy & Achieve about 15 FPS video analytic while possessing higher accuracy than that of the baseline approaches \\
\cmidrule{2-8}          & \rotatebox{90}{ \tabincell{c}{\textit{MASM} \\ \cite{MASM}} } & $\backslash$ & Simulated devices /  Cloudlet / N/A & $\backslash$ &  \multicolumn{1}{c}{$\backslash$}  & Optimize workload assignment weights and the computation capacities of the VMs hosted on the Cloudlet  & \multicolumn{1}{c}{$\backslash$} \\
\cmidrule{2-8}          & \rotatebox{90}{ \tabincell{c}{\textit{EdgeEye} \\ \cite{EdgeEye}} } & DetectNet, FaceNet & Cameras / Server with Intel i7-6700, GTX 1060 and 24GB RAM / N/A & Wi-Fi & TensorRT, ParaDrop, Kurento & Offload the live video analytics tasks to the edge using EdgeEye API, instead of using DL framework specific APIs, to provide higher inference performance & \multicolumn{1}{c}{$\backslash$} \\
      \midrule
      {\multirow{2}[1]{*}[0em]{ \rotatebox{90}{\textbf{Partial Offloading}} }} & \rotatebox{90}{ \tabincell{c}{\textit{DeepWear} \\ \cite{DeepWear}} } & MobileNet, GoogLeNet, DeepSense, etc. & Commodity smartwatches running Android Wear OS / Commodity smartphone running Android / N/A & Bluetooth & TensorFlow & Provide context-aware offloading, strategic model partition, and pipelining support to efficiently utilize the processing capacity of the edge & Bring up to 5.08$\times$ and 23.0$\times$ execution speedup, as well as 53.5\% and 85.5\% energy saving against wearable-only and handheld-only strategies, respectively \\
  \cmidrule{2-8}          & \rotatebox{90}{ \tabincell{c}{\textit{IONN} \\ \cite{IONN}} } & AlexNet & Embedded board Odroid XU4 /  Server with an quad-core CPU at 3.6GHz, GTX 1080 Ti and 32GB RAM / Unspecified & WLAN  & Caffe & Partitions the DNN layers and incrementally uploads the partitions to allow collaborative execution by the end and the edge (or cloud) to improves both the query performance and the energy consumption & Maintain almost the same uploading latency  as integral uploading while largely improving query execution time \\
      \midrule
      {\multirow{5}[1]{*}[-9em]{ \rotatebox{90}{\textbf{Vertical Collaboration}} }} & \rotatebox{90}{\cite{Huang2017}} & CNN, LSTM & Google Nexus 9 / Server with an quad-core CPU and 16GB RAM / 3 desktops, each with i7-6850K and 2$\times$GTX 1080 Ti & WLAN \& LAN & Apache Spark, TensorFlow & Perform data pre-processing and preliminary learning at the edge to reduce the network traffic, so as to speed up the computation in the cloud & Achieve 90\% accuracy while reducing the execution time and the data transmission \\
  \cmidrule{2-8}          & \rotatebox{90}{ \tabincell{c}{\textit{Neurosurgeon} \\ \cite{Neurosurgeon}} } & AlexNet, VGG, Deepface, MNIST, Kaldi, SENNA & Jetson TK1 mobile platform / Server with Intel Xeon E5$\times$2,  NVIDIA Tesla K40 GPU and 256GB RAM / Unspecified & Wi-Fi, LTE \& 3G & Caffe & Adapt to various DNN architectures, hardware platforms, wireless connections, and server load levels, and choose the partition point for best latency and best mobile energy consumption & Improve end-to-end latency by 3.1$\times$ on average and up to 40.7$\times$, reduce mobile energy consumption by 59.5\% on average and up to 94.7\%, and improve data-center throughput by 1.5$\times$ on average and up to 6.7$\times$ \\
  \cmidrule{2-8}          & \rotatebox{90}{\cite{Teerapittayanon2017}} & BranchyNet & \multicolumn{1}{c}{$\backslash$} & \multicolumn{1}{c}{$\backslash$} & \multicolumn{1}{c}{$\backslash$} & Minimize communication and resource usage for devices while allowing low-latency classification via EEoI & Reduce the communication cost by a factor of over 20$\times$ while achieving 95\% overall accuracy \\
  \cmidrule{2-8}          & \rotatebox{90}{\cite{Ren2018}} & Faster R-CNN  & Xiaomi 6 / Server with i7 6700, GTX 980Ti and 32GB RAM / Work station with E5-2683 V3,  GTX TitanXp$\times$4 and 128GB RAM & WLAN \& LAN & \multicolumn{1}{c}{$\backslash$} & Achieve efficient object detection via wireless communications by interactions between the end, the edge and the cloud & Lose only 2.5\% detection accuracy under the image compression ratio of 60\% while significantly improving image transmission efficiency \\
  \cmidrule{2-8}          & \rotatebox{90}{ \tabincell{c}{\textit{VideoEdge} \\ \cite{VideoEdge}} } & AlexNet, DeepFace, VGG16 & 10 Azure nodes emulating Cameras / 2 Azure nodes / 12 Azure nodes & Emulated hierarchical networks & \multicolumn{1}{c}{$\backslash$} & Introduce dominant demand to identify the best tradeoff between multiple resources and accuracy & Improve accuracy by 5.4$\times$ compared to VideoStorm and only lose 6\% accuracy of the optimum \\
      \midrule
      {\multirow{5}[1]{*}[-6.5em]{ \rotatebox{90}{\textbf{Horizontal Collaboration}} }} & \rotatebox{90}{ \tabincell{c}{\textit{MoDNN} \\ \cite{MoDNN}} } & VGG-16 & Multiple  LG Nexus 5 / N/A / N/A & WLAN  & MXNet & Partition already trained DNN models onto several mobile devices to accelerate DNN computations by alleviating device-level computing cost and memory usage & When the number of worker nodes increases from 2 to 4, MoDNN can speedup the DNN computation by 2.17-4.28$\times$ \\
  \cmidrule{2-8}          & \rotatebox{90}{\cite{Alwani2016}} & VGGNet-E, AlexNet & Xilinx Virtex-7 FPGA simulating multiple end devices / N/A / N/A & On-chip simulation & Torch, Vivado HLS  & Fuse the processing of multiple CNN layers and enable caching of intermediate data to save data transfer (bandwidth) & Reduce the total data transfer by 95\%, from 77MB down to 3.6MB per image \\
  \cmidrule{2-8}          & \rotatebox{90}{ \tabincell{c}{\textit{DeepThings} \\ \cite{DeepThings}} } & YOLOv2  & Perfromance-limited Raspberry Pi 3 Model B / Raspberry Pi 3 Model B as gateway / N/A & WLAN & Darknet  & Employ a scalable Fused Tile Partitioning of CNN layers to minimize memory footprint while exposing parallelism and a novel work scheduling process to reduce overall execution latency & Reduce memory footprint by more than 68\% without sacrificing accuracy, improve throughput by 1.7$\times$-2.2$\times$ and speedup CNN inference by 1.7$\times$-3.5$\times$ \\
  \cmidrule{2-8}          & \rotatebox{90}{ \tabincell{c}{\textit{DeepCham} \\ \cite{DeepCham}} } & AlexNet  & Multiple LG G2 / Wi-Fi router connected with a Linux server / N/A & WLAN \& LAN & Android Caffe, OpenCV, EdgeBoxes & Coordinate participating mobile users for collaboratively training a domain-aware adaptation model to improve object recognition accuracy & Improve the object recognition accuracy by 150\% when compared to that achieved merely using a generic DL model \\
  \cmidrule{2-8}          & \rotatebox{90}{ \tabincell{c}{\textit{LAVEA} \\ \cite{LAVEA}} } & OpenALPR & Raspberry PI 2 \& Raspberry PI 3 / Servers with quad-core CPU and 4GB RAM / N/A & WLAN \& LAN & Docker, Redis  & Design various task placement schemes that are tailed for inter-edge collaboration to minimize the service response time & Have a speedup ranging from 1.3$\times$ to 4$\times$ (1.2$\times$ to 1.7$\times$) against running in local (client-cloud confguration) \\
      \bottomrule
      \end{tabular}%
    \label{tab:addlabel}%
  \end{table*}%

Though on-device DL computation, illustrated in Sec. \ref{sec:AIinEdge}, can cater for lightweight DL services. 
Nevertheless, an independent end device still cannot afford intensive DL computation tasks. }
The concept of edge computing can potentially cope with this dilemma by offloading DL computation from end devices to edge or (and) the cloud. 
Accompanied by the edge architectures, DL-centric edge nodes can become the significant extension of cloud computing infrastructure to deal with massive DL tasks. 
In this section, we classify four modes for Edge DL computation, as exhibited in Fig. \ref{fig:EdgeComputingMode}. 

\subsubsection{Integral Offloading}
\label{subsubsec:integraloffloading}

The most natural mode of DL computation offloading is similar to the existed ``end-cloud'' computing, i.e., the end device sends its computation requests to the cloud for DL inference results (as depicted in Fig. \ref{fig:EdgeComputingMode}(a)). 
This kind of offloading is straightforward by extricating itself from DL task decomposition and combinatorial problems of resource optimization, which may bring about additional computation cost and scheduling delay, and thus simple to implement. 
{\color{red}In \cite{DeepDecision}, the proposed distributed infrastructure \textit{DeepDecision} ties together powerful edge nodes with less powerful end devices. }
In \textit{DeepDecision}, DL inference can be performed on the end or the edge, depending on the tradeoffs between the inference accuracy, the inference latency, the DL model size, the battery level, and network conditions. 
With regard to each DL task, the end device decides whether locally processing or offloading it to an edge node. 

Further, the workload optimization among edge nodes should not be ignored in the offloading problem, since edge nodes are commonly resource-restrained compared to the cloud. 
In order to satisfy the delay and energy requirements of accomplishing a DL task with limited edge resources, providing DL models with different model sizes and performance in the edge can be adopted to fulfill one kind of task. 
Hence, multiple VMs or containers, undertaking different DL models separately, can be deployed on the edge node to process DL requests. 
Specifically, when a DL model with lower complexity can meet the requirements, it is selected as the serving model. 
For instance, by optimizing the workload assignment weights and computing capacities of VMs, \textit{MASM} \cite{MASM} can reduce the energy cost and delay while guaranteeing the DL inference accuracy.

\subsubsection{Partial Offloading}
\label{subsubsec:partialoffloading}


Partially offloading the DL task to the edge is also feasible (as depicted in Fig. \ref{fig:EdgeComputingMode}(b)). 
An offloading system can be developed to enable online fine-grained partition of a DL task, and determine how to allocate these divided tasks to the end device and the edge node. 
As exemplified in \cite{Cuervo}, \textit{MAUI}, capable of adaptively partitioning general computer programs, can conserve an order of magnitude energy by optimizing the task allocation strategies, under the network constraints. 
More importantly, this solution can decompose the whole program at runtime instead of manually partitioning of programmers before program deploying. 

{\color{red}Further, particularly for DL computation, \textit{DeepWear} \cite{DeepWear} abstracts a DL model as a Directed Acyclic Graph (DAG), where each node represents a layer and each edge represents the data flow among those layers. 
To efficiently determine partial offloading decisions, \textit{DeepWear} first prunes the DAG by keeping only the computation-intensive nodes, and then grouping the repeated sub-DAGs.
In this manner, the complex DAG can be transformed into a linear and much simpler one, thus enabling a linear complexity solution for selecting the optimal partition to offload.

Nevertheless, uploading a part of the DL model to the edge nodes may still seriously delay the whole process of offloading DL computation.}
To deal with this challenge, an incremental offloading system \textit{IONN} is proposed in \cite{IONN}. 
Differ from packing up the whole DL model for uploading, \textit{IONN} divides a DL model, prepared for uploading, into multiple partitions, and uploads them to the edge node in sequential. 
The edge node, receiving the partitioned models, incrementally builds the DL model as each partitioned model arrives, while being able to execute the offloaded partial DL computation even before the entire DL model is uploaded. 
Therefore, the key lies in the determination concerning the best partitions of the DL model and the uploading order. 
Specifically, on the one hand, DNN layers, performance benefit and uploading overhead of which are high and low, respectively, are preferred to be uploaded first, and thus making the edge node quickly build a partial DNN to achieve the best-expected query performance. 
On the other hand, unnecessary DNN layers, which cannot bring in any performance increase, are not uploaded and hence avoiding the offloading.

\subsubsection{Vertical Collaboration}
\label{subsubsec:verticalcollaboration}

Expected offloading strategies among ``End-Edge'' architecture, as discussed in Section \ref{subsubsec:integraloffloading} and \ref{subsubsec:partialoffloading}, are feasible for supporting less computation-intensive DL services and small-scale concurrent DL queries. 
However, when a large number of DL queries need to be processed at one time, a single edge node is certainly insufficient. 

A natural choice of collaboration is the edge performs data pre-processing and preliminary learning, when the DL tasks are offloaded. 
Then, the intermediate data, viz., the output of edge architectures, are transmitted to the cloud for further DL computation \cite{Huang2017}. 
Nevertheless, the hierarchical structure of DNNs can be further excavated for fitting the vertical collaboration. 
In \cite{Neurosurgeon}, all layers of a DNN are profiled on the end device and the edge node in terms of the data and computation characteristics, in order to generate performance prediction models.
Based on these prediction models, wireless conditions and server load levels, the proposed \textit{Neurosurgeon} evaluates each candidate point in terms of end-to-end latency or mobile energy consumption and partition the DNN at the best one. 
Then, it decides the allocation of DNN partitions, i.e., which part should be deployed on the end, the edge or the cloud, while achieving best latency and energy consumption of end devices. 

By taking advantages of EEoI (Section \ref{subsec:EarlyExitofInference}), vertical collaboration can be more adapted. 
Partitions of a DNN can be mapped onto a distributed computing hierarchy (i.e., the end, the edge and the cloud) and can be trained with multiple early exit points \cite{Teerapittayanon2017}. 
Therefore, the end and the edge can perform a portion of DL inference on themselves rather than directly requesting the cloud. Using an exit point after inference, results of DL tasks, the local device is confident about, can be given without sending any information to the cloud. 
For providing more accurate DL inference, the intermediate DNN output will be sent to the cloud for further inference by using additional DNN layers. 
Nevertheless, the intermediate output, e.g., high-resolution surveillance video streams, should be carefully designed much smaller than the raw input, therefore drastically reducing the network traffic required between the end and the edge (or the edge and the cloud).

Though vertical collaboration can be considered as an evolution of cloud computing, i.e., ``end-cloud'' strategy. 
Compared to the pure ``end-edge'' strategy, the process of vertical collaboration may possibly be delayed, due to it requires additional communication with the cloud. 
However, vertical collaboration has its own advantages. 
One side, when edge architectures cannot afford the flood of DL queries by themselves, the cloud architectures can share partial computation tasks and hence ensure servicing these queries. 
On the other hand, the raw data must be preprocessed at the edge before they are transmitted to the cloud. 
If these operations can largely reduce the size of intermediate data and hence reduce the network traffic, the pressure of backbone networks can be alleviated.

\subsubsection{Horizontal Collaboration}
\label{subsubsec:horizontalcollaboration}

In Section \ref{subsubsec:verticalcollaboration}, vertical collaboration is discussed. 
However, devices among the edge or the end can also be united without the cloud to process resource-hungry DL applications, i.e., horizontal collaboration. 
By this means, the trained DNN models or the whole DL task can be partitioned and allocated to multiple end devices or edge nodes to accelerate DL computation by alleviating the resource cost of each of them. 
\textit{MoDNN}, proposed in \cite{MoDNN}, executes DL in a local distributed mobile computing system over a Wireless Local Area Network (WLAN). 
Each layer of DNNs is partitioned into slices to increase parallelism and to reduce memory footprint, and these slices are executed layer-by-layer. 
By the execution parallelism among multiple end devices, the DL computation can be significantly accelerated. 

With regard to specific DNN structures, e.g., CNN, a finer grid partitioning can be applied to minimize communication, synchronization, and memory overhead \cite{Alwani2016}. 
In \cite{DeepThings}, a Fused Tile Partitioning (FTP) method, able to divide each CNN layer into independently distributable tasks, is proposed. 
In contrast to only partitioning the DNN by layers as in \cite{Neurosurgeon}, FTP can fuse layers and partitions them vertically in a grid fashion, hence minimizing the required memory footprint of participated edge devices regardless of the number of partitions and devices, while reducing communication and task migration cost as well. 
Besides, to support FTP, a distributed work-stealing runtime system, viz., idle edge devices stealing tasks from other devices with active work items \cite{DeepThings}, can adaptively distribute FTP partitions for balancing the workload of collaborated edge devices. 

{\color{red}
\subsection{Tailoring Edge Frameworks for DL}
\label{subsebsec:edgearchitecturefordl}

Though there are gaps between the computational complexity and energy efficiency required by DL and the capacity of edge hardware \cite{Xu2018d}, customized edge DL frameworks can help efficiently 
1) match edge platform and DL models;
2) exploit underlying hardware in terms of performance and power; 
3) orchestrate and maintain DL services automatically. 

First, where to deploy DL services in edge computing (cellular) networks should be determined. 
The RAN controllers deployed at edge nodes are introduced in \cite{Polese2018} to collect the data and run DL services, while the network controller, placed in the cloud, orchestrates the operations of the RAN controllers. 
In this manner, after running and feeding analytics and extract relevant metrics to DL models, these controllers can provide DL services to the users at the network edge. 

Second, as the deployment environment and requirements of DL models can be substantially different from those during model development, customized operators, adopted in developing DL models with (Py)Torch, TensorFlow, etc., may not be directly executed with the DL framework at the edge. 
To bridge the gap between deployment and development, the authors of \cite{Lai2018a} propose to specify DL models in development using the deployment tool with an operator library from the DL framework deployed at the edge.
Furthermore, to automate the selection and optimization of DL models, \textit{ALOHA} \cite{ALOHA} formulates a toolflow: 
1) Automate the model design. 
It generates the optimal model configuration by taking into account the target task, the set of constraints and the target architecture;
2) Optimize the model configuration. 
It partitions the DL model and accordingly generates architecture-aware mapping information between different inference tasks and the available resources.
3) Automate the model porting. 
It translates the mapping information into adequate calls to computing and communication primitives exposed by the target architecture.

Third, the orchestration of DL models deployed at the edge should be addressed. 
\textit{OpenEI} \cite{OpenEI} defines each DL algorithm as a four-element tuple <Accuracy, Latency, Energy, Memory Footprint> to evaluate the Edge DL capability of the target hardware platform.  
Based on such tuple, \textit{OpenEI} can select a matched model for a specific edge platform based on different Edge DL capabilities in an online manner.
\textit{Zoo} \cite{Zoo} provides a concise Domain-specific Language (DSL) to enable easy and type-safe composition of DL services. 
Besides, to enable a wide range of geographically distributed topologies, analytic engines, and DL services, \textit{ECO} \cite{ECOHotEdge} uses a graph-based overlay network approach to 1) model and track pipelines and dependencies and then 
2) map them to geographically distributed analytic engines ranging from small edge-based engines to powerful multi-node cloud-based engines. 
By this means, DL computation can be distributed as needed to manage cost and performance, while also supporting other practical situations, such as engine heterogeneity and discontinuous operations.

Nevertheless, these pioneer works are not ready to natively support valuable and also challenging features discussed in Section \ref{subsec:edgecomputingmodefordl}, such as computation offloading and collaboration, which still calls for further development.

\subsection{Performance Evaluation for Edge DL}
\label{subsec:performanceevaluationforedgedl}

Throughout the process of selecting appropriate edge hardware and associated software stacks for deploying different kinds of Edge DL services, it is necessary to evaluate their performance. 
Impartial evaluation methodologies can point out possible directions to optimize software stacks for specific edge hardware. 
In \cite{Zhang2018f}, for the first time, the performance of DL libraries is evaluated by executing DL inference on resource-constrained edge devices, pertaining to metrics like latency, memory footprint, and energy. 
In addition, particularly for Android smartphones, as one kind of edge devices with mobile CPUs or GPUs, \textit{AI Benchmark} \cite{Ignatov2018} extensively evaluates DL computation capabilities over various device configurations. 
Experimental results show that no single DL library or hardware platform can entirely outperform others, and loading the DL model may take more time than that of executing it. 
These discoveries imply that there are still opportunities to further optimize the fusion of edge hardware, edge software stacks, and DL libraries. 

Nonetheless, a standard testbed for Edge DL is missing, which hinders the study of edge architectures for DL. 
To evaluate the end-to-end performance of Edge DL services, not only the edge computing architecture but also its combination with end devices and the cloud shall be established, such as \textit{openLEON} \cite{AndresRamiro2018} and \textit{CAVBench} \cite{Wang2018l} particularly for vehicular scenarios. 
Furthermore, simulations of the control panel of managing DL services are still not dabbled. 
An integrated testbed, consisting of wireless links and networking models, service requesting simulation, edge computing platforms, cloud architectures, etc., is ponderable in facilitating the evolution of ``Edge Computing for DL''.
}
\section{Deep Learning Training at Edge}
\label{sec:AIatEdge}

Present DL training (distributed or not) in the cloud data center, namely cloud training, or cloud-edge training \cite{Huang2018b}, viz., training data are preprocessed at the edge and then transmitted to cloud, are not appropriate for all kind of DL services, especially for DL models requiring locality and persistent training. 
Besides, a significant amount of communication resources will be consumed, and hence aggravating wireless and backbone networks if massive data are required to be continually transmitted from distributed end devices or edge nodes to the cloud. 
For example, with respect to surveillance applications integrated with object detection and target tracking, if end devices directly send a huge amount of real-time monitoring data to the cloud for persistent training, it will bring about high networking costs. 
In addition, merging all data into the cloud might violate privacy issues. 
All these challenges put forward the need for a novel training scheme against existing cloud training. 

Naturally, the edge architecture, which consists of a large number of edge nodes with modest computing resources, can cater for alleviating the pressure of networks by processing the data or training at themselves. 
Training at the edge or potentially among ``end-edge-cloud'', treating the edge as the core architecture of training, is called ``DL Training at Edge''. 
Such kind of DL training may require significant resources to digest distributed data and exchange updates.
Nonetheless, FL is emerging and is promised to address these issues.
We summarize select works on FL in Table \ref{tab:flatedge}.

\subsection{Distributed Training at Edge}
\label{subsec:trainingunderidealizedtrust}

\begin{figure}[!!!!!!!!!!!!!!hhhhhhhhhht]
    \centering
    \includegraphics[width=8 cm]{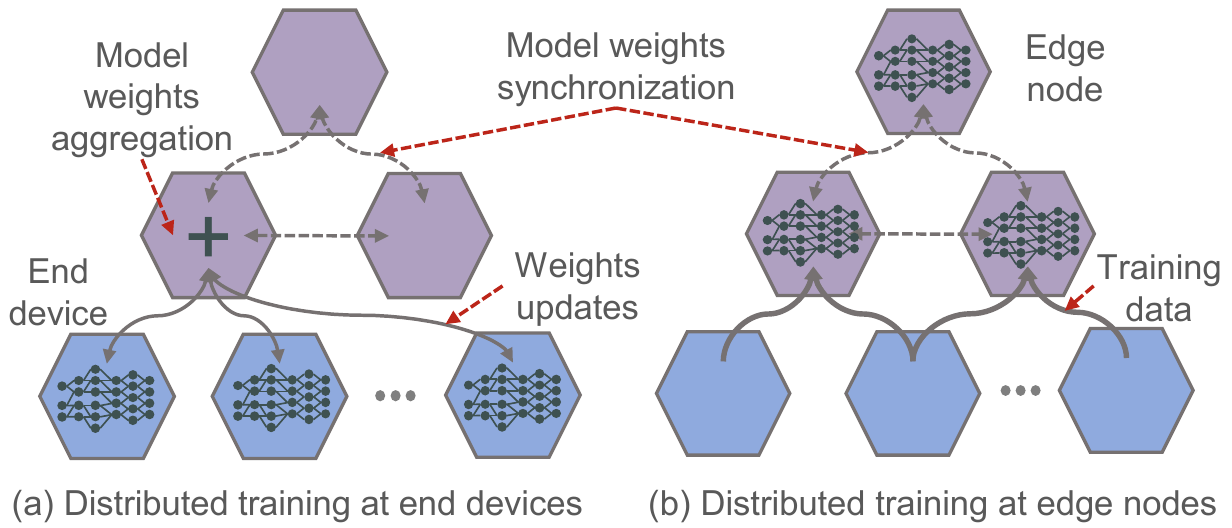}
    \caption{Distributed DL training at edge environments.}
    \label{fig:distributedtrainingatedge}
\end{figure}

Distributed training at the edge can be traced back to the work of \cite{Kamath2016}, where a decentralized Stochastic Gradient Descent (SGD) method is proposed for the edge computing network to solve a large linear regression problem. However, this proposed method is designed for seismic imaging application and can not be generalized for future DL training, since the communication cost for training large scale DL models is extremely high. In \cite{Valerio2017}, two different distributed learning solutions for edge computing environments are proposed. As depicted in Fig. \ref{fig:distributedtrainingatedge}, one solution is that each end device trains a model based on local data, and then these model updates are aggregated at edge nodes. Another one is edge nodes train their own local models, and their model updates are exchanged and refined for constructing a global model. Though large-scale distributed training at edge evades transmitting bulky raw dataset to the cloud, the communication cost for gradients exchanging between edge devices is inevitably introduced. Besides, in practical, edge devices may suffer from higher latency, lower transmission rate and intermittent connections, and therefore further hindering the gradients exchanging between DL models belong to different edge devices. 

Most of the gradient exchanges are redundant, and hence updated gradients can be compressed to cut down the communication cost while preserving the training accuracy (such as \textit{DGC} in \cite{Dally2018}). \textit{First}, \textit{DGC} stipulates that only important gradients are exchanged, i.e., only gradients larger than a heuristically given threshold are transmitted. In order to avoid the information losing, the rest of the gradients are accumulated locally until they exceed the threshold. To be noted, gradients whether being immediately transmitted or accumulated for later exchanging will be coded and compressed, and hence saving the communication cost. \textit{Second}, considering the sparse update of gradients might harm the convergence of DL training, momentum correction and local gradient clipping are adopted to mitigate the potential risk. By momentum correction, the sparse updates can be approximately equivalent to the dense updates. Before adding the current gradient to previous accumulation on each edge device locally, gradient clipping is performed to avoid the exploding gradient problem possibly introduced by gradient accumulation. Certainly, since partial gradients are delayed for updating, it might slow down the convergence. Hence, \textit{finally}, for preventing the stale momentum from jeopardizing the performance of training, the momentum for delayed gradients is stopped, and less aggressive learning rate and gradient sparsity are adopted at the start of training to reduce the number of extreme gradients being delayed.  

With the same purpose of reducing the communication cost of synchronizing gradients and parameters during distributed training, two mechanisms can be combined together \cite{Taoa}. The first is transmitting only important gradients by taking advantage of sparse training gradients \cite{strom2015scalable}. Hidden weights are maintained to record times of a gradient coordinate participating in gradient synchronization, and gradient coordinates with large hidden weight value are deemed as important gradients and will be more likely be selected in the next round training. On the other hand, the training convergence will be greatly harmed if residual gradient coordinates (i.e., less important gradients) are directly ignored, hence, in each training round, small gradient values are accumulated. Then, in order to avoid that these outdated gradients only contribute little influence on the training, momentum correction, viz., setting a discount factor to correct residual gradient accumulation, is applied.

Particularly, when training a large DL model, exchanging corresponded model updates may consume more resources. Using an online version of KD can reduce such kind of communication cost \cite{Jeong2018}. In other words, the model outputs rather the updated model parameters on each device are exchanged, making the training of large-sized local models possible. Besides communication cost, privacy issues should be concerned as well. For example, in \cite{Fredrikson:2015:MIA:2810103.2813677}, personal information can be purposely obtained from training data by making use of the privacy leaking of a trained classifier. The privacy protection of training dataset at the edge is investigated in \cite{Du2018a}. Different from \cite{Valerio2017, Dally2018, Taoa}, in the scenario of \cite{Du2018a}, training data are trained at edge nodes as well as be uploaded to the cloud for further data analysis. Hence, Laplace noises \cite{10.1007/11681878_14} are added to these possibly exposed training data for enhancing the training data privacy assurance.

\subsection{Vanilla Federated Learning at Edge}
\label{subsec:flatedge}

In Section \ref{subsec:trainingunderidealizedtrust}, the holistic network architecture is explicitly separated, specifically, training is limited at the end devices or the edge nodes independently instead of among both of them.
Certainly, by this means, it is simple to orchestrate the training process since there is no need to deal with heterogeneous computing capabilities and networking environments between the end and the edge. 
Nonetheless, DL training should be ubiquitous as well as DL inference. Federated Learning (FL) \cite{mcmahan2016communication, Bonawitz2017} is emerged as a practical DL training mechanism among the end, the edge and the cloud. 
Though in the framework of native FL, modern mobile devices are taken as the clients performing local training. 
Naturally, these devices can be extended more widely in edge computing \cite{Samarakoon2018, Xie2019a}. 
End devices, edge nodes and servers in the cloud can be equivalently deemed as clients in FL. 
These clients are assumed capable of handling different levels of DL training tasks, and hence contribute their updates to the global DL model. 
In this section, fundamentals of FL are discussed. 

\begin{figure}[!!!!!!!!!!!!!!hhhhhhhhhht]
    \centering
    \includegraphics[width=8.85 cm]{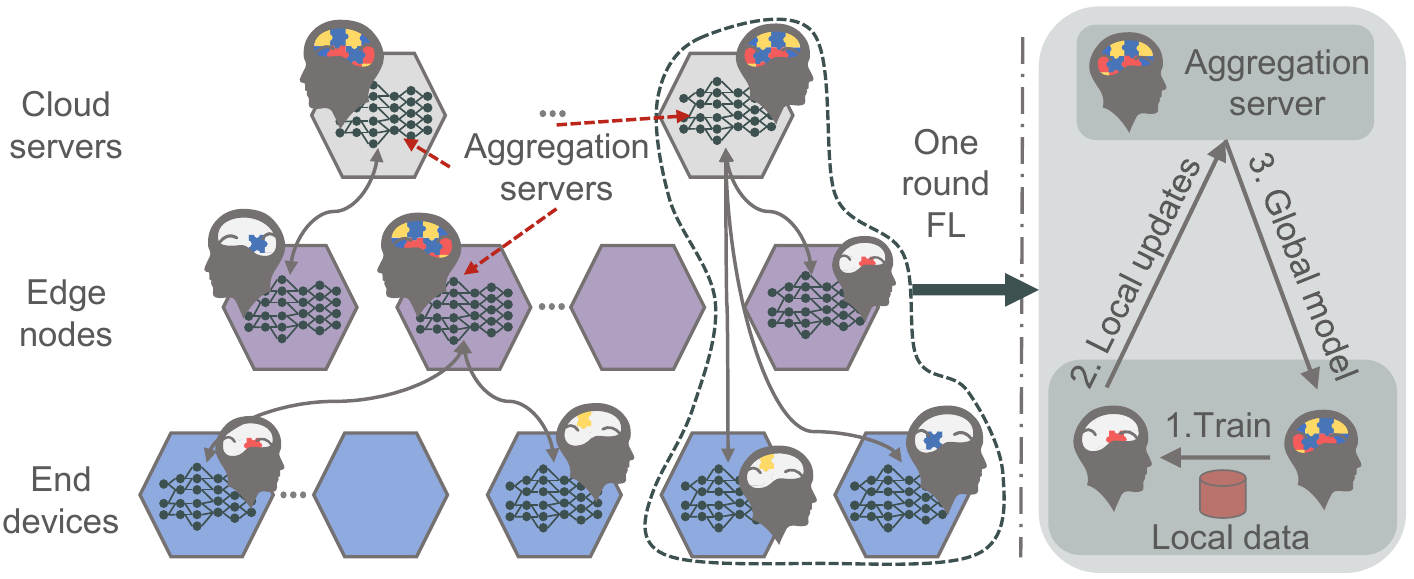}
    \caption{Federated learning among hierarchical network architectures.}
    \label{fig:federatedlearning}
\end{figure}

Without requiring uploading data for central cloud training, FL \cite{mcmahan2016communication, Bonawitz2017} can allow edge devices to train their local DL models with their own collected data and upload only the updated model instead. As depicted in Fig. \ref{fig:federatedlearning}, FL iteratively solicits a random set of edge devices to 1) download the global DL model from an aggregation server (use ``server'' in following), 2) train their local models on the downloaded global model with their own data, and 3) upload only the updated model to the server for model averaging. Privacy and security risks can be significantly reduced by restricting the training data to only the device side, and thus avoiding the privacy issues as in \cite{Fredrikson:2015:MIA:2810103.2813677}, incurred by uploading training data to the cloud. Besides, FL introduces \textit{FederatedAveraging} to combine local SGD on each device with a server performing model averaging. Experimental results corroborate \textit{FederatedAveraging} is robust to unbalanced and non-IID data and can facilitate the training process, viz., reducing the rounds of communication needed to train a DL model. 

To summarize, FL can deal with several key challenges in edge computing networks: 1) \textbf{Non-IID training data}. Training data on each device is sensed and collected by itself. Hence, any individual training data of a device will not be able to represent the global one. In FL, this can be met by \textit{FederatedAveraging}; 2) \textbf{Limited communication}. Devices might potentially off-line or located in a poor communication environment. Nevertheless, performing more training computation on resource-sufficient devices can cut down communication rounds needed for global model training. In addition, FL only selects a part of devices to upload their updates in one round, therefore successfully handling the circumstance where devices are unpredictably off-line; 3) \textbf{Unbalanced contribution}. It can be tackled by \textit{FederatedAveraging}, specifically, some devices may have less free resources for FL, resulting in varying amounts of training data and training capability among devices; 4) \textbf{Privacy and security}. The data need to be uploaded in FL is only the updated DL model. Further, secure aggregation and differential privacy \cite{10.1007/11681878_14}, which are useful in avoiding the disclosure of privacy-sensitive data contained in local updates, can be applied naturally.


\subsection{{\color{red}Communication-efficient FL}}
\label{subsubsec:FederatedLearningOptimization}

In FL, raw training data are not required to be uploaded, thus largely reducing the communication cost. 
However, FL still needs to transmit locally updated models to the central server. 
Supposing the DL model size is large enough, uploading updates, such as model weights, from edge devices to the central server may also consume nonnegligible communication resources. 
To meet this, we can let FL clients communicate with the central server periodically (rather continually) to seek consensus on the shared DL model \cite{Abad2019}.
In addition, \textit{structured update}, \textit{sketched update} can help enhance the communication efficiency when clients uploading updates to the server as well. 
\textit{Structured update} means restricting the model updates to have a pre-specified structure, specifically, 
1) low-rank matrix; or 
2) sparse matrix \cite{Konecny2016, Abad2019}. 
On the other hand, for \textit{sketched update}, full model updates are maintained. 
But before uploading them for model aggregation, combined operations of subsampling, probabilistic quantization, and structured random rotations are performed to compress the full updates \cite{Konecny2016}. 
\textit{FedPAQ} \cite{FedPAQ} simultaneously incorporates these features and provides near-optimal theoretical guarantees for both strongly convex and non-convex loss functions, while empirically demonstrating the communication-computation tradeoff.

Different from only investigating on reducing communication cost on the uplink, \cite{Caldas2018a} takes both server-to-device (downlink) and device-to-server (uplink) communication into consideration. 
For the downlink, the weights of the global DL model are reshaped into a vector, and then subsampling and quantization are applied \cite{Konecny2016}. 
Naturally, such kind of model compression is lossy, and unlike on the uplink (multiple edge devices are uploading their models for averaging), the loss cannot be mitigated by averaging on the downlink. 
\textit{Kashin's representation} \cite{Kas77} can be utilized before subsampling as a basis transform to mitigate the error incurred by subsequent compression operations. 
{\color{red}Furthermore, for the uplink, each edge device is not required to train a model based on the whole global model locally, but only to train a smaller sub-model or pruned model \cite{Jiang2019b} instead. 
Since sub-models and pruned models are more lightweight than the global model, the amount of data in updates uploading is reduced. }

Computation resources of edge devices are scarce compared to the cloud. 
Additional challenges should be considered to improve communication efficiencies: 
1) Computation resources are heterogeneous and limited at edge devices; 
2) Training data at edge devices may be distributed non-uniformly \cite{Wang2018f, Wang, Tuor2018}. 
For more powerful edge devices, \textit{ADSP} \cite{Hu2019c} lets them continue training while committing model aggregation at strategically decided intervals. 
For general cases, based on the deduced convergence bound for distributed learning with non-IID data distributions, the aggregation frequency under given resource budgets among all participating devices can be optimized with theoretical guarantees \cite{Wang2018f}.
{\color{red}\textit{Astraea} \cite{Astraea} reduces $92\%$ communication traffic by designing a mediator-based multi-client rescheduling strategy. 
On the one hand, \textit{Astraea} leverages data augmentation [5] to alleviate the defect of non-uniformly distributed training data. 
On the other hand, \textit{Astraea} designs a greedy strategy for mediator-based rescheduling, in order to assign clients to the mediators. 
Each mediator traverses the data distribution of all unassigned clients to select the appropriate participating clients, aiming to make the mediator's data distribution closest to the uniform distribution, i.e., minimizing the KullbackLeibler divergence \cite {KLDivergence} between mediator's data distribution and uniform distribution. 
When a mediator reaches the max assigned clients limitation, the central server will create a new mediator and repeat the process until all clients have been assigned with training tasks.}

Aiming to accelerate the global aggregation in FL, \cite{Yang2018d} takes advantage of over-the-air computation \cite{Nazer2007, Chen2018l, Zhu2018d}, of which the principle is to explore the superposition property of a wireless multiple-access channel to compute the desired function by the concurrent transmission of multiple edge devices. 
The interferences of wireless channels can be harnessed instead of merely overcoming them. 
During the transmission, concurrent analog signals from edge devices can be naturally weighed by channel coefficients.
Then the server only needs to superpose these reshaped weights as the aggregation results, nonetheless, without other aggregation operations.

\begin{table*}[htbp!!!!!!!]
    \centering
    \scriptsize
    \caption{{\color{red}Summary of the Selected Works on FL}}
      \begin{tabular}{cm{3em}<{\centering}m{4em}<{\centering}m{10em}<{\centering}m{5em}<{\centering}m{19em}m{17em}}
      \toprule
            & \textbf{Ref.} & \textbf{DL Model} & \textbf{Scale} & \textbf{Dependency} & \multicolumn{1}{c}{\textbf{Main Idea}} & \multicolumn{1}{c}{\textbf{Key Metrics or Performance}} \\
      \midrule
      {\multirow{2}[1]{*}[0.2em]{ {\rotatebox{90}{\textbf{Vanilla FL}}} }} &  \cite{mcmahan2016communication} & FCNN, CNN, LSTM & Up to $5\mathrm{e}{5}$ clients & TensorFlow & Leave the training data distributed on the mobile devices, and learns a shared model by aggregating locally-training updates & Communication rounds reduction: 10-100$\times$  \\
  \cmidrule{2-7}          & \cite{Bonawitz2017} & RNN   & Up to $1.5\mathrm{e}{6}$ clients & TensorFlow & Pace steering for scalable FL & Scalability improvement: up to 1.5e6 clients \\
      \midrule
      {\multirow{8}[1]{*}[-8em]{ {\rotatebox{90}{\textbf{Communication-efficient FL}}} }} & \cite{Abad2019} & ResNet18 & 4 clients per cluster / 7 clusters & $\backslash$     & Gradient sparsification; Periodic averaging  & Top 1 accuracy; Communication latency reduction \\
  \cmidrule{2-7}          & \cite{Konecny2016} & CNN, LSTM & Up to $1\mathrm{e}{3}$ clients & $\backslash$     & Sketched updates & Communication cost reduction: by two orders of magnitude \\
  \cmidrule{2-7}          & \cite{Caldas2018a} & CNN   & Up to 500 clients & TensorFlow & Lossy compression on the global model; Federated Dropout  & Downlink reduction: 14$\times$; Uplink reduction: 28$\times$; Local computation reduction: 1.7$\times$ \\
  \cmidrule{2-7}          & \cite{Hu2019c} & CNN, RNN & Up to 37 clients & TensorFlow & Let faster clients continue with their mini-batch training to keep overall synchronization & Convergence acceleration: 62.4\% \\
  \cmidrule{2-7}          & \cite{Wang2018f} & CNN   & 5-500 clients (simulation); 3 Raspberry Pi and 2 laptops (testbed) & $\backslash$     & Design a control algorithm that determines the best trade-off between local update and global aggregation & Training accuracy under resource budget \\
  \cmidrule{2-7}          & \cite{FedPAQ} & FCNN  & 50 clients & $\backslash$     & Periodic averaging; Partial device participation; Quantized message-passing & Total training loss and time \\
  \cmidrule{2-7}          & \cite{Astraea} & CNN   & 500 clients & $\backslash$    & Global data distribution based data augmentation; Mediator based multi-client rescheduling & Top 1 accuracy imrpovement: 5.59\%-5.89\%; Communication traffic reduction: 92\% \\
  \cmidrule{2-7}          & \cite{Jiang2019b} &  LeNet, CNN,  VGG11 & 10 Raspberry Pi & Py(Torch) & Jointly trains and prunes the model in a federated manner & Communication and computation load reduction \\
      \midrule
      {\multirow{4}[1]{*}[-1em]{ {\rotatebox{90}{\textbf{ \tabincell{c}{Resource \\ -optimized FL} }}} }} & \cite{Xu2019f} & AlexNet, LeNet & Multiple Nvidia Jetson Nano & $\backslash$    & Partially train the model by masking a particular number of resource-intensive neurons & Training acceleration: 2$\times$; Model accuracy improvement: 4\% \\
  \cmidrule{2-7}          & \cite{Dinh2019a} & $\backslash$     & Up to 50 clients & TensorFlow & Jointly optimize FL parameters and resources of user equipments & Convergence rate; Test accuracy \\
  \cmidrule{2-7}          & \cite{Chen2019m} & $\backslash$    & 20 clients / 1 BS & $\backslash$     & Jointly optimize wireless resource allocation and client selection & Reduction of the FL loss function value: up to 16\% \\
  \cmidrule{2-7}          & \cite{Li2019} & LSTM  & 23-1,101 clients & TensorFlow &  Modify FL training objectives with $\alpha$-fairness & Fairness; Training accuracy \\
      \midrule
      {\multirow{3}[1]{*}[0.3em]{ {\rotatebox{90}{ \textbf{ \tabincell{c}{Security \\ -enhanced FL} } }} }} & \cite{Xie2019a} & CNN   & 100 clients & MXNET & Use the trimmed mean as a robust aggregation & Top 1 accuracy against data poisoning \\
  \cmidrule{2-7}          & \cite{Bonawitz2016} & $\backslash$    & $2\mathrm{e}{10}$-$2\mathrm{e}{14}$ clients & $\backslash$     & Use Secure Aggregation to protect the privacy of each client’s model gradient &  Communication expansion: 1.73$\times$-1.98$\times$ \\
  \cmidrule{2-7}          & \cite{Kim2018a} & $\backslash$     & 10 clients & $\backslash$    & Leverage blockchain to exchange and verify model updates of local training & Learning completion latency \\
      \bottomrule
      \end{tabular}%
    \label{tab:flatedge}%
\end{table*}%

{\color{red}
\subsection{Resource-optimized FL}

When FL deploys the same neural network model to heterogeneous edge devices, devices with weak computing power (stragglers) may greatly delay the global model aggregation. 
Although the training model can be optimized to accelerate the stragglers, due to the limited resources of heterogeneous equipment, the optimized model usually leads to diverged structures and severely defect the collaborative convergence.
\textit{ELFISH} \cite{Xu2019f} first analyzes the computation consumption of the model training in terms of the time cost, memory usage, and computation workload. 
Under the guidance of the model analysis, which neurons need to be masked in each layer to ensure that the computation consumption of model training meets specific resource constraints can be determined. 
Second, unlike generating a deterministically optimized model with diverged structures, 
different sets of neurons will be dynamically masked in each training period and recovered and updated during the subsequent aggregation period, thereby ensuring comprehensive model updates overtime. 
It is worth noting that although \textit{ELFISH} improves the training speed by 2$\times$ through resource optimization, the idea of \textit{ELFISH} is to make all stragglers work synchronously, the synchronous aggregation of which may not able to handle extreme situations.

When FL is deployed in a mobile edge computing scenario, the wall-clock time of FL will mainly depend on the number of clients and their computing capabilities. 
Specifically, the total wall-clock time of FL includes not only the computation time but also the communication time of all clients. 
On the one hand, the computation time of a client depends on the computing capability of the clients and local data sizes.
On the other hand, the communication time correlates to clients' channel gains, transmission power, and local data sizes.
Therefore, to minimize the wall-clock training time of the FL, appropriate resource allocation for the FL needs to consider not only FL parameters, such as accuracy level for computation-communication trade-off, but also the resources allocation on the client side, such as power and CPU cycles.

However, minimizing the energy consumption of the client and the FL wall-clock time are conflicting. 
For example, the client can save energy by always maintain its CPU at low frequency, but this will definitely increase training time. 
Therefore, in order to strike a balance between energy cost and training time, the authors of \cite{Dinh2019a} first design a new FL algorithm \textit{FEDL} for each client to solve its local problem approximately till a local accuracy level achieved. 
Then, by using Pareto efficiency model \cite{ParetoEfficiency}, they formulate a non-convex resource allocation problem for \textit{FEDL} over wireless networks to capture the trade-off between the clients' energy cost and the FL wall-clock time). 
Finally, by exploiting the special structure of that problem, they decompose it into three sub-problems, and accordingly derive closed-form solutions and characterize the impact of the Pareto-efficient controlling knob to the optimal.

Since the uplink bandwidth for transmitting model updates is limited, the BS must optimize its resource allocation while the user must optimize its transmit power allocation to reduce the packet error rates of each user, thereby improving FL performance. 
To this end, the authors of \cite{Chen2019m} formulate resource allocation and user selection of FL into a joint optimization problem, the goal of which is to minimize the value of the FL loss function while meeting the delay and energy consumption requirements. 
To solve this problem, they first derive a closed-form expression for the expected convergence rate of the FL in order to establish an explicit relationship between the packet error rates and the FL performance. 
Based on this relationship, the optimization problem can be reduced to a mixed-integer nonlinear programming problem, and then solved as follows: 
First, find the optimal transmit power under a given user selection and resource block allocation; 
Then, transform the original optimization problem into a binary matching problem; 
Finally, using Hungarian algorithm \cite{HungarianAlgo} to find the best user selection and resource block allocation strategy.

The number of devices involved in FL is usually large, ranging from hundreds to millions. 
Simply minimizing the average loss in such a large network may be not suited for the required model performance on some devices. 
In fact, although the average accuracy under vanilla FL is high, the model accuracy required for individual devices may not be guaranteed. 
To this end, based on the utility function $\alpha$-fairness \cite{COMSTFairnessinWirelessNetworks} used in fair resource allocation in wireless networks, the authors of \cite{Li2019} define a fair-oriented goal $q$-FFL for joint resource optimization. 
$q$-FFL minimizes an aggregate re-weighted loss parameterized by $q$, so that devices with higher loss are given higher relative weight, thus encouraging less variance (i.e., more fairness) in the accuracy distribution. 
Adaptively minimizing $q$-FFL avoids the burden of hand-crafting fairness constraints, and can adjust the goal according to the required fairness dynamically, achieving the effect of reducing the variance of accuracy distribution among participated devices.
}

\subsection{{\color{red}Security-enhanced FL}}
\label{subsubsec:SecurityEnhancement}

In vanilla FL, local data samples are processed on each edge device. Such a manner can prevent the devices from revealing private data to the server. However, the server also should not trust edge devices completely, since devices with abnormal behavior can forge or poison their training data, which results in worthless model updates, and hence harming the global model. To make FL capable of tolerating a small number of devices training on the poisoned dataset, \textit{robust federated optimization} \cite{Xie2019a} defines a trimmed mean operation. By filtering out not only the the values produced by poisoned devices but also the natural outliers in the normal devices, robust aggregation protecting the global model from data poisoning is achieved.

Other than intentional attacks, passive adverse effects on the security, brought by unpredictable network conditions and computation capabilities, should be concerned as well. FL must be robust to the unexpectedly drop out of edge devices, or else once a device loses its connection, the synchronization of FL in one round will be failed. To solve this issue, \textit{Secure Aggregation} protocol is proposed in \cite{Bonawitz2016} to achieve the robustness of tolerating up to one-third devices failing to timely process the local training or upload the updates. 

In turn, malfunctions of the aggregation server in FL may result in inaccurate global model updates and thereby distorting all local model updates. Besides,  edge devices (with a larger number of data samples) may be less willing to participate FL with others (with less contribution). Therefore, in \cite{Kim2018a}, combining Blockchain and FL as \textit{BlockFL} is proposed to realize 1) locally global model updating at each edge device rather a specific server, ensuring device malfunction cannot affect other local updates when updating the global model; 2) appropriate reward mechanism for stimulating edge devices to participate in FL. 

\section{Deep Learning for Optimizing Edge}
\label{sec:AIforEdge}

DNNs (general DL models) can extract latent data features, while DRL can learn to deal with decision-making problems by interacting with the environment. Computation and storage capabilities of edge nodes, along with the collaboration of the cloud, make it possible to use DL to optimize edge computing networks and systems. With regard to various edge management issues such as edge caching, offloading, communication, security protection, etc., 1) DNNs can process user information and data metrics in the network, as well as perceiving the wireless environment and the status of edge nodes, and based on these information 2) DRL can be applied to learn the long-term optimal resource management and task scheduling strategies, so as to achieve the intelligent management of the edge, viz., intelligent edge as shown in Table \ref{tab:DLforEdge}. 



\subsection{{\color{red}DL for Adaptive Edge Caching}}
\label{subsec:edgecaching}

From Content Delivery Network (CDN) \cite{HOFMANN200553} to caching contents in cellular networks, caching in the network have been investigated over the years to deal with soaring demand for multimedia services \cite{Wang2014}. 
Aligned with the concept of pushing contents near to users, edge caching \cite{Zeydan2016}, is deemed as a promising solution for further reducing the redundant data transmission, easing the pressure of cloud data centers and improving the QoE.

\begin{figure}[!!!!!!!!!!!!!!hhhhhhhhhht]
    \centering
    \includegraphics[width=7 cm]{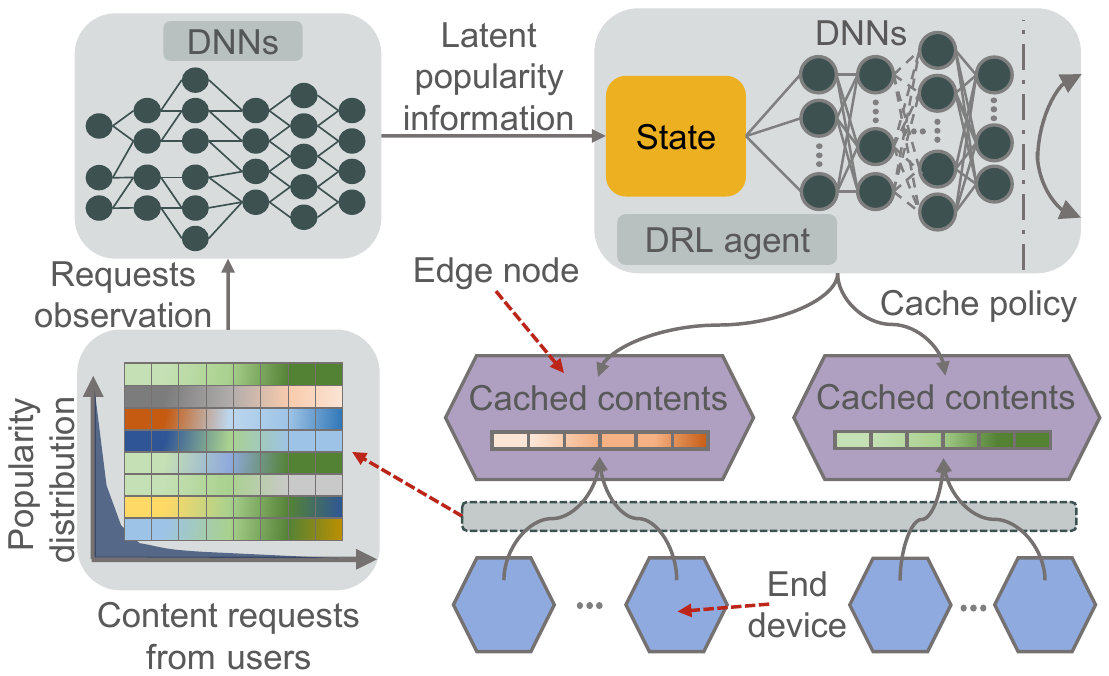}
    \caption{DL and DRL for optimizing the edge caching policy.}
    \label{fig:EdgeCaching}
\end{figure}

Edge caching meets two challenges: 
1) the content popularity distribution among the coverage of edge nodes is hard to estimate, since it may be different and change with spatio-temporal variation \cite{Song2017}; 
2) in view of massive heterogeneous devices in edge computing environments, the hierarchical caching architecture and complex network characteristics further perplex the design of content caching strategy \cite{Li2018d}. 
Specifically, the optimal edge caching strategy can only be deduced when the content popularity distribution is known. 
However, users' predilection for contents is actually unknown since the mobility, personal preference and connectivity of them may vary all the time. 
In this section, DL for determining edge caching policies, as illustrated in Fig. \ref{fig:EdgeCaching}, are discussed. 

\subsubsection{Use Cases of DNNs}
\label{subsubsec:edgecachingusecasesofdnns}

Traditional caching methods are generally with high computational complexity since they require a large number of online optimization iterations to determine 
1) the features of users and contents and 
2) the strategy of content placement and delivery. 

{\color{red}For the first purpose, DL can be used to process raw data collected from the mobile devices of users and hence extract the features of the users and content as a feature-based content popularity matrix. 
By this means, the popular content at the core network is estimated by applying feature-based collaborative filtering to the popularity matrix \cite{DeepCachNet}. 

For the second purpose, when using DNNs to optimize the strategy of edge caching, online heavy computation iterations can be avoided by offline training. }
A DNN, which consists of an encoder for data regularization and a followed hidden layer, can be trained with solutions generated by optimal or heuristic algorithms and be deployed to determine the cache policy \cite{Chang2018a}, hence avoiding online optimization iterations. 
Similarly, in \cite{Yang2019b}, inspired by the fact that the output of optimization problem about partial cache refreshing has some patterns, an MLP is trained for accepting the current content popularity and the last content placement probability as input to generate the cache refresh policy. 

As illustrated in \cite{Chang2018a}\cite{Yang2019b}, the complexity of optimization algorithms can be transferred to the training of DNNs, and thus breaking the practical limitation of employing them. 
In this case, DL is used to learn input-solution relations, and DNN-based methods are only available when optimization algorithms for the original caching problem exist. 
Therefore, the performance of DNN-based methods bounds by fixed optimization algorithms and is not self-adapted. 

In addition, DL can be utilized for customized edge caching. 
For example, to minimize content-downloading delay in the self-driving car, an MLP is deployed in the cloud to predict the popularity of contents to be requested, and then the outputs of MLP are delivered to the edge nodes (namely MEC servers at RSUs in \cite{Ndikumana2018}). 
According to these outputs, each edge node caches contents that are most likely to be requested. 
On self-driving cars, CNN is chosen to predict the age and gender of the owner. 
Once these features of owners are identified, $k$-means clustering \cite{Kanungo2002} and binary classification algorithms are used to determine which contents, already cached in edge nodes, should be further downloaded and cached from edge nodes to the car. 
Moreover, concerning taking full advantage of users' features, \cite{Tang2019} points out that the user's willing to access the content in different environments is varying. 
Inspired by this, RNN is used to predict the trajectories of users. 
And based on these predictions, all contents of users' interests are then prefetched and cached in advance at the edge node of each predicted location.

\subsubsection{Use Cases of DRL}
\label{subsubsec:edgecachingusecasesofdrl}

The function of DNNs described in Section \ref{subsubsec:edgecachingusecasesofdnns} can be deemed as a part of the whole edge caching solution, i.e., the DNN itself does not deal with the whole optimization problem. Different from these DNNs-based edge caching, DRL can exploit the context of users and networks and take adaptive strategies for maximizing the long-term caching performance \cite{doi:10.1287/opre.1070.0445} as the main body of the optimization method. Traditional RL algorithms are limited by the requirement for handcrafting features and the flaw that hardly handling high-dimensional observation data and actions \cite{Zhu2018b}. Compared to traditional RL irrelevant to DL, such as $Q$-learning \cite{Guo2017} and Multi-Armed Bandit (MAB) learning \cite{Song2017}, the advantage of DRL lies in that DNNs can learn key features from the raw observation data. The integrated DRL agent combining RL and DL can optimize its strategies with respect to cache management in edge computing networks directly from high-dimensional observation data. 

In \cite{Zhong2018}, DDPG is used to train a DRL agent, in order to maximize the long-term cache hit rate, to make proper cache replacement decisions. This work considers a scenario with a single BS, in which the DRL agent decides whether to cache the requested contents or replace the cached contents. While training the DRL agent, the reward is devised as the cache hit rate. In addition, \textit{Wolpertinger} architecture \cite{Dulac-arnold} is utilized to cope with the challenge of large action space. In detail, a primary action set is first set for the DRL agent and then using $k$NN to map the practical action inputs to one out of this set. In this manner, the action space is narrowed deliberately without missing the optimal caching policy. Compared DQL-based algorithms searching the whole action space, the trained DRL agent with DDPG and \textit{Wolpertinger} architecture is able to achieve competitive cache hit rates while reducing the runtime.

\subsection{{\color{red}DL for Optimizing Edge Task Offloading}}
\label{subsec:taskoffloading}

\begin{figure}[!!!!!!!!!!!!!!hhhhhhhhhht]
    \centering
    \includegraphics[width=6.5 cm]{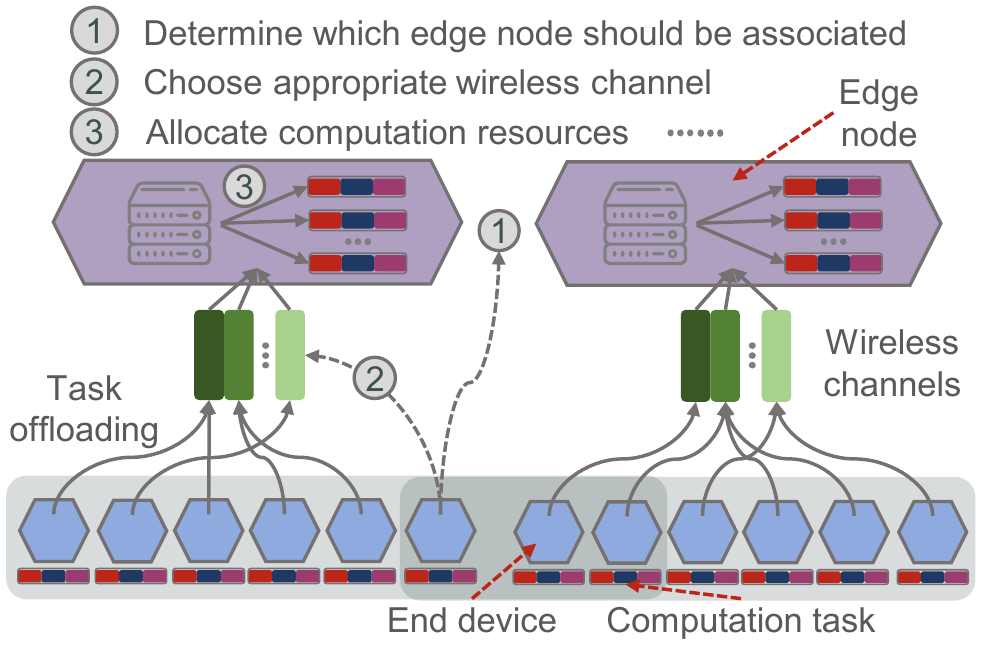}
    \caption{Computation offloading problem in edge computing.}
    \label{fig:ComputationOffload}
\end{figure}

Edge computing allows edge devices offload part of their computing tasks to the edge node \cite{Mach2017}, under constraints of energy, delay, computing capability, etc. 
As shown in Fig. \ref{fig:ComputationOffload}, these constraints put forward challenges of identifying 1) which edge nodes should receive tasks, 2) what ratio of tasks edge devices should offload and 3) how many resources should be allocated to these tasks. 
To solve this kind of task offloading problem is NP-hard \cite{Chen2016f}, since at least combination optimization of communication and computing resources along with the contention of edge devices is required. 
Particularly, the optimization should concern both the time-varying wireless environments (such as the varying channel quality) and requests of task offloading, hence drawing the attention of using learning methods \cite{Xu2017b,Dinh2018a,Chen2017c,Zhang2019m,Yu2018b,Yang2018c,Chen2018k,Luong2018a,Li2018h,Min2017,Chenc}. 
Among all these works related to learning-based optimization methods, DL-based approaches have advantages over others when multiple edge nodes and radio channels are available for computation offloading. 
At this background, large state and action spaces in the whole offloading problem make the conventional learning algorithms \cite{Xu2017b}\cite{Chen2018j}\cite{Chen2017c} infeasible actually. 

\begin{table*}[htbp!!!!]
    \centering
    \scriptsize
    \caption{{\color{red}DL for Optimizing Edge Application Scenarios}}
    \label{tab:DLforEdge}%
    \begin{tabular}{ccm{1em}<{\centering}m{2em}<{\centering}m{6.5em}<{\centering}m{11em}m{11em}m{11em}m{12em}}
      \toprule
      \multicolumn{2}{c}{} & \textbf{Ref.} & \textbf{DL} & \textbf{Comm. Scale} & \textbf{Inputs - DNN (States - DRL)} & \textbf{Outputs - DNN (Action - DRL)} & \textbf{Loss func. - DL (Reward - DRL)} & \textbf{Performance} \\
      \midrule
      \multicolumn{2}{c}{\multirow{6}[1]{*}[-4em]{ \rotatebox{90}{\textbf{DL for Adaptive Edge Caching}} }} & \rotatebox{90}{\cite{DeepCachNet}} &  \rotatebox{90}{SDAE}  & 60 users / 6 SBSs & User features, content features & Feature-based content popularity matrix & Normalized differences between input features and the consequent reconstruction & QoE improvement: up to 30\%; Backhaul offloading: 6.2\% \\
  \cmidrule{3-9}    \multicolumn{2}{c}{} & \rotatebox{90}{\cite{Chang2018a}} & \rotatebox{90}{FCNN}  & 100-200 UEs per cell / 7 BSs & Channel conditions, file requests & Caching decisions & Normalized differences between prediction decisions and the optimum & Prediction accuracy: up to 92\%; Energy saving: 8\% gaps to the optimum  \\
  \cmidrule{3-9}    \multicolumn{2}{c}{} & \rotatebox{90}{\cite{Yang2019b}} & \rotatebox{90}{FCNN}  & UEs with density 25-30 / Multi-tier BSs & Current content popularity, last content placement probability & Content placement probability & Statistical average of the error between the model outputs and the optimal CVX solution & Prediction accuracy: slight degeneration to the optimum \\
  \cmidrule{3-9}    \multicolumn{2}{c}{} & \rotatebox{90}{\cite{Ndikumana2018}} & \rotatebox{90}{\tabincell{c}{FCNN \\ CNN}} & Cars / 6 RSUs with MEC servers  & Facial images - CNN; Content features - FCNN &   Gender and age prediction - CNN; Content request probability - FCNN & N/A - CNN; Cross entropy error - FCNN & Caching accuracy: up to 98.04\% \\
  \cmidrule{3-9}    \multicolumn{2}{c}{} & \rotatebox{90}{\cite{Tang2019}} & \rotatebox{90}{RNN}   & 20 UEs / 10 servers & User historical traces & User location prediction & Cross entropy error & Caching accuracy: up to 75\% \\
  \cmidrule{3-9}    \multicolumn{2}{c}{} & \rotatebox{90}{\cite{Zhong2018}} & \rotatebox{90}{DDPG}  & Multiple UEs / Single BS & Features of cached contents, current requests & Content replacement & Cache hit rate & Cache hit rate: about 50\% \\
      \midrule
      \multicolumn{2}{c}{\multirow{7}[1]{*}[-5em]{ \rotatebox{90}{\textbf{DL for Optimizing Edge Task Offloading}} }} &  \rotatebox{90}{\cite{Luong2018a}} & \rotatebox{90}{FCNN}  & 20 miners / Single edge node & Bidder valuation profiles of miners & Assignment probabilities, conditional payments & Expected, negated revenue of the service provider & Revenue increment \\
  \cmidrule{3-9}    \multicolumn{2}{c}{} & \rotatebox{90}{\cite{Zhang2018g}} & \rotatebox{90}{\tabincell{c}{Double- \\ DQL}} & Single UE & System utilization states, dynamic slack states & DVFS algorithm selection & Average energy consumption & Energy saving: 2\%-4\% \\
  \cmidrule{3-9}    \multicolumn{2}{c}{} & \rotatebox{90}{\cite{Li2018h}} &  \rotatebox{90}{DQL}  & Multiple UEs / Single eNodeB & Sum cost of the entire system, available capacity of the MEC server & Offloading decision, resource allocation & Negatively correlated to the sum cost & System cost reduction \\
  \cmidrule{3-9}    \multicolumn{2}{c}{} & \rotatebox{90}{\cite{Chenc}} & \rotatebox{90}{DDPG}  & Multiple UEs / Single BS with an MEC server & Channel vectors, task queue length & Offloading decision, power allocation & Negative wighted sum of the power consumption and task queue length & Computation cost reduction \\
  \cmidrule{3-9}    \multicolumn{2}{c}{} & \rotatebox{90}{\cite{Min2017}} &  \rotatebox{90}{DQL}  & Single UE / Multiple MEC servers & Previous radio bandwidth, predicted harvested energy, current battery level & MEC server selection, offloading rate & Composition of overall data sharing gains, task drop loss, energy consumption and delay & Energy saving; Delay improvement \\
  \cmidrule{3-9}    \multicolumn{2}{c}{} & \rotatebox{90}{\cite{Chen2018k}} & \rotatebox{90}{\tabincell{c}{Double- \\ DQL}} & Single UE / 6 BSs with MEC servers & Channel gain states, UE-BS association state, energy queue length, task queue length & Offloading decision,  energy units allocation & Composition of task execution delay, task drop times, task queuing delay, task failing penalty and service payment  & Offloading performance improvement \\
  \cmidrule{3-9}    \multicolumn{2}{c}{} & \rotatebox{90}{\cite{Huang}} & \rotatebox{90}{DROO} & Multiple UEs / Single MEC server & Channel gain states & Offloading action & Computation rate & Algorithn execution time: less than 0.1s in 30-UE network \\
      \midrule
      \multicolumn{1}{c}{\multirow{6}[1]{*}[-2.5em]{ \rotatebox{90}{\textbf{DL for Edge Management and Maintenance}} }} & \multicolumn{1}{c}{\multirow{2}[1]{*}{ \rotatebox{90}{\textbf{Communication}} }} & \rotatebox{90}{\cite{Memon2018}} & \rotatebox{90}{\tabincell{c}{RNN \& \\ LSTM}} & 53 vehicles / 20 fog servers & Coordinates of vehicles and interacting fog nodes, time, service cost & Cost prediction & Mean absolute error & Prediction accuracy: 99.2\% \\
  \cmidrule{3-9}          &       & \rotatebox{90}{\cite{Sun2019}} & \rotatebox{90}{DQL} & 4 UEs / Multiple RRHs & Current on-off states of processors, current communication modes of UEs, cache states & Processor state control, communication mode selection & Negative of system energy consumption & System power consumption \\
  \cmidrule{2-9}          & \multicolumn{1}{c}{ \multirow{1}[1]{*}[1.6em]{\rotatebox{90}{ {\textbf{Security}} }} } & \rotatebox{90}{\cite{Xiao2018a}} & \rotatebox{90}{DQL}   & Multiple UEs / Multiple edge nodes & Jamming power, channel bandwidth, battery levels, user density & Edge node and channel selection, offloading rate, transmit power & Composition of defense costs and secrecy capacity & Signal SINR increasement \\
  \cmidrule{2-9}          & \multicolumn{1}{c}{\multirow{3}[1]{*}[-1em]{ \rotatebox{90}{\textbf{Joint Optimization}} }} & \rotatebox{90}{\tabincell{c}{\cite{He2018e}}} & \rotatebox{90}{\tabincell{c}{Double- \\ Dueling \\ DQL}} & Multiple UEs / 5 BSs and 5 MEC servers & Status from each BS, MEC server and content cache & BS allocation, caching decision, offloading decision & Composition of received SNRs, computation capabilities and cache states & System utility increasement \\
  \cmidrule{3-9}          &       & \rotatebox{90}{\cite{Wei2018b}} & \rotatebox{90}{\tabincell{c}{AC \\ DRL}} & 20 UEs per router / 3 fog nodes & States of requests, fog nodes, tasks, contents and SINR & Decisions about fog node, channel, resource allocation, offloading and caching & Composition of computation offloading delay and content delivery delay & Average service latency: 1.5-4.0s \\
  \cmidrule{3-9}          &       & \rotatebox{90}{\cite{Le2018a}} & \rotatebox{90}{DQL}   & 50 vehicles / 10 RSUs & States of RSUs, vehicles and caches, contact rate, contact times & RSU assignment, caching control and control & Composition of communication, storage and computation cost & Backhaul capacity mitigation; Resource saving \\
      \bottomrule
      \end{tabular}%
    \label{tab:addlabel}%
  \end{table*}%

\subsubsection{Use Cases of DNNs}
\label{subsubsec:offloadingusecasesofdnns}

In \cite{Yu2018b}, the computation offloading problem is formulated as a multi-label classification problem. By exhaustively searching the solution in an offline way, the obtained optimal solution can be used to train a DNN with the composite state of the edge computing network as the input, and the offloading decision as the output. By this means, optimal solutions may not require to be solved online avoiding belated offloading decision making, and the computation complexity can be transferred to DL training.
	
Further, a particular offloading scenario with respect to Blockchain is investigated in \cite{Luong2018a}. The computing and energy resources consumption of mining tasks on edge devices may limit the practical application of Blockchain in the edge computing network. Naturally, these mining tasks can be offloaded from edge devices to edge nodes, but it may cause unfair edge resource allocation. Thus, all available resources are allocated in the form of auctions to maximize the revenue of the Edge Computing Service Provider (ECSP). Based on an analytical solution of the optimal auction, an MLP can be constructed \cite{Luong2018a} and trained with valuations of the miners (i.e., edge devices) for maximizing the expected revenue of ECSP.

\subsubsection{Use Cases of DRL}
\label{subsubsec:offloadingusecasesofdrl}

Though offloading computation tasks to edge nodes can enhance the processing efficiency of the computation tasks, the reliability of offloading suffers from the potentially low quality of wireless environments. 
{\color{red}In \cite{Zhang2019m}, to maximize offloading utilities, the authors first quantify the influence of various communication modes on the task offloading performance and accordingly propose applying DQL to online select the optimal target edge node and transmission mode.
For optimizing the total offloading cost, a DRL agent that modifies Dueling- and Double-DQL \cite{Nguyen2019c} can allocate edge computation and bandwidth resources for end devices. }

Besides, offloading reliability should also be concerned. 
The coding rate, by which transmitting the data, is crucial to make the offloading meet the required reliability level. 
Hence, in \cite{Yang2018c}, effects of the coding block-length are investigated and an MDP concerning resource allocation is formulated and then solved by DQL, in order to improve the average offloading reliability. 
Exploring further on scheduling fine-grained computing resources of the edge device, in \cite{Zhang2018g}, Double-DQL \cite{Hasselt20016} is used to determine the best Dynamic Voltage and Frequency Scaling (DVFS) algorithm. 
Compared to DQL, the experiment results indicate that Double-DQL can save more energy and achieve higher training efficiency. 
Nonetheless, the action space of DQL-based approaches may increase rapidly with increasing edge devices. 
Under the circumstances, a pre-classification step can be performed before learning \cite{Li2018h} to narrow the action space. 

IoT edge environments powered by Energy Harvesting (EH) is investigated in \cite{Min2017, Chen2018k}. 
In EH environments, the energy harvesting makes the offloading problem more complicated, since IoT edge devices can harvest energy from ambient radio-frequency signals. 
Hence, CNN is used to compress the state space in the learning process \cite{Min2017}. 
Further, in \cite{Chen2018k}, inspired by the additive structure of the reward function, $Q$-function decomposition is applied in Double-DQL, and it improves the vanilla Double-DQL.
However, value-based DRL can only deal with discrete action space. 
To perform more fine-grained power control for local execution and task offloading, policy-gradient-based DRL should be considered. 
For example, compared tot he discrete power control strategy based on DQL, DDPG can adaptively allocate the power of edge devices with finer granularity \cite{Chenc}.

Freely letting DRL agents take over the whole process of computation offloading may lead to huge computational complexity. 
Therefore, only employing DNN to make partial decisions can largely reduce the complexity. 
For instance, in \cite{Huang}, the problem of maximizing the weighted sum computation rate is decomposed into two sub-problems, viz., offloading decision and resource allocation. 
By only using DRL to deal with the NP-hard offloading decision problem rather than both, the action space of the DRL agent is narrowed, and the offloading performance is not impaired as well since the resource allocation problem is solved optimally.

\subsection{{\color{red}DL for Edge Management and Maintenance}}
\label{subsec:edgemanagement}

Edge DL services are envisioned to be deployed on BSs in cellular networks, as implemented in \cite{Li2018k}. 
Therefore, edge management and maintenance require optimizations from multiple perspectives (including communication perspective). 
Many works focus on applying DL in wireless communication \cite{Mao2018, Li2017d, Chen2015}. 
Nevertheless, management and maintenance at the edge should consider more aspects.

\subsubsection{Edge Communication}
\label{subsubsec:edgecommunication}

When edge nodes are serving mobile devices (users), mobility issues in edge computing networks should be addressed. 
DL-based methods can be used to assist the smooth transition of connections between devices and edge nodes. 
{\color{red}To minimize energy consumption per bit, in \cite{Dong2019c}, the optimal device association strategy is approximated by a DNN.
Meanwhile, a digital twin of network environments is established at the central server for training this DNN off-line. 
To minimize the interruptions of a mobile device moving from an edge node to the next one throughout its moving trajectory, 
the MLP can be used to predict available edge nodes at a given location and time \cite{Memon2018}. 
Moreover, determining the best edge node, with which the mobile device should associate, still needs to evaluate the cost (the latency of servicing a request) for the interaction between the mobile device and each edge node.}
Nonetheless, modeling the cost of these interactions requires a more capable learning model. 
Therefore, a two-layer stacked RNN with LSTM cells is implemented for modeling the cost of interaction. 
At last, based on the capability of predicting available edge nodes along with corresponding potential cost, the mobile device can associate with the best edge node, and hence the possibility of disruption is minimized. 

{\color{red}Aiming at minimizing long-term system power consumption in the communication scenario with multiple modes (to serve various IoT services), i.e.,  Cloud-Radio Access Networks (C-RAN) mode, Device-to-Device (D2D) mode, and Fog radio Access Point (FAP) mode, DQL can be used to control communication modes of edge devices and on-off states of processors throughout the communicating process \cite{Sun2019}. 
After determining the communication mode and the processors' on-off states of a given edge device, the whole problem can be degraded into an Remote Radio Head (RRH) transmission power minimization problem and solved.
Further, TL is integrated with DQL to reduce the required interactions with the environment in the DQL training process while maintaining a similar performance without TL.}

\subsubsection{Edge Security}
\label{subsubsec:edgesecurity}

Since edge devices generally equipped with limited computation, energy and radio resources, the transmission between them and the edge node is more vulnerable to various attacks, such as jamming attacks and Distributed Denial of Service (DDoS) attacks, compared to cloud computing. 
Therefore, the security of the edge computing system should be enhanced. 
{\color{red}First, the system should be able to actively detect unknown attacks, for instance, using DL techniques to extract features of eavesdropping and jamming attacks \cite{Chen2019g}.
According to the attack mode detected, the system determines the strategy of security protection.}
Certainly, security protection generally requires additional energy consumption and the overhead of both computation and communication. 
Consequently, each edge device shall optimize its defense strategies, viz., choosing the transmit power, channel and time, without violating its resource limitation. 
The optimization is challenging since it is hard to estimate the attack model and the dynamic model of edge computing networks. 

DRL-based security solutions can provide secure offloading (from the edge device to the edge node) to against jamming attacks \cite{Xiao2018a} or protect user location privacy and the usage pattern privacy \cite{Min2018}.
The edge device observes the status of edge nodes and the attack characteristics and then determines the defense level and key parameters in security protocols. 
By setting the reward as the anti-jamming communication efficiency, such as the signal-to-interference-plus-noise ratio of the signals, the bit error rate of the received messages, and the protection overhead, the DQL-based security agent can be trained to cope with various types of attacks.



\subsubsection{Joint Edge Optimization}
\label{subsubsec:jointedgeopt}

Edge computing can cater for the rapid growth of smart devices and the advent of massive computation-intensive and data-consuming applications. 
Nonetheless, it also makes the operation of future networks even more complex \cite{Bogale}. 
To manage the complex networks with respect to comprehensive resource optimization \cite{Wang2017e} is challenging, particularly under the premise of considering key enablers of the future network, including Software-Defined Network (SDN) \cite{SDNsurvey}, IoTs, Internet of Vehicles (IoVs).

In general, SDN is designed for separating the control plane from the data plane, and thus allowing the operation over the whole network with a global view. 
Compared to the distributed nature of edge computing networks, SDN is a centralized approach, and it is challenging to apply SDN to edge computing networks directly. 
In \cite{He2017a}, an SDN-enabled edge computing network catering for smart cities is investigated. To improve the servicing performance of this prototype network, DQL is deployed in its control plane to orchestrate networking, caching, and computing resources.

Edge computing can empower IoT systems with more computation-intensive and delay-sensitive services but also raises challenges for efficient management and synergy of storage, computation, and communication resources. 
For minimizing the average end-to-end servicing delay, policy-gradient-based DRL combined with AC architecture can deal with the assignment of edge nodes, the decision about whether to store the requesting content or not, the choice of the edge node performing the computation tasks and the allocation of computation resources \cite{Wei2018b}. 

IoVs is a special case of IoTs and focuses on connected vehicles. 
Similar to the consideration of integrating networking, caching and computing as in \cite{Wei2018b}, Double-Dueling DQL (i.e., combining Double DQL and Dueling DQL) with more robust performance, can be used to orchestrate available resources to improve the performance of future IoVs \cite{He2018e}. 
In addition, considering the mobility of vehicles in the IoVs, the hard service deadline constraint might be easily broken, and this challenge is often either neglected or tackled inadequately because of high complexities. 
To deal with the mobility challenge, in \cite{Le2018a}, the mobility of vehicles is first modeled as discrete random jumping, and the time dimension is split into epochs, each of which comprises several time slots. 
Then, a small timescale DQL model, regarding the granularity of time slot, is devised for incorporating the impact of vehicles' mobility in terms of the carefully designed immediate reward function. 
At last, a large timescale DQL model is proposed for every time epoch. 
By using such multi-timescale DRL, issues about both immediate impacts of the mobility and the unbearable large action space in the resource allocation optimization are solved. 

\section{Lessons Learned and Open Challenges}
\label{sec:challenges}


{\color{red}To identify existing challenges and circumvent potential misleading directions, we briefly introduce the potential scenario of ``\textit{DL application on Edge}'', and separately discuss open issues related to four enabling technologies that we focus on, i.e., ``\textit{DL inference in Edge}'', ``\textit{Edge Computing for DL}'', ``\textit{DL training at Edge}'' and ``\textit{DL for optimizing Edge}''.}


\subsection{{\color{red}More Promising Applications}}
\label{subsec:OpenChallengesforDLonEdge}


if DL and edge are well-integrated, they can offer great potential for the development of innovative applications. 
There are still many areas to be explored to provide operators, suppliers and third parties with new business opportunities and revenue streams.

For example, with more DL techniques are universally embedded in these emerged applications, the introduced processing delay and additional computation cost make the cloud gaming architecture struggle to meet the latency requirements. 
Edge computing architectures, near to users, can be leveraged with the cloud to form a hybrid gaming architecture. 
Besides, intelligent driving involves speech recognition, image recognition, intelligent decision making, etc. 
Various DL applications in intelligent driving, such as collision warning, require edge computing platforms to ensure millisecond-level interaction delay. 
In addition, edge perception is more conducive to analyze the traffic environment around the vehicle, thus enhancing driving safety.

\subsection{{\color{red}General DL Model for Inference}}
\label{subsec:openDLinEdge}

When deploying DL in edge devices, it is necessary to accelerate DL inference by model optimization. 
In this section, lessons learned and future directions for ``\textit{DL inference in Edge}'', with respect to model compression, model segmentation, and EEoI, used to optimize DL models, is discussed. 

{\color{red}
\subsubsection{Ambiguous Performance Metrics}

For an Edge DL service for a specific task, there are usually a series of DL model candidates that can accomplish the task.
However, it is difficult for service providers to choose the right DL model for each service.
Due to the uncertain characteristics of edge computing networks (varying wireless channel qualities, unpredictable concurrent service requests, etc.),
commonly used standard performance indicators (such as top-$k$ accuracy \cite{Chameleon} or mean average accuracy \cite{DeepMon}) cannot reflect the runtime performance of DL model inference in the edge.
For Edge DL services, besides model accuracy, inference delay, resource consumption, and service revenue are also key indicators.
Therefore, we need to identify the key performance indicators of Edge DL, quantitatively analyze the factors affecting them, and explore the trade-offs between these indicators to help improve the efficiency of Edge DL deployment.
}

\subsubsection{Generalization of EEoI}

Currently, EEoI can be applied to classification problems in DL \cite{Mcdanel2016}, but there is no generalized solution for a wider range of DL applications. Furthermore, in order to build an intelligent edge and support edge intelligence, not only DL but also the possibility of applying EEoI to DRL should be explored, since applying DRL to real-time resource management for the edge, as discussed in Section \ref{sec:AIforEdge}, requires stringent response speed. 


\subsubsection{Hybrid model modification}

Coordination issues with respect to model optimization, model segmentation, and EEoI should be thought over. These customized DL models are often used independently to enable ``end-edge-cloud'' collaboration. Model optimizations, such as model quantification and pruning, may be required on the end and edge sides, but because of the sufficient computation resources, the cloud does not need to take the risk of model accuracy to use these optimizations. Therefore, how to design a hybrid precision scheme, that is, to effectively combine the simplified DL models in the edge with the raw DL model in the cloud is important.

\subsubsection{Coordination between training and inference}

Pruning, quantizing and introducing EEoI into trained raw DL models require retraining to give them the desired inference performance. In general, customized models can be trained offline in the cloud. However, the advantage of edge computing lies in its response speed and might be neutralized because of belated DL training. Moreover, due to a large number of heterogeneous devices in the edge and the dynamic network environment, the customization requirements of DL models are not monotonous. Then, is this continuous model training requirement reasonable, and will it affect the timeliness of model inference? How to design a mechanism to avoid these side-effects?

\subsection{{\color{red}Complete Edge Architecture for DL}}
\label{subsec:challengesedgefordl}

Edge intelligence and intelligent edge require a complete system framework, covering data acquisition, service deployment and task processing. 
In this section, we discuss challenges for ``\textit{Edge Computing for DL}'' to build a complete edge computing framework for DL.



\subsubsection{Edge for Data Processing}
\label{subsubsec:EdgeforDataProcessing}

Both pervasively deployed DL services on the edge and DL algorithms for optimizing edge cannot be realized without data acquiring. Edge architecture should be able to efficiently acquire and process the original data, sensed or collected by edge devices, and then feed them to DL models. 


Adaptively acquiring data at the edge and then transmitting them to cloud (as done in \cite{Mudassar2018}) is a natural way to alleviate the workload of edge devices and to reduce the potential resource overhead. In addition, it is better to further compress the data, which can alleviate the bandwidth pressure of the network, while the transmission delay can be reduced to provide better QoS. Most existed works focus only on vision applications \cite{Ren2018}. However, the heterogeneous data structures and characteristics of a wide variety of DL-based services are not addressed well yet. Therefore, developing a heterogeneous, parallel and collaborative architecture for edge data processing for various DL services will be helpful. 

\subsubsection{Microservice for Edge DL Services}
\label{subsubsec:ContainerManagementforDLServices}

Edge and cloud services have recently started undergoing a major shift from monolithic entities to graphs of hundreds of loosely-coupled microservices \cite{gan:asplos:2019:microservices}. 
Executing DL computations may need a series of software dependencies, and it calls for a solution for isolating different DL services on the shared resources. 
At present, the microservice framework, deployed on the edge for hosting DL services, is in its infant \cite{Alam2018a}, due to several critical challenges: 
1) Handling DL deployment and management flexibly; 
2) Achieving live migration of microservices to reduce migration times and unavailability of DL services due to user mobilities; 
3) Orchestrating resources among the cloud and distributed edge infrastructures to achieve better performance, as illustrated in Section \ref{subsubsec:verticalcollaboration}.



\subsubsection{Incentive and trusty offloading mechanism for DL}
\label{subsubsec:BlockchainedEdgeforDL}

Heavy DL computations on resource-limited end devices can be offloaded to nearby edge nodes (Section \ref{subsec:edgecomputingmodefordl}). However, there are still several issues, 1) an incentive mechanism should be established for stimulating edge nodes to take over DL computations; 2) the security should be guaranteed to avoid the risks from anonymous edge nodes \cite{8632682}.

Blockchain, as a decentralized public database storing transaction records across participated devices, can avoid the risk of tampering the records \cite{8029379}. By taking advantage of these characteristics,  incentive and trust problems with respect to computation offloading can potentially be tackled. To be specific, all end devices and edge nodes have to first put down deposits to the blockchain to participate. The end device request the help of edge nodes for DL computation, and meantime send a ``require'' transaction to the blockchain with a bounty. Once an edge nodes complete the computation, it returns results to the end device with sending a ``complete'' transaction to the blockchain. After a while, other participated edge nodes also execute the offloaded task and validate the former recorded result. At last, for incentives, firstly recorded edge nodes win the game and be awarded \cite{Kim2018b}. However, this idea about blockchained edge is still in its infancy. Existing blockchains such as Ethereum \cite{wood2014ethereum} do not support the execution of complex DL computations, which raises the challenge of adjusting blockchain structure and protocol in order to break this limitation.

{\color{red}
\subsubsection{Integration with ``DL for optimizing Edge''}

End devices, edge nodes, and base stations in edge computing networks are expected to run various DL models and deploy corresponding services in the future.
In order to make full use of decentralized resources of edge computing, and to establish connections with existing cloud computing infrastructure, dividing the computation-intensive DL model into sub-tasks and effectively offloading these tasks between edge devices for collaboration are essential.
Owing to deployment environments of Edge DL are usually highly dynamic, edge computing frameworks need excellent online resource orchestration and parameter configuration to support a large number of DL services.
Heterogeneous computation resources, real-time joint optimization of communication and cache resources, and high-dimensional system parameter configuration are critical.
We have introduced various theoretical methods to optimize edge computing frameworks (networks) with DL technologies in Section \ref{sec:AIforEdge}.
Nonetheless, there is currently no relevant work to deeply study the performance analysis of deploying and using these DL technologies for long-term online resource orchestration in practical edge computing networks or testbeds.
We believe that ``\textit{Edge Computing for DL}'' should continue to focus on how to integrate ``\textit{DL for optimizing Edge}'' into the edge computing framework to realize the above vision.
}

\subsection{{\color{red}Practical Training Principles at Edge}}

Compared with DL inference in the edge, DL training at the edge is currently mainly limited by the weak performance of edge devices and the fact that most Edge DL frameworks or libraries still do not support training. At present, most studies are at the theoretical level, i.e., simulating the process of DL training at the edge. 
In this section, we point out the lessons learned and challenges in ``\textit{DL Training at Edge}''.

\subsubsection{Data Parallelism versus Model Parallelism}

DL models are both computation and memory intensive. 
When they become deeper and larger, it is not feasible to acquire their inference results or train them well by a single device. 
Therefore, large DL models are trained in distributed manners over thousands of CPU or GPU cores, in terms of data parallelism, model parallelism or their combination (Section \ref{subsec:DistributedTraining}). 
However, differing from parallel training over bus-or switch-connected CPUs or GPUs in the cloud, perform model training at distributed edge devices should further consider wireless environments, device configurations, privacies, etc. 

\begin{figure}[!!!!!!!!!!!!!!hhhhhhhhhht]
    \centering
    \includegraphics[width=8.85 cm]{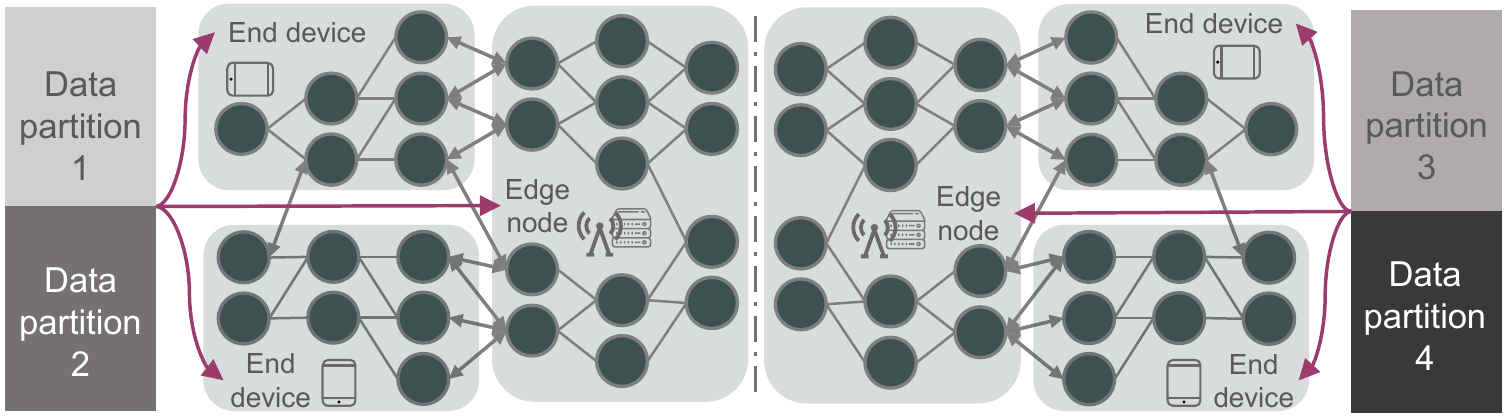}
    \caption{DL training at the edge by both data and model parallelism.}
    \label{fig:DPandMP}
\end{figure}

At present, FL only copies the whole DL model to every participated edge devices, namely in the manner of data parallelism. Hence, taking the limited computing capabilities of edge devices (at least for now) into consideration, partitioning a large-scale DL model and allocating these segments to different edge devices for training may be a more feasible and practical solution. Certainly, this does not mean abandoning the native data parallelism of FL, instead, posing the challenge of blending data parallelism and model parallelism particularly for training DL models at the edge, as illustrated in Fig. \ref{fig:DPandMP}. 

\subsubsection{Where is training data from?}

{\color{red}
Currently, most of the DL training frameworks at the edge are aimed at supervised learning tasks, and test their performance with complete data sets. 
However, in practical scenarios, we cannot assume that all data in the edge computing network are labeled and with a correctness guarantee. 
For unsupervised learning tasks such as DRL, we certainly do not need to pay too much attention to the production of training data. 
For example, the training data required for DRL compose of the observed state vectors and rewards obtained by interacting with the environment. 
These training data can generate automatically when the system is running. 
But for a wider range of supervised learning tasks, how edge nodes and devices find the exact training data for model training?
The application of vanilla FL is using RNN for next-word-prediction \cite{Bonawitz2017}, in which the training data can be obtained along with users' daily inputs. 
Nonetheless, for extensive Edge DL services concerning video analysis, where are their training data from.
If all training data is manually labeled and uploaded to the cloud data center, and then distributed to edge devices by the cloud, the original intention of FL is obviously violated. 
One possible solution is to enable edge devices to construct their labeled data by learning ``labeled data'' from each other.  
We believe that the production of training data and the application scenarios of DL models training at the edge should first be clarified in the future, and the necessity and feasibility of DL model training at the edge should be discussed as well.
}

\subsubsection{Asynchronous FL at Edge}

Existing FL methods \cite{Bonawitz2017, mcmahan2016communication} focus on synchronous training, and can only process hundreds of devices in parallel. 
However, this synchronous updating mode potentially cannot scale well, and is inefficient and inflexible in view of two key properties of FL, specifically, 
1) infrequent training tasks, since edge devices typically have weaker computing power and limited battery endurance and thus cannot afford intensive training tasks; 
2) limited and uncertain communication between edge devices, compared to typical distributed training in the cloud.

Thus, whenever the global model is updating, the server is limited to selecting from a subset of available edge devices to trigger a training task. 
In addition, due to limited computing power and battery endurance, task scheduling varies from device to device, making it difficult to synchronize selected devices at the end of each epoch. 
Some devices may no longer be available when they should be synchronized, and hence the server must determine the timeout threshold to discard the laggard. 
If the number of surviving devices is too small, the server has to discard the entire epoch including all received updates. 
These bottlenecks in FL can potentially be addressed by asynchronous training mechanisms \cite{Zheng:2017:ASG:3305890.3306107, Xie2019b, Wu2019f}.
Adequately selecting clients in each training period with resource constraints may also help. 
{\color{red}By setting a certain deadline for clients to download, update, and upload DL models, the central server can determine which clients to perform local training such that it can aggregate as many client updates as possible in each period, thus allowing the server to accelerate performance improvement in DL models \cite{Nishio2018}.}

\subsubsection{Transfer Learning-based Training}

Due to resource constraints, training and deploying computation-intensive DL models on edge devices such as mobile phones is challenging. In order to facilitate learning on such resource-constrained edge devices, TL can be utilized. For instance, in order to reduce the amount of training data and speeding up the training process, using unlabeled data to transfer knowledge between edge devices can be adopted \cite{Xing2018}. By using the cross-modal transfer in the learning of edge devices across different sensing modalities, required labeled data and the training process can be largely reduced and accelerated, respectively. 

Besides, KD, as a method of TL, can also be exploited thanks to several advantages \cite{Sharma2018}: 1) using information from well-trained large DL models (teachers) to help lightweight DL models (students), expected to be deployed on edge devices, converge faster; 2) improving the accuracy of students; 3) helping students become more general instead of being overfitted by a certain set of data. Although results of \cite{Sharma2018, Xing2018} show some prospects, further research is needed to extend the TL-based training method to DL applications with different types of perceptual data.

\subsection{{\color{red}Deployment and Improvement of Intelligent Edge}}





There have been many attempts to use DL to optimize and schedule resources in edge computing networks. 
In this regard, there are many potential areas where DL can be applied, including online content streaming \cite{Yoon2016}, routing and traffic control \cite{K2017}\cite{Fadlullah2017}, etc. 
However, since DL solutions do not rely entirely on accurate modeling of networks and devices, finding a scenario where DL can be applied is not the most important concern. 
Besides, if applying DL to optimize real-time edge computing networks, the training and inference of DL models or DRL algorithms may bring certain side effects, such as the additional bandwidth consumed by training data transmission and the latency of DL inference.

\begin{figure}[!!!!!!!!!!!!!!hhhhhhhhhht]
    \centering
    \includegraphics[width=7.5 cm]{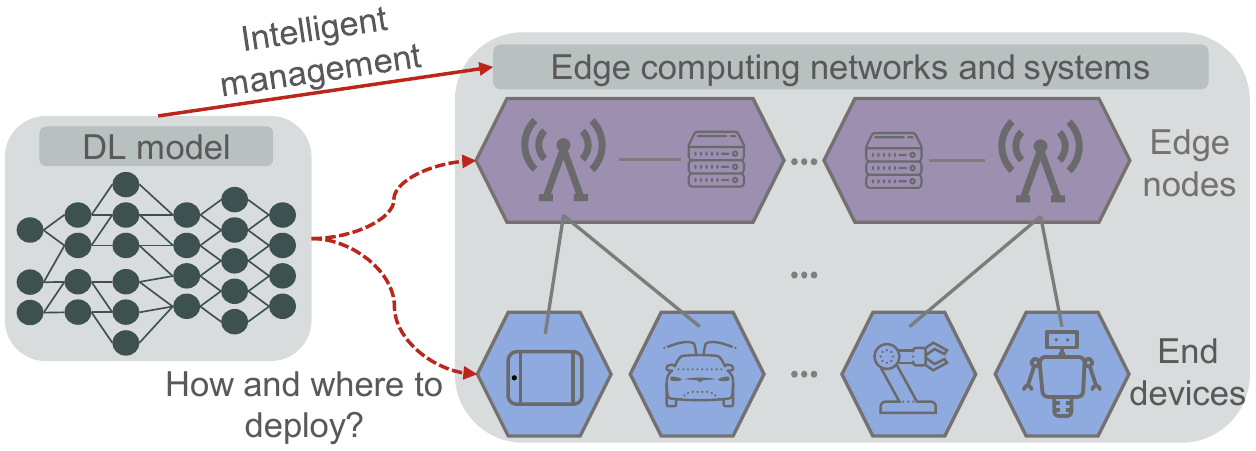}
    \caption{Deployment issues of intelligent edge, i.e., how and where to deploy DL models for optimizing edge computing networks (systems).}
    \label{fig:IEDeploy}
\end{figure}

Existing works mainly concern about solutions of ``\textit{DL for optimizing Edge}'' at the high level, but overlook the practical feasibility at the low level. 
Though DL exhibits its theoretical performance, the deployment issues of DNNs/DRL should be carefully considered (as illustrated in Fig. \ref{fig:IEDeploy}): 
\begin{itemize}
    \item Where DL and DRL should be deployed, in view of the resource overhead of them and the requirement of managing edge computing networks in real time?
    \item When using DL to determine caching policies or optimize task offloading, will the benefits of DL be neutralized by the bandwidth consumption and the processing delay brought by DL itself?
    \item How to explore and improve edge computing architectures in Section \ref{sec:EdgeforAI} to support ``\textit{DL for optimizing Edge}''?
    \item Are the ideas of customized DL models, introduced in Section \ref{sec:AIinEdge}, can help to facilitate the practical deployment?
    \item How to modify the training principles in Section \ref{sec:AIatEdge} to enhance the performance of DL training, in order to meet the timeliness of edge management?
\end{itemize}

Besides, the abilities of the state-of-the-art DL or DRL, such as Multi-Agent Deep Reinforcement Learning \cite{NIPS2016_6042, omidshafiei2017deep, lowe2017multi}, Graph Neural Networks (GNNs) \cite{zhou2018graph, zhang2018deep}, can also be exploited to facilitate this process. 
For example, end devices, edge nodes, and the cloud can be deemed as individual agents.
By this means, each agent trains its own strategy according to its local imperfect observations, and all participated agents work together for optimizing edge computing networks. 
In addition, the structure of edge computing networks across the end, the edge, and the cloud is actually an immense graph, which comprises massive latent structure information, e.g., the connection and bandwidth between devices. 
For better understanding edge computing networks, GNNs, which focuses on extracting features from graph structures instead of two-dimensional meshes and one-dimensional sequences, might be a promising method.

\section{Conclusions}
\label{sec:conclusion}

{\color{red}DL, as a key technique of artificial intelligence, and edge computing are expected} to benefit each other. 
This survey has comprehensively introduced and discussed various applicable scenarios and fundamental enabling techniques for \textit{edge intelligence} and \textit{intelligent edge}.  
In summary, the key issue of extending DL from the cloud to the edge of the network is: 
under the multiple constraints of networking, communication, computing power, and energy consumption, {\color{red}how to devise and develop} edge computing architecture to achieve the best performance of DL training and inference. 
As the computing power of the edge increases, edge intelligence will become common, and intelligent edge will play an important supporting role to improve the performance of edge intelligence. 
{\color{red}We hope that this survey will increase discussions and research efforts on DL/Edge integration that will advance future communication applications and services.}

\section*{Acknowledgement}
\label{sec:acknowledgement}

This work was supported by the National Key R\&D Program of China (No.2019YFB2101901 and No.2018YFC0809803), National Science Foundation of China (No.61702364, No.61972432 and No.U1711265), the Program for Guangdong Introducing Innovative and Enterpreneurial Teams (No.2017ZT07X355), Chinese National Engineering Laboratory for Big Data System Computing Technology and Canadian Natural Sciences and Engineering Research Council.
It was also supported in part by Singapore NRF National Satellite of Excellence, Design Science and Technology for Secure Critical Infrastructure NSoE DeST-SCI2019-0007, A*STAR-NTU-SUTD Joint Research Grant Call on Artificial Intelligence for the Future of Manufacturing RGANS1906, WASP/NTU M4082187 (4080), Singapore MOE Tier 1 2017-T1-002-007 RG122/17, MOE Tier 2 MOE2014-T2-2-015 ARC4/15, Singapore NRF2015-NRF-ISF001-2277, and Singapore EMA Energy Resilience NRF2017EWT-EP003-041.
Especially, we would like to thank the editors of IEEE COMST and the reviewers for their help and support in making this work possible.




%


\begin{IEEEbiography}
[{\includegraphics[width=1in,height=1.25in,clip,keepaspectratio]{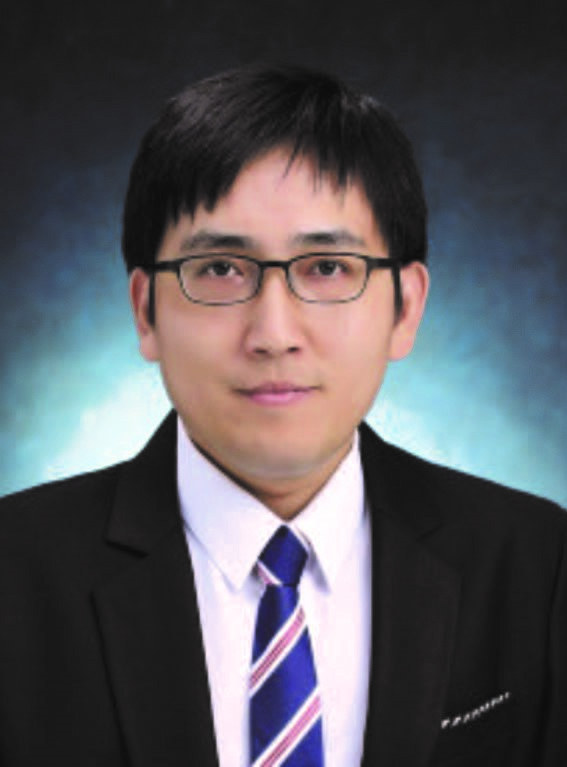}}]
    {Xiaofei Wang} [S'06, M'13, SM'18] is currently a Professor with the Tianjin Key Laboratory of Advanced Networking, School of Computer Science and Technology, Tianjin University, China. He got master and doctor degrees in Seoul National University from 2006 to 2013, and was a Post-Doctoral Fellow with The University of British Columbia from 2014 to 2016. Focusing on the research of social-aware cloud computing, cooperative cell caching, and mobile traffic offloading, he has authored over 100 technical papers in the IEEE JSAC, the IEEE TWC, the IEEE WIRELESS COMMUNICATIONS, the IEEE COMMUNICATIONS, the IEEE TMM, the IEEE INFOCOM, and the IEEE SECON. He was a recipient of the National Thousand Talents Plan (Youth) of China. He received the ``Scholarship for Excellent Foreign Students in IT Field'' by NIPA of South Korea from 2008 to 2011, the ``Global Outstanding Chinese Ph.D. Student Award'' by the Ministry of Education of China in 2012, and the Peiyang Scholar from Tianjin University. In 2017, he received the ``Fred W. Ellersick Prize'' from the IEEE Communication Society.
\end{IEEEbiography}


\begin{IEEEbiography}
[{\includegraphics[width=1in,height=1.25in,clip,keepaspectratio]{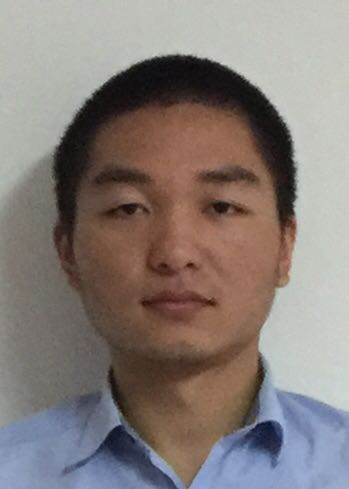}}]
{Yiwen Han} [S'18] received his B.S. degree from Nanchang University, China, and M.S. degree from Tianjin University, China, in 2015 and 2018, respectively, both in communication engineering. He received the Outstanding B.S. Graduates in 2015 and M.S. National Scholarship of China in 2016.
He is currently pursuing the Ph.D. degree in computer science at Tianjin University. His current research interests include edge computing, reinforcement learning, and deep learning.
\end{IEEEbiography}


\begin{IEEEbiography}
[{\includegraphics[width=1in,height=1.25in,clip,keepaspectratio]{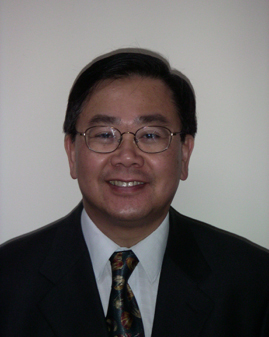}}]
{Victor C. M. Leung} [S'75, M'89, SM'97, F'03] is a Distinguished Professor of Computer Science and Software Engineering at Shenzhen University. He was a Professor of Electrical and Computer Engineering and holder of the TELUS Mobility Research Chair at the University of British Columbia (UBC) when he retired from UBC in 2018 and became a Professor Emeritus.  His research is in the broad areas of wireless networks and mobile systems. He has co-authored more than 1300 journal/conference papers and book chapters. Dr. Leung is serving on the editorial boards of the IEEE Transactions on Green Communications and Networking, IEEE Transactions on Cloud Computing, IEEE Access, IEEE Network, and several other journals. He received the IEEE Vancouver Section Centennial Award, 2011 UBC Killam Research Prize, 2017 Canadian Award for Telecommunications Research, and 2018 IEEE TCGCC Distinguished Technical Achievement Recognition Award. He co-authored papers that won the 2017 IEEE ComSoc Fred W. Ellersick Prize, 2017 IEEE Systems Journal Best Paper Award, 2018 IEEE CSIM Best Journal Paper Award, and 2019 IEEE TCGCC Best Journal Paper Award. He is a Fellow of IEEE, the Royal Society of Canada, Canadian Academy of Engineering, and Engineering Institute of Canada. He is named in the current Clarivate Analytics list of ``Highly Cited Researchers''.
\end{IEEEbiography}

\vskip -20pt plus -1fil

\begin{IEEEbiography}
[{\includegraphics[width=1in,height=1.25in,clip,keepaspectratio]{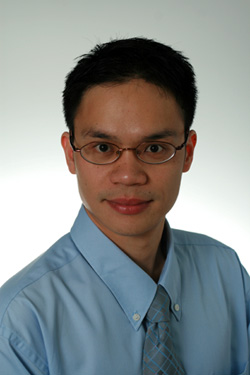}}]
{Dusit Niyato} [M'09, SM'15, F'17] is currently a Professor in the School of Computer Science and Engineering, at Nanyang Technological University,Singapore. He received B.Eng. from King Mongkuts Institute of Technology Ladkrabang (KMITL), Thailand in 1999 and Ph.D. in Electrical and Computer Engineering from the University of Manitoba, Canada in 2008. His research interests are in the area of Internet of Things (IoT) and network resource pricing.
\end{IEEEbiography}

\vskip -10pt plus -1fil

\begin{IEEEbiography}
[{\includegraphics[width=1in,height=1.25in,clip,keepaspectratio]{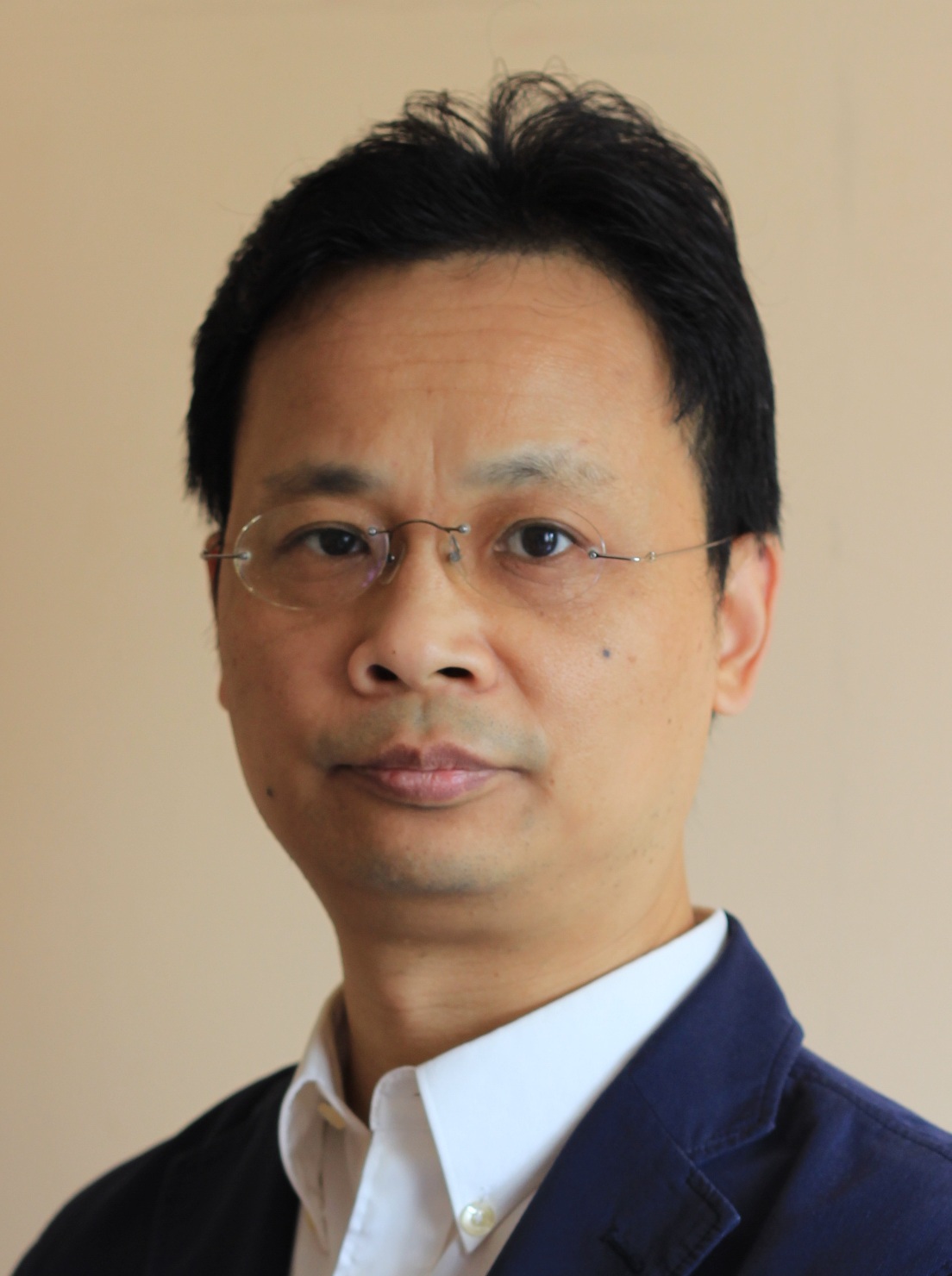}}]
{Xueqiang Yan} is currently a technology expert with Wireless Technology Lab at Huawei Technologies. He was a member of technical staff of Bell Labs from 2000 to 2004. From 2004 to 2016 he was a director of Strategy Department of Alcatel-Lucent Shanghai Bell. His current research interests include wireless networking, Internet of Things, edge AI, future mobile network architecture, network convergence and evolution.
\end{IEEEbiography}

\vskip -10pt plus -1fil

\begin{IEEEbiography}
[{\includegraphics[width=1in,height=1.25in,clip,keepaspectratio]{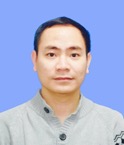}}]
{Xu Chen} [M'12] is a Full Professor with Sun Yat-sen University, Guangzhou, China, and the vice director of National and Local Joint Engineering Laboratory of Digital Home Interactive Applications. He received the Ph.D. degree in information engineering from the Chinese University of Hong Kong in 2012, and worked as a Postdoctoral Research Associate at Arizona State University, Tempe, USA from 2012 to 2014, and a Humboldt Scholar Fellow at Institute of Computer Science of University of Goettingen, Germany from 2014 to 2016. He received the prestigious Humboldt research fellowship awarded by Alexander von Humboldt Foundation of Germany, 2014 Hong Kong Young Scientist Runner-up Award, 2016 Thousand Talents Plan Award for Young Professionals of China, 2017 IEEE Communication Society Asia-Pacific Outstanding Young Researcher Award, 2017 IEEE ComSoc Young Professional Best Paper Award, Honorable Mention Award of 2010 IEEE international conference on Intelligence and Security Informatics (ISI), Best Paper Runner-up Award of 2014 IEEE International Conference on Computer Communications (INFOCOM), and Best Paper Award of 2017 IEEE Intranational Conference on Communications (ICC). He is currently an Associate Editor of IEEE Internet of Things Journal and IEEE Transactions on Wireless Communications, and Area Editor of IEEE Open Journal of the Communications Society.
\end{IEEEbiography}



\end{document}